\documentclass[longauth,20pt]{aa}
\usepackage{graphicx}
\usepackage{txfonts}
\usepackage{lipsum}
\usepackage{subcaption}         
\usepackage{lscape}            
\usepackage{placeins}          
\usepackage{amsmath}
\usepackage{threeparttable}
\usepackage{longtable}
\usepackage{gensymb}
\usepackage{mathtools}
\usepackage{isotope} 
\usepackage[title,toc]{appendix}
\usepackage{natbib}
\usepackage{hyperref}
\usepackage{amsmath}
\usepackage{graphicx}
\usepackage{capt-of}
\usepackage{longtable}
\usepackage{multirow}
\usepackage{xcolor}
\usepackage[title,toc]{appendix}
\usepackage{CJK}
\usepackage[utf8]{inputenc}   
\usepackage[T1]{fontenc} 
\usepackage[normalem]{ulem} 
\usepackage{footnotehyper}
\usepackage{textcomp}

\defcitealias{Piro_2021}{P21}
\defcitealias{Waxman2017}{SW17}
\makeatletter

\newcommand{\Rmnum}[1]{\expandafter\@slowromancap\romannumeral #1@}
\makeatother

\begin{document}

   \title{SN\,2024iss: A Double-peaked Type IIb Supernova with Evidence of Circumstellar Interaction}



            
\author{
Liyang Chen \inst{\ref{thu}\thanks{E-mail: chenly23@mails.tsinghua.edu.cn}}
\and Xiaofeng Wang \inst{\ref{thu}\thanks{E-mail: wang\_xf@mail.tsinghua.edu.cn}} 
\and Qinyu Wu \inst{\ref{NAO}, \ref{UCAS}}
\and Moira~Andrews \inst{\ref{LCO},\ref{California}}
\and Joseph~Farah \inst{\ref{LCO},\ref{California}}
\and Paolo Ochner \inst{\ref{UP},\ref{INAFa}}
\and Andrea Reguitti \inst{\ref{INAFb},\ref{INAFa}}
\and Thomas~G.~Brink \inst{\ref{Berkeley}}
\and Jujia~Zhang \inst{\ref{YOCA},\ref{ICOS},\ref{KLSEC}}
\and Cuiying~Song \inst{\ref{thu}}
\and Jialian~Liu \inst{\ref{thu}}
\and Alexei~V.~Filippenko \inst{\ref{Berkeley}}
\and David J. Sand\inst{\ref{UA}}
\and Irene~Albanese \inst{\ref{UP}}
\and Kate D. Alexander \inst{\ref{UA}}
\and Jennifer Andrews \inst{\ref{GeminiNorth}}
\and K. Azalee Bostroem \inst{\ref{UA},\ref{Catalyst}}
\and Yongzhi Cai \begin{CJK}{UTF8}{gbsn}(蔡永志)\end{CJK} \inst{\ref{YOCA},\ref{ICOS},\ref{KLSEC}}
\and Collin Christy \inst{\ref{UA}}
\and Ali Esamdin \inst{\ref{Xinjiang}, \ref{UCAS}}
\and Andrea Farina \inst{\ref{UP1}}
\and Noah Franz \inst{\ref{UA}}
\and D.~Andrew~Howell \inst{\ref{LCO},\ref{California}}
\and Brian Hsu \inst{\ref{UA}}
\and Maokai~Hu \inst{\ref{thu}}
\and Abdusamatjan Iskandar \inst{\ref{Xinjiang}, \ref{UCAS}}
\and Liping~Li \inst{\ref{YOCA},\ref{ICOS},\ref{KLSEC}}
\and Gaici~Li \inst{\ref{thu}}
\and Dongyue Li \inst{\ref{NAO}}
\and Wenxiong Li \inst{\ref{NAO}}
\and Jinzhong Liu \inst{\ref{Xinjiang}, \ref{UCAS}}
\and Curtis~McCully \inst{\ref{LCO}}
\and Megan~Newsome \inst{\ref{inst21}}
\and Yuan~Qi~Ni \inst{\ref{Kavli}}
\and Andrea Pastorello \inst{\ref{INAFa}}
\and Estefania~Padilla~Gonzalez \inst{\ref{JHU}}
\and Jeniveve Pearson\inst{\ref{UA}}
\and Haowei~Peng \inst{\ref{thu}}
\and Conor Ransome\inst{\ref{UA}}
\and Manisha Shrestha\inst{\ref{Monash},\ref{OzGrav}}
\and Nathan Smith\inst{\ref{UA}}
\and Bhagya Subrayan\inst{\ref{UA}}
\and Giacomo~Terreran \inst{\ref{AP}}
\and Giorgio Valerin \inst{\ref{INAFa}}
\and J. Vink\'o \inst{\ref{CSFK},\ref{Szeged},\ref{ELTE},\ref{inst21}}
\and Sergiy~S.Vasylyev \inst{\ref{Berkeley}}
\and Letian Wang \inst{\ref{Xinjiang}}
\and Zhenyu~Wang \inst{\ref{YOCA},\ref{UCAS}}
\and Hao Wang \inst{\ref{IHEP}}
\and J.\ Craig Wheeler \inst{\ref{inst21}}
\and Kathryn~Wynn \inst{\ref{LCO},\ref{California}}
\and Danfeng~Xiang \inst{\ref{thu},\ref{BP}}
\and Shengyu~Yan \inst{\ref{thu}}
\and Weimin Yuan \inst{\ref{NAO},\ref{UCAS}}
\and Juan Zhang \inst{\ref{IHEP}}
\and WeiKang Zheng \inst{\ref{Berkeley}}
\and Yu Zhang \inst{\ref{Xinjiang}, \ref{UCAS}}
}

\institute
{Department of Physics, Tsinghua University, Beijing, 100084, China \label{thu}
\and National Astronomical Observatories, Chinese Academy of Sciences, Beijing 100101, China\label{NAO}
\and School of Astronomy and Space Science, University of Chinese Academy of Sciences, Beijing 100049,1408, People's Republic of China \label{UCAS}
\and Las Cumbres Observatory \label{LCO}
\and University of California, Santa Barbara, Department of Physics \label{California}
\and Universit\`a degli Studi di Padova, Dipartimento di Fisica e Astronomia, Vicolo dell'Osservatorio 2, 35122 Padova, Italy \label{UP}
\and INAF - Osservatorio Astronomico di Padova, Vicolo dell'Osservatorio 5, 35122 Padova, Italy \label{INAFa}
\and INAF - Osservatorio Astronomico di Brera, Via Bianchi 46, 23807 Merate (LC), Italy \label{INAFb}
\and Department of Astronomy, University of California, Berkeley, CA 94720-3411, USA \label{Berkeley}
\and Yunnan Observatories, Chinese Academy of Sciences, Kunming 650216, China 10 \label{YOCA}
\and International Centre of Supernovae, Yunnan Key Laboratory, Kunming 650216, China 11 \label{ICOS}
\and Key Laboratory for the Structure and Evolution of Celestial Objects, Chinese Academy of Sciences, Kunming 650216, China 12 \label{KLSEC}
\and Steward Observatory, University of Arizona, 933 North Cherry Avenue, Tucson, AZ 85721-0065, USA\label{UA}
\and Gemini Observatory, 670 North A`ohoku Place, Hilo, HI 96720-2700, USA\label{GeminiNorth}
\and LSSTC Catalyst Fellow\label{Catalyst}
\and Xinjiang Astronomical Observatory, Chinese Academy of Sciences, Urumqi, Xinjiang, 830011, China\label{Xinjiang}
\and Universit\`a degli Studi di Padova, Dipartimento di Fisica e Astronomia, Vicolo dell'Osservatorio 3, 35122 Padova, Italy \label{UP1}
\and Department of Astronomy, University of Texas, Austin, TX, 78712, USA\label{inst21}
\and Kavli Institute for Theoretical Physics \label{Kavli}
\and Johns Hopkins University \label{JHU}
\and School of Physics and Astronomy, Monash University, Clayton, Australia\label{Monash}
\and OzGrav: The ARC Center of Excellence for Gravitational Wave Discovery, Australia\label{OzGrav}
\and Adler Planetarium \label{AP}
\and HUN-REN CSFK Konkoly Observatory, MTA Center of Excellence, Konkoly Thege ut 15-17, Budapest, 1121, Hungary \label{CSFK}
\and Department of Experimental Physics, University of Szeged, D\'om t\'er 9, Szeged, 6720, Hungary \label{Szeged}
\and ELTE E\"otv\"os Lor\'and University, Institute of Physics and Astronomy, P\'azm\'any P\'eter s\'et\'any 1/A, Budapest, 1117 Hungary \label{ELTE}
\and Institute of High Energy Physics, Chinese Academy of Sciences, Beijing, 100049, China\label{IHEP}
\and Department of Scientific Research, Beijing Planetarium, Beijing 100044, China \label{BP}
}

\date{October 27, 2025}

\abstract
   {}
{We present optical, ultraviolet, and X-ray observations of supernova (SN) 2024iss, a Type IIb SN that shows a prominent double-peaked light curve. 
    } 
{We modeled the first peak with a semianalytical shock-cooling model and the X-ray emission with a free-free model. We also compare the envelope radius and mass-loss rate with those of other Type IIb SNe to explore the relationships between the progenitor envelope and the circumstellar material (CSM).
   }
   {The shock-cooling peak in the $V$-band light curve reached $M_V = -17.33 \pm 0.26$ mag, while the $^{56}$Ni-powered second peak attained $M_V = -17.43 \pm 0.26$ mag. Early spectra show an photospheric velocity of $\sim 19,400$\,km\,s$^{-1}$ at 3.82 days from the H$\alpha$ P~Cygni profile. The Balmer lines persist at least +87 days after the explosion, characterizing hydrogen-rich ejecta. Modeling the first light-curve peak with the shock-cooling model suggests an extended hydrogen envelope with a mass of $0.11 \pm0.04\ M_{\odot}$ and a radius of $244 \pm 43~R_{\odot}$. Fitting the second light-curve peak with an Arnett-like model indicates a typical $^{56}$Ni mass of $ 0.117 \pm 0.013~M_{\odot}$ and a relatively low ejecta mass of $1.272 \pm 0.343\,M_{\odot}$. X-ray observations reveal bright thermal bremsstrahlung emission and indicate a mass-loss rate of $1.6 \times 10^{-5} \ M_{\odot} \ \rm{yr}^{-1}$, which is similar to that of SN\,1993J.
   }
   {SN\,2024iss occupies a transitional position between the two subclasses of extended (eIIb) and compact (cIIb) Type IIb SNe. Its envelope radius and pre-explosion mass-loss rate appear to be correlated, in agreement with theoretical predictions. The observational properties of SN\,2024iss are compatible with a binary interaction scenario being the dominant mechanism for envelope stripping. Furthermore, the low column density of neutral hydrogen suggests a compact CSM with an outer radius of $\lesssim 1.3 \times 10^{14}$ cm, indicating that the progenitor star experienced  eruptive mass loss within $\sim 4$\,yr of its terminal explosion.
   }


   \maketitle

\nolinenumbers
\section{Introduction}

Type IIb supernovae (SNe) are generally thought to form a transitional subclass whose spectrophotometic properties lie between those of the Type II and Ib SNe \citep{Filippenko1997,Gal_Yam_2017}. The spectra of Type IIb SNe initially exhibit strong hydrogen features, with no evidence of helium. The hydrogen lines persist only for the first few weeks, which is much shorter compared to that observed in Type II/IIP SNe. P~Cygni features of He{\sc\,I} appear in the spectra at about two weeks after explosion, while hydrogen lines become significantly weakening at the same time (see, e.g.,~\citealp{Filippenko1997}).

The rather prompt emergence of helium lines indicates that the outermost envelope of the progenitors of Type IIb SNe has been partially stripped, thus providing an important approach to test the mechanisms that can dramatically strip the outer envelopes of massive progenitor stars. The remarkable transformation between the hydrogen- and helium-dominated spectroscopic phases was initially revealed by the extensive observing campaign on SN\,1993J~\citep{1993ApJ...415L.103F, Filippenko1994, Woosley1994, Barbon1995, Richmond1996, Matheson2000}. Thanks to the transient-alert stream produced by high-cadence wide field sky surveys such as the Palomar Transient Factory (PTF,~\citealp{2009PASP..121.1334R, 2009PASP..121.1395L}) and later the Zwicky Transient Facilities (ZTF,~\citealp{2019PASP..131a8002B, 2019PASP..131g8001G}), large samples of SNe IIb have been studied, including SNe\,2011dh~\citep{Arcavi2011,Ergon_2014}, 2011fu~\citep{Kumar2013,Morales_Garoffolo_2015}, 2013df~\citep{Morales2014,Szalai2016}, 2016gkg~\citep{Tartaglia2017,Bersten_2018,Sravan2018}, 2020acat~\citep{Medler2022} and 2024uwq~\citep{Subrayan2025}. Detailed analyses of the homogeneity and diversity of SNe IIb have also been preformed based on a larger sample~\citep[e.g.,][]{Shivvers2019}.

One way to investigate the nature of Type IIb SNe can be facilitated by analyzing the archival pre-explosion images. By fitting stellar evolutionary tracks to the spectral energy distribution (SED) extracted at the exact location of the SN on the archival images taken by the {\it Hubble Space Telescope} ({\it HST}), various studies suggest that the progenitor star of SNe IIb tend to be yellow (or even cooler) supergiants, for example, SNe\,1993J~\citep{Maund2004}, 2013df~\citep{Van_Dyk_2014}, 2016gkg~\citep{Kilpatrick2017,Tartaglia2017}, 2017gkk~\citep{Niu_2024} and 2024abfo~\citep{reguitti2025, Niu_2025}.

A more general approach to investigate the nature of the progenitor can be achieved by fitting the early-time emissions of the SN explosion with a shock-cooling model. This process characterizes the energy emitted by the shock breakout deposited in the ejecta and describes how such energy diffuses out as the ejecta expands and cools (see, e.g., \citealp{2017hsn..book..967W} for a review). The early luminosity evolution of SNe IIb often manifests itself as a double-peaked light curve, where the first peak is powered by the cooling of the extended envelope of the progenitor star heated by the shock, namely the shock-cooling emission (SCE,~\citealp{Soderberg_2012}). Various semi-analytic models have been developed within the framework of shock cooling and were used to fit the early light curves of core-collapse (CC) SNe. 

\citet{Piro_2021} introduced a two-zone model with a broken power law to characterize the radial density structure of the envelope. This model builds on the earlier works of \citet{Piro_2015} and \citet{Nakar2014}. Meanwhile, \citet{Waxman2017} extended the earlier work of \citet{Rabinak2011} to a longer regime of validity, introducing two choices of the polytropic index: $n=3$ for radiative envelopes (e.g., BSGs) and $n=3/2$ for convective envelopes (e.g., RSGs). \citet{Morag2023} further developed the framework established in~\citet{Waxman2017} by considering the effects of strong blanketing of numerous iron absorption lines in the near ultraviolet (UV) to optical wavelength ranges
and the multizone dynamics, increasing the feasibility of the model at early phases. \citet{Farah_2025} developed a framework for shock-cooling fitting of SNe IIb, employing various models to constrain the envelope mass and radius.
Moreover, \citet{Nagy_2016} developed a semi-analytical model to constrain the nature of the envelope, by considering a dense inner region and an extended low-mass envelope. 

These models of the early luminosity evolution of SNe IIb generally suggest that the extended hydrogen envelope associated with their first light curve peak have masses of $10^{-2}$-$10^{-1} M_\odot$ and radii of 100$-$500\,$R_\odot$, which are both broadly consistent with the results of pre-explosion imaging, although some discrepancies remain unresolved. Alternative interpretations of the early light curve peak have also been proposed, including a bimodal radial distribution of \isotope[56]{Ni}~\citep{Orellana2022} and thermal energy deposited by the shock as it propagates through the envelope \citep{liu2025}. 

Owing to pre-explosion mass loss and partial envelope stripping, some SNe IIb also exhibit signatures of circumstellar material (CSM), offering valuable insights into pre-explosion mass-loss history of their progenitor systems \citep{Smith_2017}.

X-ray emission provides another important trace of the ejecta-CSM interaction. \citet{Fransson1996} analyzed the X-ray emission of SN\,1993J and attributed it to the thermal bremsstrahlung produced 
by the forward shock. This framework was extended in studies of SNe\,2011dh \citep{Soderberg2012} and 2013df \citep{Kamble2016}, further supporting the role of X-ray observations in providing critical constraints on the electron number density of the CSM around Type IIb SNe. \citet{Vikram2025} compiled the early X-ray light curves of interacting CCSNe and found no compelling evidence for non-thermal X-ray emission in SNe IIb. 
 
The mechanism stripping the envelopes of the progenitor star remains uncertain. The removal
of the progenitor's envelope and the formation of circumstellar material (CSM) generally result from two categories: single-star or binary interaction driven mechanisms. In the framework of wind-driven mass loss (e.g., \citealp{Woosley1993, Georgy2012, Yoon2017a}), the stripping of a star can be rather intense and may indicate a massive progenitor such as a Wolf-Rayet star, as seen in events like the first hypernova \citep{Hamuy2009}. Additionally, eruptive mass loss just prior to the terminal explosion may be triggered by processes including pulsation-driven superwinds \citep{Yoon_2010} or core neutrino emission \citep{Moriya2014}.

Alternatively, binary interaction has gained increasing support in recent years as a more prevalent explanation for envelope stripping \citep{Podsiadlowski_1992, Smith2014, Sravan_2020, Dessart2024}. \citet{Sana2012} indicates that more than 70\% massive stars undergo mass exchange with a companion, with about one-third resulting in binary mergers. Direct imaging has confirmed companion stars in some Type IIb progenitors, such as those of SN 1993J \citep{Maund2009} and SN\,2001ig \citep{Ryder_2018}. Moreover, binary scenarios naturally explain the observed correlations between CSM properties and the residual envelope of the progenitor.

\citet{Chevalier_2010} were the first to classify SNe IIb into two subtypes based on the prominence of their shock-cooling emission, and investigated their associated X-ray properties. The extended subtype (Type eIIb) exhibits a pronounced shock-cooling peak in the early light curve, indicating a relatively massive hydrogen envelope ($\geq$0.1$M_{\odot}$) and a large progenitor radius ($\sim$100--1000\,$R_{\odot}$). These events typically show thermal X-ray emission. In contrast, objects belonging to the compact subtype (Type cIIb) display a weaker or no signature of shock cooling, indicating smaller sizes of their progenitors.

Following this classification, \citet{Soderberg_2012} and \citet{Kamble2016} found that Type eIIb SNe generally exhibit stronger X-ray emission and are surrounded by denser CSM. The distinction between Type eIIb and Type cIIb is further supported by \citet{Maeda2015} and \citet{Ouchi_2017}, which show that the initial orbital separation affects the residual envelope mass and pre-explosion mass loss. In this framework, closer binaries tend to produce more stripped, compact progenitors with lower mass-loss rates. A parallel study by \citet{Yoon_2017} proposed that Type eIIb SNe originate from systems undergoing late Case~B mass transfer and evolve into red supergiants (RSGs), while Type cIIb events would result from early Case B transfer and evolve into blue or yellow supergiants (BSGs or YSGs). Nevertheless, the sample size of SNe IIb is still limited, and events with comprehensive optical and X-ray observations are even sparse.  Further studies of the relationship between the envelope and the CSM will require a larger, well-characterized sample.

In this article, we present and analyze the spectrophotometric and X-ray observations of SN\,2024iss,
a nearby type IIb SN with a prominent double peak in optical bands, and it exhibits bright thermal X-ray emission comparable with SN\,1993J. Basic information about SN\,2024iss is provided in Section \ref{sec:basic_information}. The details of the observations and data reduction are described in Section \ref{sec:data}. In Section \ref{sec:optical_results}, we analyze the multiband light curves and the evolution of the bolometric luminosity. The spectral evolution and comparisons with other well-observed SNe IIb are discussed in Section \ref{sec:spectroscopy}. In Section \ref{sec:shock_cooling_model}, we model the early shock-cooling emission peak. The X-ray properties are inferred in Section \ref{sec:X_ray}. In Section \ref{sec:Discussion}, we discuss possible mass-loss mechanisms for SN\,2024iss. Finally, the main conclusions are summarized in Section \ref{sec:Conclusion}.


\section{Observations and Data Reduction} \label{sec:data}

\subsection{Discovery and Host Galaxy} \label{sec:basic_information}

SN\,2024iss was discovered on UT 2024 May 12 21:37 (MJD\,60442.901) by the Gravitational-wave Optical Transient Observer \citep[GOTO;][]{Steeghs_2022}. 
Approximately six hours before the reported discovery, the SN emission was recorded by a public outreach telescope located at the Xinglong Observatory, as detailed in Section~\ref{sec:data}.
The SN exploded at celestial coordinates $\rm \alpha = 12^{h}59^{m}06\farcs{130}$, $\rm \delta = +28^{\circ}48'42\farcs{62}$, in the outskirts of the nearby dwarf galaxy WISEA~J125906.48+284842.6 (redshift $z = 0.003334$).

Assuming a flat $\Lambda$ cold dark matter ($\Lambda$CDM) model with $H_0 = 73.0~\rm{km\,s^{-1}\,Mpc^{-1}}$, $\Omega_{\rm{m}} = 0.27$, and $\Omega_{\Lambda} = 0.73$ \citep{Spergel_2007}, we adopt a distance of $13.4 \pm 1.6$~Mpc, based on the Local Group velocity \citep{Fixsen_1996}, as inquired from the NASA/IPAC Extragalactic Database (NED).
 
Considering the low redshift of SN\,2024iss, we adopt an additional 10\% uncertainty in distance to account for the effect of the peculiar velocity. We add this systematic uncertainty and the error in the Local Group distance in quadrature to obtain the final uncertainty in the distance to SN\,2024iss.
This yields a distance modulus of $30.64 \pm 0.26$~mag, which is used throughout this paper.

The Galactic reddening toward the line sight of SN\,2024iss was estimated as $ \rm{E(B{-}V)} = 0.0084$\,mag based on the extinction map derived by \citet{Schlafly_2011}. 
We neglect the reddening of the host galaxy, as no strong Na{\sc\,I} D absorption features can be identified at the redshift of the host (see Section~\ref{sec:spectroscopy}). 
Adopting the $R_{V}=3.1$ extinction law \citep{Cardelli_1989}, this results in an extinction of A$_{V}$ = 0.026 mag.

The last non-detection was reported by the All-Sky Automated Survey for Supernovae \citep[ASAS-SN;][]{Kochanek2017} with a $g$-band limiting magnitude of 18.24~mag, on 2024 May 11 18:29 (MJD 60441.770),
$\sim$ 0.87~days before the first detection in Xinglong (MJD = 60442.645). Throughout this paper, all phases are given relative to the estimated explosion time of SN\,2024iss, defined as the midpoint between the last non-detection and the first detection, i.e., MJD~$60442.208 \pm 0.438$. A more detailed estimation of $t_{0}$ by fitting a shock-cooling model to the early light curves of SN\,2024iss will be presented in Section~\ref{sec:shock_cooling_model}. 
Basic observational properties of SN\,2024iss are summarized in Table \ref{table:basic property}.

\begin{table}
\caption{Basic properties of SN\,2024iss}\label{table:basic property}
\centering
\begin{tabular}{l c}
\cline{1-2}
\textbf{ } & \textbf{ } \\
\hline
Host galaxy & WISEA J125906.48+284842.6 \\
RA (J2000) & $12^{h}59^{m}06.^{s}130$ \\
DEC (J2000) & $+28^{\circ}48'42.''62$ \\
Distance & 13.4 $\pm$ 1.6 Mpc \\
Distance modulus & 30.64 $\pm$ 0.26 mag \\
Redshift & 0.003334 \\
E(B-V)$_{\text{MW}}$ & 0.0084 $\pm$ 0.0002 mag \\
E(B-V)$_{\text{host}}$ & 0.0 \\
Explosion time & MJD 60442.21 $\pm$ 0.44 (2024/05/12) \\
V$_{\text{max}}$ & MJD 60460.63 (2024/05/30) \\
\hline
\end{tabular}
\end{table}

\begin{figure*}
    \centering
    \includegraphics[width=\textwidth]{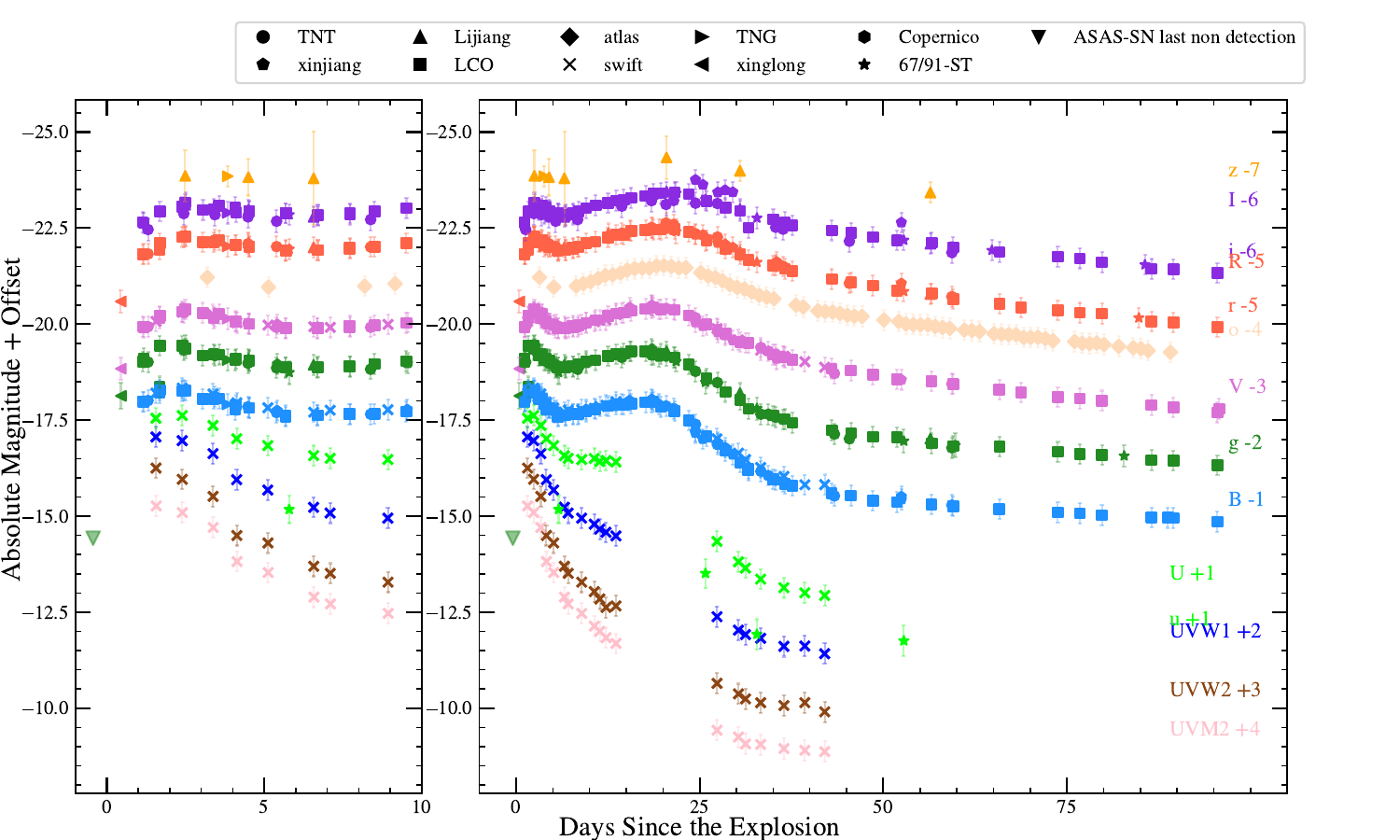}
    \caption{
    Optical light curves of SN\,2024iss. Left panel: the early-time multiband photometry within the first ten days. 
    Right panel: the multiband light curves of SN\,2024iss within the first $\sim$100 days. All magnitudes were corrected for the Galactic extinction. For better display, light curves in different bandpasses were shifted vertically by arbitrary numbers as labeled on the right. Facilities used to obtain the photometry are indicated by the legend. The last non-detection from ASAS-SN is marked by the green inverted triangle.
    }
    \label{fig:photometry_all}  
\end{figure*}

\subsection{Optical Photometry}  

The observing campaign on SN\,2024iss started at approximately six hours before the reported discovery. Follow-up photometry was obtained by the 0.8~m Tsinghua University-NAOC telescope \citep[TNT;][]{Huang_2012} at Xinglong Station, the Lijiang 2.4 m telescope \citep[LJT;][]{Fan_2015} of Yunnan Astronomical Observatories, the Nanshan One-meter Wide-field Telescope \citep[NOWT;][]{Bai2020}, the 0.36~m reflector of the Tsinghua University (SNOVA) at Nanshan Station of Xinjiang Astronomical Observatory, the Schmidt 67/91 cm Telescope (67/91-ST), the Copernico 1.82~m telescope at the Asiago Astrophysical Observatory, and the 3.58~m Telescopio Nazionale Galileo \citep[TNG;][]{Barbieri1994} on the island of La Palma and the Las Cumbres Observatory's global network of robotic telescopes \citep[LCO;][]{Brown2013} through the Global Supernova Project.

Optical images were pre-processed with standard procedures, which include bias subtraction, flat-field correction and cosmic-ray removal. We used a custom ZURTYPHOT pipeline (Mo et al. in prep.)
to reduce the TNT images. The AUTOmated Photometry Of Transients (AutoPhOT) pipeline \citep{Brennan2022} was implemented to extract the photometry from the images obtained by the XL\_106, LJT, 67/91-ST, Copernico and TNG. As XL\_106 is equipped with $RGB$ filters, we calibrate the $R$-, $G$-, and $B$-band photometry to the $r$-, $V$-, and $g$-band under the Sloan Digital Sky Survey (SDSS) photometric system~\citep{Fukugita1996} in AB magnitudes~\citep{1983ApJ...266..713O}, respectively, following the prescriptions in \citet{Li2024}. Considering SN\,2024iss exploded in the outskirts of a faint dwarf galaxy, template subtraction is not necessary.
The instrumental magnitudes of SN\,2024iss were calibrated using the AAVSO Photometric All Sky Survey (APASS) DR9 Catalogue \citep{Henden2016}. The final $BV-$ and $gri-$ band magnitudes of SN\,2024iss were calibrated against local bright comparison stars, and were transformed to the standard Johnson \citep[Vega magnitude;][]{Johnson1966} and SDSS \citep[AB magnitude;][]{Fukugita1996} systems, respectively.
We also include the near-infrared (NIR) band photometry from \citet{Yamanaka2025} for bolometric luminosity calculation in Section~\ref{sec:bolometric_luminosity}. 

SN\,2024iss was also observed with the UVOT \citep{Roming2005} board on the Neil Gehrels Swift Observatory \citep[Swift;][]{Gehrels2004}, using a $5^{\prime\prime}$ aperture to measure magnitudes in $uvw1$, $uvw2$, $uvm2$, $u$, $b$, and $v$ band-passes. Photometric data were extracted with HEASOFT\footnote{https:// www.swift.ac.uk/analysis/software.php} and the latest Swift calibration database\footnote{https://heasarc.gsfc.nasa.gov/docs/heasarc/caldb/swift/} \citep{Breeveld2011AIPC.1358..373B}. In addition, high-cadence $o$-band photometry from ATLAS \citep{Tonry2018} was included in our dataset. We list the photometry of SN\,2024iss in Table \ref{table:photometric observation}. 
Figure~\ref{fig:photometry_all} shows the multiband light curves of SN\,2024iss.

\begin{table}
\caption{Photometric Observations of SN\,2024iss}\label{table:photometric observation}
\centering
\begin{tabular}{cccccc}
\hline
MJD\tablefootmark{a} & $\rm{Phase}$\tablefootmark{b} & Filter & Mag. & Mag. error & Instrument \\
\hline
60442.65 & 0.44  & r & 15.07 & 0.12 & XL\_106\\
60442.65 & 0.44  & V & 14.83 & 0.11 & XL\_106\\
60442.65 & 0.44  & g & 14.54 & 0.22 & XL\_106\\
60443.35 & 1.14  & B & 13.69 & 0.02 & LCO\\
60443.35 & 1.14  & V & 13.73 & 0.02 & LCO\\
60443.35 & 1.14  & g & 13.65 & 0.01 & LCO\\
60443.35 & 1.14  & r & 13.84 & 0.01 & LCO\\
60443.35 & 1.14  & i & 14.0 & 0.02 & LCO\\
... & ...  & ... & ... & ... & ...\\
\hline
\end{tabular}

\tablefoot{
\tablefoottext{a}{This table is available in its entirety in machine-readable form.}\\
\tablefoottext{b}{Time respect to MJD=60,442.21.}
}
\end{table}

\subsection{Optical Spectroscopy} 

Our spectral sequence of SN\,2024iss consists of 51 spectra, spanning the phase from $t \approx 1$ days to 87 days after the explosion.
The data set is obtained by various facilities including the Beijing Faint Object Spectrograph and Camera (BFOSC) on the Xinglong 2.16~m telescope \citep[XLT;][], the Kast double spectrograph \citep{miller_stone_1994} mounted on the Shane 3~m telescope at Lick Observatory, the Yunnan Faint Object Spectrograph and Camera (YFOSC) on the Lijiang 2.4~m telescope \citep[LJT;][]{Wang2019}, the FLOYDS spectrographs \citep{Brown2013} on the 2~m Faulkes Telescopes South and North (FTS and FTN), the Gemini Multi-Object Spectrograph \citep[GMOS;][]{Hook2004} on the Gemini North Telescope (GNT), the B\&C spectrograph on the 1.22~m Galileo Telescope (GT), the DOLORES (Device Optimised for the LOw RESolution) on the 3.5~m Telescopio Nazionale Galileo (TNG), the Low Resolution Spectrograph 2 (LRS2; \citealt{chonis_lrs2:_2014}) mounted on the Hobby-Eberly Telescope (HET; \citealt{1998SPIE.3352...34R}) located at McDonald Observatory, Binospec \citep{Fabricant_2019} at the MMT Observatory, the Boller and Chivens Spectrograph (B\&C) on the University of Arizona's Bok 2.3~m telescope located at Kitt Peak Observatory, and the Multi-Object Double Spectrographs \citep[MODS;][]{Pogge2010} on the Large Binocular Telescope (LBT) located on Mt. Graham, Arizona USA. The observation log of optical spectroscopy is presented in Table~\ref{table:Log of spectra}.

All spectra were processed following routines within IRAF \citep{Tody1986, Tody1993}, custom Python and IDL codes\footnote{https://github.com/ishivvers/TheKastShiv}, including bias subtraction, flat-field correction, and cosmic-ray removal. The FLOYDS pipeline \citep{Valenti2014} were used to reduce the FLOYDS spectra. Spectra taken with GNT + GMOS are reduced with the Data Reduction for Astronomy from Gemini Observatory North and South \citep[DRAGONS;][]{Labrie2019} package. The HET spectra was reduced by the Panacea pipeline\footnote{\url{ https://github.com/grzeimann/Panacea}}. Spectra taken with the Binospec on MMT are reduced using the Python-based package Pypeit \citep[v1.17.4;][]{pypeit}, while those taken with the MODS are bias and flat-field corrected using the modsCCDred package \citep{Pogge2019}, then they are extracted and flux-calibrated with the standard IRAF routines. Wavelengths were calibrated using Fe/Ar or Fe/Ne lamp spectra taken on the same night. Flux calibration was achieved using standard stars observed at similar airmasses. We also remove the telluric features by scaling a mean telluric spectrum constructed from multiple observations of telluric standard stars to match the SN spectrum.

\subsection{X-ray Observations}  
\label{sec:X-ray_Observations}

\begin{figure*}
    \centering
    \includegraphics[width=\textwidth]{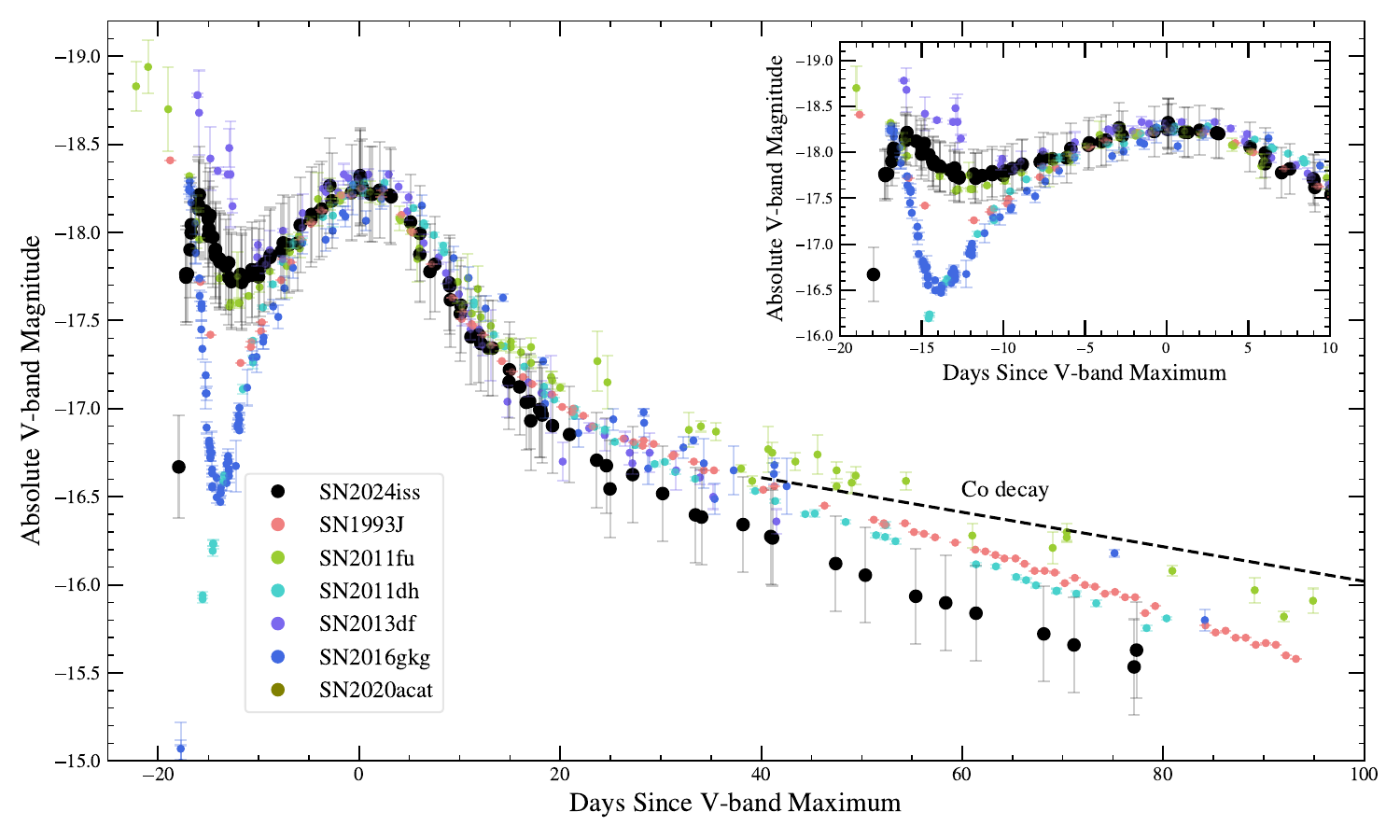}
    \caption{The $V$-band light curve of SN\,2024iss compared to those of well-sampled cases, i.e., the Type IIb SNe\,1993J, 2011dh, 2013df, 2016gkg and 2020acat. The light curves of the comparison SNe have been shifted in both magnitude and time to align with the peak magnitude and the time of the $V$-band light curve peak of SN\,2024iss. The black dashed line represents the expected decline rate of \isotope[56]{Co}~\citep[0.98 mag/100 day,][]{Woosley1989}. The upper-right inset shows a zoom-in of the first 30 days.
    }
    \label{fig:photometry_comparision}
\end{figure*}

Immediately after the explosion of SN\,2024iss, rapid follow-up observations in soft X-ray band were carried out by the Follow-up X-ray Telescope \citep[FXT;][]{2020SPIE11444E..5BC} on the Einstein Probe (EP) satellite \citep{Yuan2022}, the X-ray Telescope \citep[XRT;][]{Burrows2005} on the $Swift$ satellite \citep{Gehrels2004}, and the Nuclear Spectroscopic Telescope Array \citep[NuSTAR;][]{2013ApJ...770..103H}. 
Note that a prominent signal was detected by EP-FXT during its first epoch observations (corresponding to $t \approx +3$ days after the explosion). 
Integration of SN\,2024iss by Swift-XRT was started approximately two days after the explosion. A total of 18 observations were conducted spanning the phases from days +2 to +34.
We detected prominent signals in the first two time intervals, spanning from days +1.5 to +3.7 and +4.1 to +7.1, respectively.
Rebinning the Swift XRT integration to the third time interval centered at day +11 reveals a rapid decline of the X-ray luminosity.
We do not detect any significant signal during the fourth time interval centered at day +35 after the explosion.
Subsequent Swift observations conducted at $t \approx 207$ and 277 days did not detect any signals at the location of SN\,2024iss.
A log of the X-ray observations is provided in Table~\ref{table:Log of X-ray observation}.

The FXT data were reduced with the FXT data-analysis software (fxtsoftware v1.10)\footnote{http://epfxt.ihep.ac.cn/analysis} to produce high-level science products. Swift/XRT data were processed with the online XRT product builder\footnote{https://www.swift.ac.uk/user$\_$objects} \citep{Evans2007,Evans2009}. NuSTAR observations were analyzed with the standard NuSTAR data-analysis software (NuSTARDAS v2.1.2) to extract science products. We fit the energy spectrum with the tbabs*apec model\footnote{https://heasarc.gsfc.nasa.gov/docs/xanadu/xspec/index.html}. 
The tbabs model computes the total cross section of the gas particles along the source-Earth line of sight that induce the X-ray absorption, including the Galactic, the host, and the matter surrounding the emitting source.
The apec model calculation returns the emission spectrum produced by the collisionally-ionized diffuse gas. The best-fit model parameters of X-ray spectrum are tabulated in Table~\ref{table:Log of X-ray spectral parameter}. Due to the limited photon counts and the restricted energy coverage of FXT (0.5-10 keV), the plasma temperature ($kT$) of SN\,2024iss at each epoch cannot be reliably constrained independently. We therefore adopt a fixed plasma temperature to the temperature estimated
from the second NuSTAR observation for all other observations. 
The neutral hydrogen column density from the Milky Way towards the direction of SN\,2024iss is $7.5 \times 10^{19}$ cm$^{-2}$ \citep{Foight2016}, as derived assuming it is linearly correlated with the Galactic reddening component, namely E(B-V) = 0.0084 mag.

\section{Optical Photometry 
} \label{sec:optical_results}

\subsection{Photometric Evolution}  
Figure~\ref{fig:photometry_all} shows the UV and optical light curves of SN\,2024iss. 
We divide the temporal evolution of the photometry into two phases, namely the early shock-cooling (from days 0 to 10 after the SN explosion) and the later radioactive decay phases.
SN\,2024iss was discovered very shortly after the explosion, following the serendipitous $gVr$-band images obtained on day 0.44, we continued the high-cadence multicolor photometric followup campaign on SN\,2024iss and obtained the second epoch of observation on day 1.3.
SN\,2024iss reached the first V-band peak of $-17.33 \pm 0.26$ mag on about day\,2.4. 
This evolution is consistent with the cooling of a hot blackbody, where the peak of the emission shifts progressively to longer wavelengths as the temperature decreases. 

At $t\approx 5$ days after the explosion, as photons escape from the expanding shock heated envelope, the SN luminosity becomes progressively dominated by the radioactive decay of \isotope[56]{Ni}, which powered the second peak of SN\,2024iss. After correcting for extinction, the peak of the $V$-band light curve ($M_{V}=-17.43 \pm 0.26$) was reached on day 18.7. 
At $t \approx 50$ days after the explosion, the light curves of SN\,2024iss in all observed bandpasses enter a linear decline phase, with a characteristic $V$-band decline rate of $\sim$2.0\, mag 100$^{-1}$~day. As seen in other SNe IIb, SN\,2024iss declines faster than the radioactive decay rate of \isotope[56]{Co} $\rightarrow$ \isotope[56]{Fe}, 0.98\, mag/100~days.

\begin{table*}
\caption{Properties of some typical Type IIb SNe Compared with SN\,2024iss}
\label{table:comparison_info}
\centering
\begin{tabular}{cccccccc}
\hline
SN & Explosion Time & Time of the Second Peak & Redshift & $\rm{m} - \rm{M}$
 & $E(B-V)_{MW}$ & $E(B-V)_{Host}$ & Source\tablefootmark{a}\\
& (MJD) & (MJD) & & (mag) & (mag) & (mag)\\
\hline
1993J & 49074.0 & 49095.0 & -0.001130 & 27.31  & 0.0690 & 0.110 & 1,2,3 \\
2011dh & 55712.0 & 55732.0 & 0.001638 & 29.46  & 0.0310 & 0.040 & 4\\
2011fu & 55825.0 & 55846.9 & 0.001845 & 34.36  & 0.0680 & 0.150 & 5,6\\
2013df & 56450.0 & 56469.0 & 0.002390 & 31.10  & 0.0170 & 0.081 & 7,8\\
2016gkg & 57651.2 & 57669.0 & 0.004900 & 32.11  & 0.0166 & 0.090 & 9\\
2020acat & 59192.0 & 59208.7 & 0.007932 & 32.74  & 0.0207 & $-$ & 10 \\
\hline
\end{tabular}

\tablefoot{
\tablefoottext{a}{(1) \citet{Richmond1994}; (2) \citet{Barbon1995}; (3) \citet{Richmond1996}; (4) \citet{Ergon_2014}; (5) \citet{Kumar2013}; (6) \citet{Morales_Garoffolo_2015}; (7) \citet{Van_Dyk_2014}; (8) \citet{Morales2014}; (9) \citet{Sravan2018}; (10) \citet{Medler2022}. }

}

\end{table*}

Figure~\ref{fig:photometry_comparision} compares the absolute $V$-band light curve of SN\,2024iss with a sample of well-observed SNe IIb, including SNe\,1993J \citep{Richmond1996}, 2011fu \citep{Morales_Garoffolo_2015}, 2011dh \citep{Ergon_2014}, 2013df \citep{Morales2014}, 2016gkg \citep{Arcavi_2017,Bersten_2018} and 2020acat \citep{Medler2022}. Basic observational properties of this SN sample are provided in Table \ref{table:comparison_info}. Considering the relatively large uncertainties in the estimation of the distance modulus among various cases, we arbitrarily shifted the light curve to match the peak of the $V$-band magnitude of SN\,2024iss, and aimed to compare the morphology of the light curves of different SNe IIb.

SN\,2024iss shows a prominent first peak, which is similar to that of SNe\,1993J, 2011fu, 2013df and 2016gkg, but distinct from that of SNe\,2011dh and 2020acat.  According to \citet{Chevalier_2010}, SNe IIb can be further classified into two subtypes, namely the extended-envelope (eIIb) and compact-envelope (cIIb) events. The prominent early peak of SN\,2024iss indicates its extended-envelope nature.
Among all the known cases that belong to the eIIb group, SN\,2024iss is notable for its well-sampled first peak, with multiband photometry covering both the rise and the decline during the shock-cooling phase. The decline rate of the first peak in SN\,2024iss is comparable to that of SN\,2011fu. However, the former exhibits a shorter decline time and a shallower subsequent valley compared to those observed in SNe\,1993J and 2011fu.
At the \isotope[56]{Ni}-powered phase, as illustrated in Figure~\ref{fig:photometry_comparision}, the $V$-band light curve of SN\,2024iss displays remarkable similarities to those of other comparison SNe IIb. At the linear decline phase of the light curve (from $\sim$60 days post-explosion), SN\,2024iss exhibits a faster decline compared to that of other comparison SNe.
A more detailed modeling of the initial and the secondary peaks, and the radioactive decay tail of the light curve of SN\,2024iss, will be presented in Sections~\ref{sec:shock_cooling_model},~\ref{sec:Ni_mass_estimate}, and~\ref{subsec:low_ejecta_mass}, respectively.

\subsection{Color Evolution} 
\label{subsec:color_evolution}

\begin{figure}
    \centering
    \includegraphics[width=1\columnwidth]{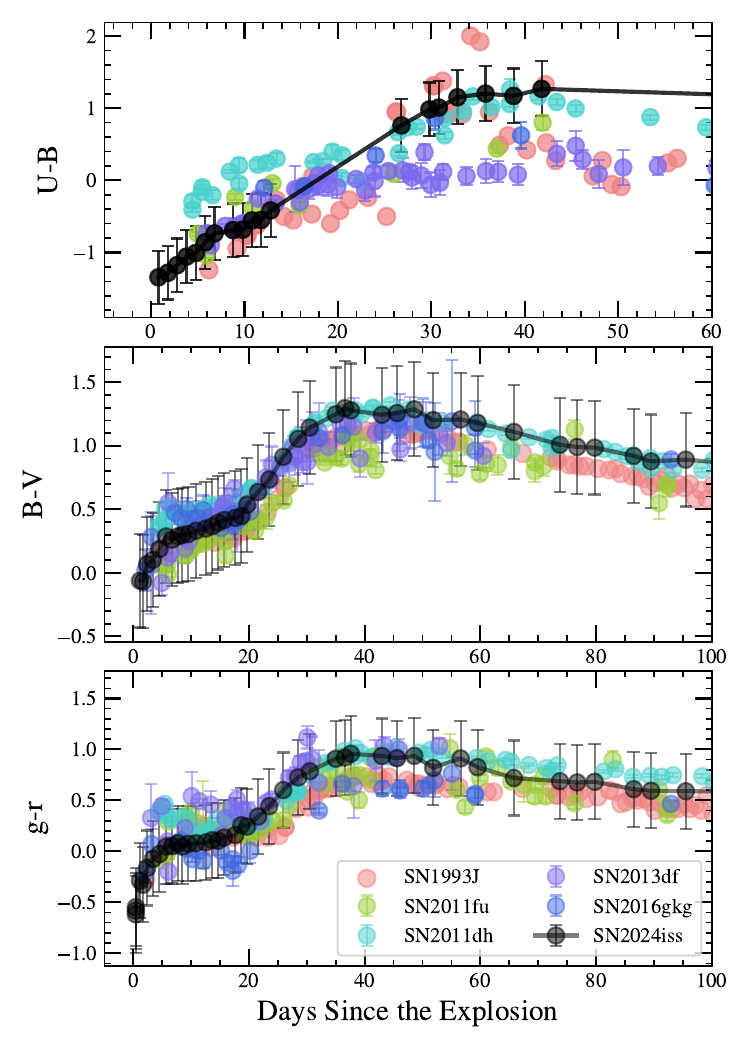}
    \caption{Galactic reddening-corrected $U-B$, $B-V$, and $g-r$ color curves of SN\,2024iss,
    compared with Galactic reddening-corrected color curves of SNe\,1993J, 2011fu, 2011dh, 2013df, and 2016gkg. For the purpose of display, the $V-R$ color curve of SNe\,1993J, 2011fu, 2013df and 2016gkg were converted to $g-r$ using the transformation from \citet{Jordi2006}.}
    \label{fig:color_comparision}
\end{figure}

Figure \ref{fig:color_comparision} compares the extinction-corrected $U-B$, $B-V$, and $g-r$ color curves of SN\,2024iss with those of SNe\,1993J, 2011fu, 2011dh, 2013df and 2016gkg. 
For better comparison, the $V-R$ colors of SNe\,1993J, 2011fu, 2013df and 2016gkg were converted to $g-r$ using the transformation relations provided by \citet{Jordi2006}. 

As shown in the top panel of Figure~\ref{fig:color_comparision}, the $U-B$ color of SN\,2024iss displays a continuous increase until setting to a plateau at a maximum of $\sim1.2\pm0.37$\,mag at $t \approx 35$ days. The $B - V$ and $g - r$ colors of SN\,2024iss show a similar evolution, with two distinct phases of increase of the color indices, which coincide with the first and second optical light curve peaks, respectively.
In particular, the $B-V$ and the $g-r$ colors of SN\,2024iss increase by $\sim 0.35$ mag and 0.67 mag within the first five days, respectively. Such an initial rise indicates a rapid decrease in photospheric temperature as fitted by a blackbody function (see Section~\ref{sec:bolometric_luminosity} and Figure~\ref{fig:bolo_result}).
From $t \approx 5$ to 20 days, both the $B-V$ and $g-r$ colors of SN\,2024iss continue to become redder, but at a much slower rate.
Such a two-phase redward evolution in the early color curves is broadly consistent with the average behavior of the comparison SNe IIb shown in Figure~\ref{fig:color_comparision}. 
In contrast, a blueward color evolution between the two rapid increases in $B-V$ and $g-r$ is reported in SNe\,2022crv \citep{Dong_2024} and 2024uwq \citep{Subrayan2025}.

As the $V$-band light curve reaches its secondary peak at around day 20, the $B-V$ and $g-r$ colors of SN\,2024iss display another redward evolution until reaching the reddest colors of $\sim 1.29\pm0.37$ and $0.97\pm0.37$ magnitudes, respectively, at $t \approx 40$ days. 
Overall, such color evolution is broadly consistent with those of other SNe IIb, with the exception of SN\,2011dh, which does not show a clear initial rapid reddening phase and only exhibits a very shallow initial decline in the $V$ band \citep{tsvetkov2012} and P48 $g$ band \citep{Arcavi2011}.

\subsection{Pseudo-bolometric Luminosity and Temperature Evolution
}  
\label{sec:bolometric_luminosity}

\begin{figure}
    \centering
    \includegraphics[width=1\columnwidth]{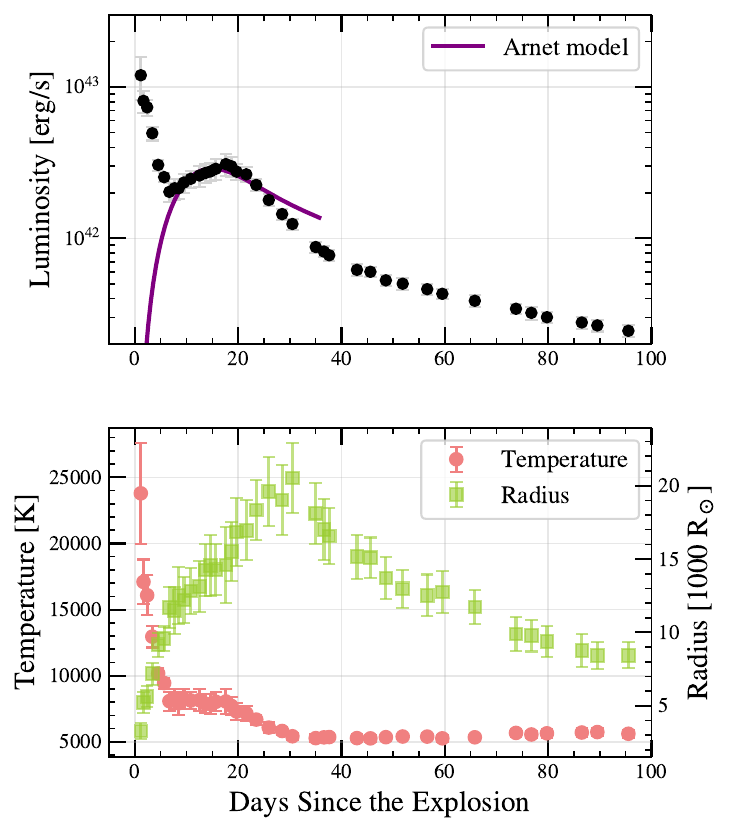}
    \caption{Top: Pseudo-bolometric light curve of SN\,2024iss. The purple line shows the best-fit Arnett model to the secondary peak. Bottom: Evolution of the blackbody temperature and the photospheric radius of SN\,2024iss, as indicated by the left- and right-hand ordinates, respectively.} 
    \label{fig:bolo_result}
\end{figure}

Using extinction-corrected photometry, we constructed the pseudo-bolometric light curve of SN\,2024iss with the \texttt{Light Curve Fitting} package \citep{Hosseinzadeh2020}. The SED at each epoch was fitted with a blackbody function, employing a Markov Chain Monte Carlo (MCMC) routine with the \texttt{emcee} package \citep{Foreman_Mackey_2013}. We also included the NIR photometry published in \citet{Yamanaka2025}, as the NIR contribution at later phase at a few months after the explosion may reach up to $\sim$35\% \citep{Medler2022}.
Temporal evolutions of the constructed pseudo-bolometric luminosity, the blackbody temperature, and the radius of SN\,2024iss are shown in Figure~\ref{fig:bolo_result}. The latter two parameters were derived by fitting a blackbody spectrum to the SED constructed from the UV, optical, and NIR ($uvw2$, $uvm2$, $uvw1$, $B$, $V$, $g$, $r$, $i$, $J$, $H$, and $K$-band) photometry.

As shown in the top panel of Figure~\ref{fig:bolo_result}, the maximum pseudo-bolometric luminosity of SN\,2024iss yields $L = (1.2 \pm 0.4 )\times 10^{43} \rm{erg~s^{-1}}$, which is observed on day 1.2. 
Owing to the lack of multiband photometry at even earlier phases, the constructed pseudo-bolometric light curve does not cover the phase before day 1.2.
The pseudo-bolometric light curve subsequently declines until reaching a turning point on day 6.7. The light curve then rises again until a secondary peak of $L = 3.1 \times 10^{42} \rm{erg~s^{-1}}$, $log(L/\rm{erg~s^{-1}}) = 42.49 \pm 0.07 $ is attained on day 17.6, which is coincident with the time of the $V$-band maximum. 
The pseudo-bolometric luminosity of SN\,2024iss observed at its secondary peak is slightly above the mean peak luminosity of SNe IIb given by \citet{Prentice_2016}, namely $log(L/\rm{erg~s^{-1}}) = 42.36 \pm_{0.11}^{0.26}$. This may result from either a larger amount of \isotope[56]{Ni} synthesized during the SN explosion or a more thorough mixing of \isotope[56]{Ni} throughout the ejecta~\citep{liu2025}.

Both the temperature and photospheric radius of SN\,2024iss show similar evolution to those of other SESNe \citep{Taddia2018}. The effective temperature drops rapidly during the early phase, from $\sim 23,000$~K to $\sim 8,200$~K within the first $\sim$7 days, until entering a plateau phase.
It remains constant until $t\approx 18$ days and then decreases, which is roughly coincident with the $V$-band maximum light. 
The temperature evolution of SN\,2024iss derived by fitting a blackbody to the SED mirrors the trend observed in the $B-V$ and $g-r$ color curves (see, Section~\ref{subsec:color_evolution} and Figure~\ref{fig:color_comparision}).

\subsection{Estimating the \isotope[56]{Ni} Mass
} \label{sec:Ni_mass_estimate}

\begin{figure*}
    \centering
    \includegraphics[width=\textwidth]{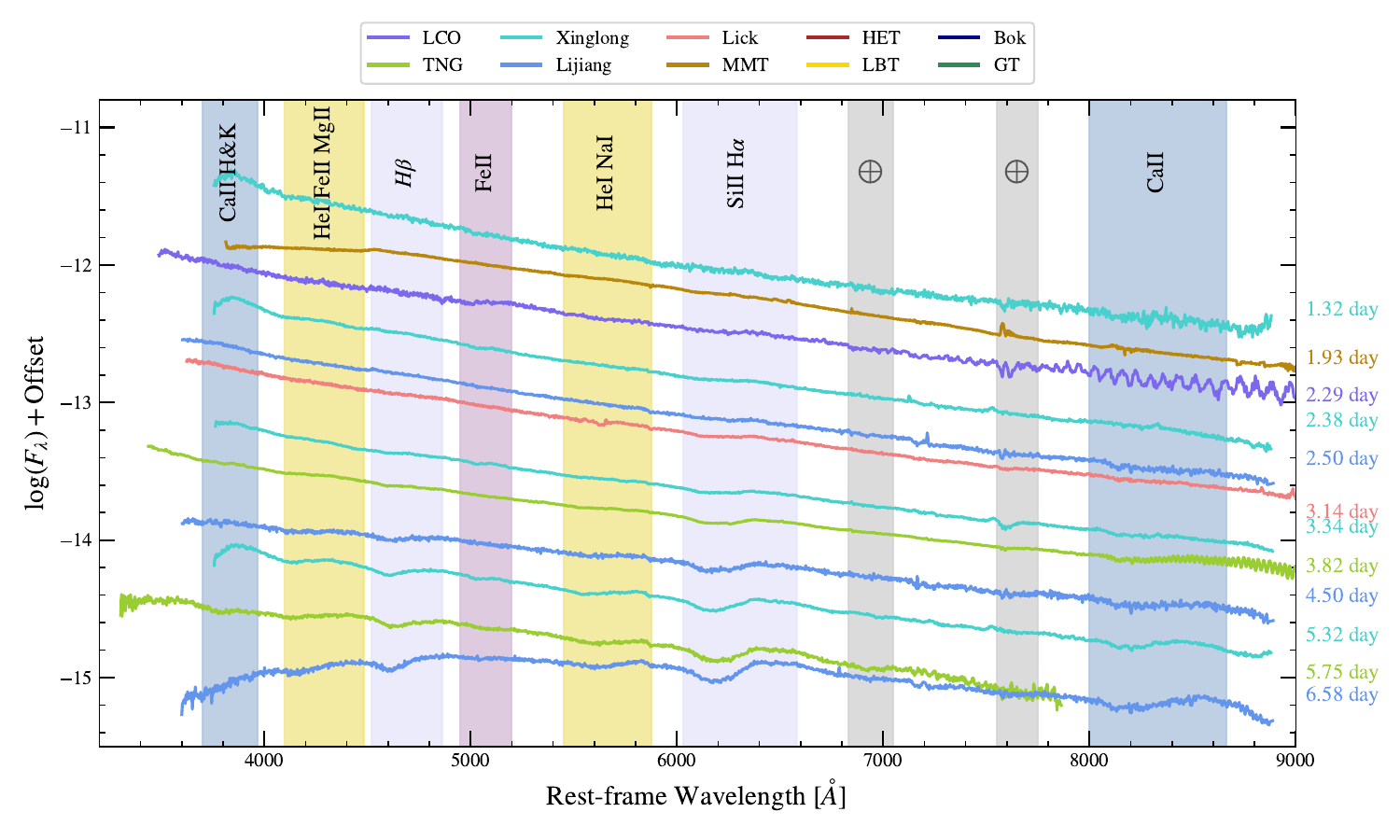}
    \caption{Spectral sequence of  SN\,2024iss, spanning the first 7 days after the explosion.
    Phases are marked on the right.
    Different colors distinguish the different spectrographs used in the observations, as shown at the top.
    A log of the spectroscopy of SN\,2024iss is given  in Table~\ref{table:Log of spectra}.
    }
    \label{fig:spectra_early}
\end{figure*}

We fit the second peak of the bolometric light curve with the Arnett law, which was first proposed as an oversimplified expanding fireball model to explain the early light curve of Type Ia supernovae (SNe Ia) by \citet{Arnett1982}. \citet{Valenti2008} extended this model to SESNe by separating their photometric evolution into the photospheric and nebular phases, and introducing a $\gamma-$ray leakage term. 
This model involves two key parameters to describe the secondary peak of the bolometric light curve, namely the synthesized nickel mass in the SN ejecta ($M_{\rm{Ni}}$), which determines the peak bolometric luminosity, and the ejecta mass ($M_{\rm{ej}}$), which is related to the photon diffusion time $\tau_m$ and the width of the bolometric light curve. According to \citet{Arnett1982}, $\tau_m$ can be expressed in terms of $M_{\rm{ej}}$ and the photospheric velocity $v_{\rm{ph}}$:

\begin{equation}
    \tau_m = \sqrt{2} \left( \frac{\kappa_{\rm{opt}}}{\beta c} \right)^{1/2} \left( \frac{M_{\rm{ej}}}{v_{\rm{ph}}} \right)^{1/2}
    \label{eq:ni_decay}
\end{equation}

where $\beta \approx 13.8$ is a constant, $c$ denotes the speed of light and $v_{\rm{ph}}$ is the photospheric velocity. Following common assumptions for SESNe \citep{Taddia2018, Dong_2024}, we adopt a constant optical opacity $\kappa_{\rm{opt}} = 0.07~\rm{cm}^2~\rm{g}^{-1}$ and a $\gamma-$ray opacity $\kappa_{\gamma} = 0.027~\rm{cm}^2~\rm{g}^{-1}$. This model assumes all \isotope[56]{Ni} is placed at the center of the SN ejecta without any outward mixing, and that the opacity remains constant throughout the evolution. For a spherical SN ejecta, $v_{\rm{ph}}$ is linked to the kinetic energy and $M_{\rm{ej}}$ through the relation: 

\begin{equation}
    E_k \approx \frac{3}{5} \frac{M_{\rm{ej}} v_{\rm{ph}}^2}{2}
    \label{eq:kinetic_energy}
\end{equation}

Following the prescriptions described by~\citet{Dessart2005}, we adopt a photospheric velocity of $7500~\rm{km}~\rm{s}^{-1}$, as measured from absorption minimum of the Fe{\sc\,II} $\lambda$5169 in the near-maximum-light spectrum.
This velocity is comparable to those of SN\,1993J ($8000~\rm{km}~\rm{s}^{-1}$) and other SNe IIb\citep{Medler2022}, but is significantly lower than the velocity measured from the H$\alpha$ P~Cygni profile in the spectrum of SN\,2024iss, which is $15,\!000~\rm{km}~\rm{s}^{-1}$~\citep{Yamanaka2025}.

We fit the bolometric light curve of SN\,2024iss spanning the phase from $t \approx +7$ to +30 days with the \texttt{emcee} package \citep{Foreman_Mackey_2013}.
The time interval was chosen to encompass the secondary peak but exclude the nebular phase when the observed flux becomes dominated by various emission lines from iron-group elements.
The best-fit result is shown by the purple line in the upper panel of Fig.~\ref{fig:bolo_result}, indicating $M_{\rm{Ni}} = 0.117 \pm 0.013~M_{\odot}$, $M_{\rm{ej}} = 1.272 \pm 0.343~M_{\odot}$ and $E_{k} = 0.427 \pm 0.115 \times 10^{51} \rm{erg}$. 
The nickel mass derived for SN\,2024iss is similar to those reported for other Type IIb SNe, for example, $0.11~M_{\odot}$ for SN\,2013df \citep{Morales2014}. However, the ejecta mass of SN\,2024iss appears to be lower than all of the SN IIb sample presented in \citet{Medler_2021}. Additionally, we measure the magnitude decline from maximum light to 15 days after that in the V-band light curve, i.e., $\Delta m_{15}(V)$ = 1.13\,mag, which is larger compared to an average value of 0.93~mag derived for a sample of Type IIb SNe \citep{Taddia2018}. Such a rapid decline may not solely result 
from a low ejecta mass, but may also be caused by other effects such as vigorous mixing of radioactive \isotope[56]{Ni} throughout the SN ejecta, or variations in the density profile of the ejecta \citep{liu2025}. A more detailed analysis will be discussed in Section \ref{subsec:low_ejecta_mass}.

\section{Optical Spectroscopy
} \label{sec:spectroscopy}

\subsection{Spectral evolution 
}  

\begin{figure*}
    \centering
    \includegraphics[width=\textwidth]{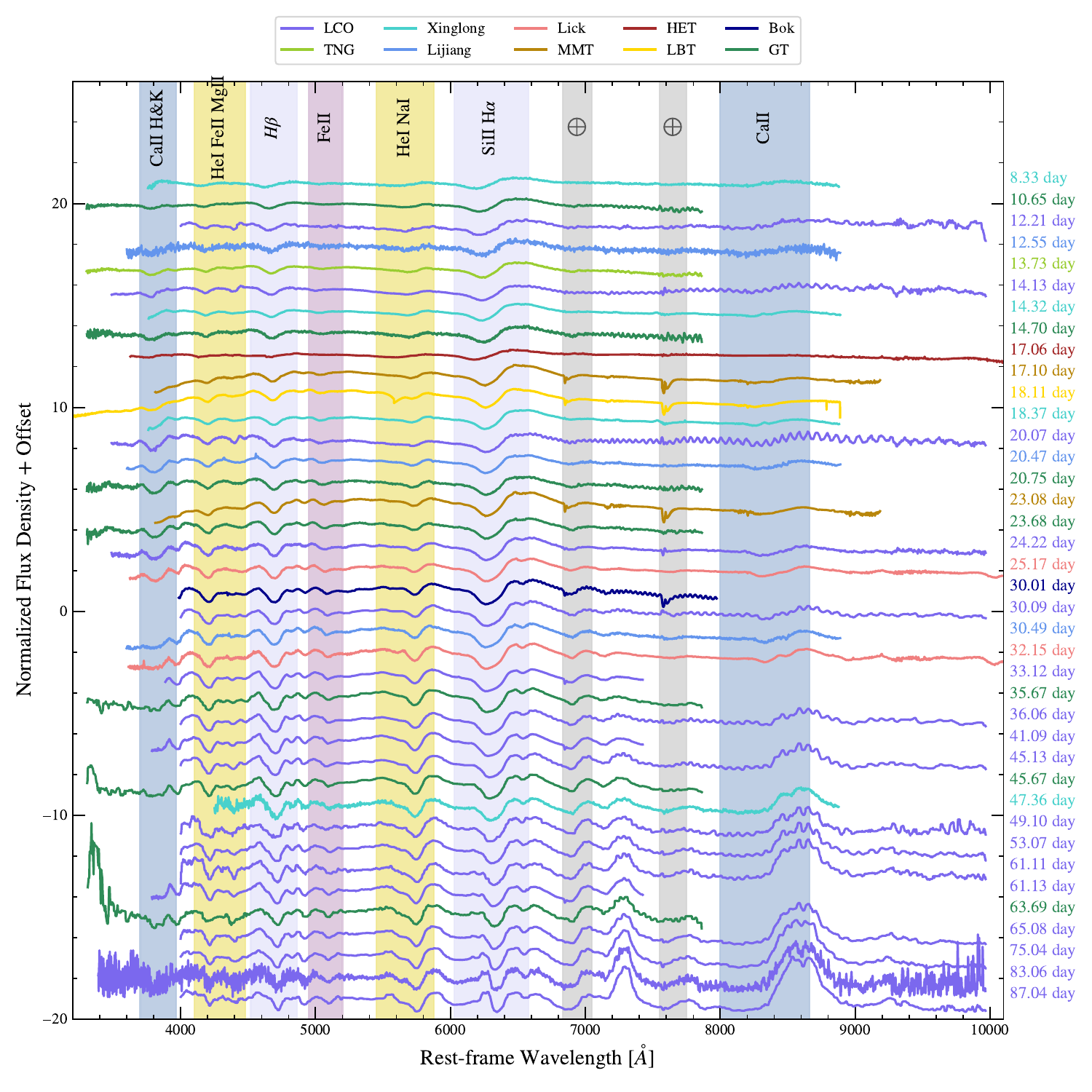}
    \caption{Spectral consequence of SN\,2024iss obtained during the phase from $t \approx 8$ to 87 days after the explosion. The layout of the figure is similar to those of Figure~\ref{fig:spectra_early}.
    For the purpose of presentation, the best-fit blackbody function was subtracted from each spectrum.
    A log of the spectroscopy of SN\,2024iss is presented in Table~\ref{table:Log of spectra}.
    }
    \label{fig:spectra_late}
\end{figure*}

We collected a total of 51 spectra of SN\,2024iss, spanning the phases from $t \approx +1.3$ to +87.0 days. 
We present the spectral sequences before and after $t \approx 7.0$ days in Figures~\ref{fig:spectra_early} and \ref{fig:spectra_late}, respectively.
All spectra were corrected for the Galactic reddening and are shown in the rest frame.
Within the first 7 days after the estimated explosion, the spectra of SN\,2024iss are characterized by blue and featureless continuums.
At this stage, the SN emission is dominated by the shock cooling. The outermost envelope of the exploding progenitor star manifests sufficiently high temperature at the location of the photosphere, where the high-opacity line-forming regions were not often present.
We see no narrow emission features in the first few spectra of SN\,2024iss, as shown in Figure~\ref{fig:spectra_early}. However, narrow emission features can be identified in the spectra of SN\,1993J~\citep{Benetti1994} and 2013cu~\citep{Gal_Yam_2017} within the first days after the explosion.

The absence of such narrow lines may indicate either a relatively low density or less radially extended CSM content around the progenitor of SN 2024iss. 
As the ejecta cool down, the characteristic P~Cygni profiles of the Balmer lines start to emerge in the $t \approx 3.8$ day spectrum.
Features of He{\sc\,I} $\lambda$5876, $\rm{H}_{\gamma}$ and Ca{\sc\,II} NIR $\lambda\lambda$8498, 8542, 8662 can be also identified in the spectra starting from $t \approx 5.3$ days.

As illustrated in Figure~\ref{fig:spectra_late}, the spectroscopic evolution of SN\,2024iss after the shock-cooling phase show close resembles to those of other SNe IIb \citep{Filippenko1997, Morales2014, Medler2022}.
All spectrum are normalized by the fitted blackbody continuum.
After $t \approx 18.4$ days, Fe{\sc\,II} lines develop P Cygni profiles.
Meanwhile, He{\sc\,I} $\lambda$5876, 6678 features become progressively prominent after the shock cooling.
We attribute the absorption troughs at $\sim$4300\,\AA\ to the blending of several lines, namely He{\sc\,I} $\lambda$4472, Mg{\sc\,I} $\lambda$4481 and Fe{\sc\,II} $\lambda$4549, consistent with the line identification in the literature (e.g., \citealp{Hachinger2012,Silverman2009,Morales2014}).

Before entering the nebular phase, the Ca{\sc\,II} H\&K $\lambda\lambda$3943, 3969 and Ca{\sc\,II} $\lambda\lambda$8498, 8542, 8662 lines manifest as prominent blue-shifted absorption features.
As the ejecta cool, the [Ca{\sc\,II}] $\lambda\lambda$7291, 7323 doublet appear around day 32, followed by a series of oxygen lines, i.e., [O{\sc\,I}] $\lambda\lambda$6300, 6364 and O{\sc\,I} $\lambda$7774. Such signatures can be clearly identified in the $t \approx 61$ day spectrum. 
By $t \approx 87$ days, as the ejecta cool, various forbidden lines become prominent in the spectrum, 
i.e., double peak in the Ca{\sc\,II} $\lambda\lambda$8498, 8542, 8662 feature (see further discussions in Section \ref{subsec:spectra_com}).

\begin{figure}
    \centering
    \includegraphics[width=1\columnwidth]{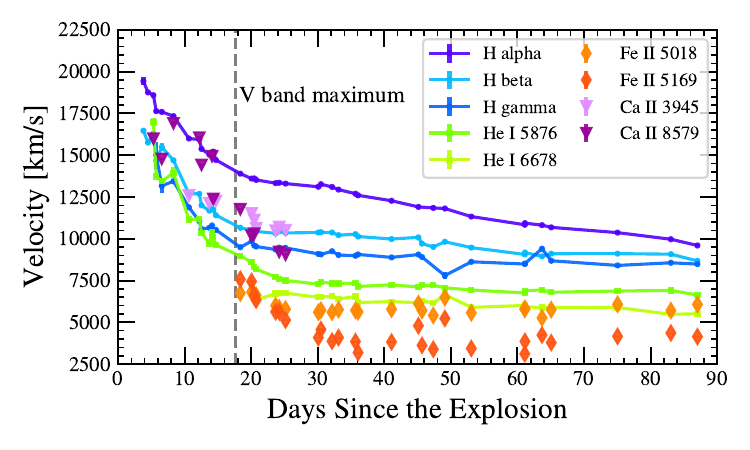}
    \caption{Evolution of the expansion velocity of SN\,2024iss measured from spectral features of
    $\rm{H}_{\alpha}$, $\rm{H}_{\beta}$, $\rm{H}_{\gamma}$, He{\sc\,I}, Fe{\sc\,II} and Ca{\sc\,II}. 
    The vertical gray-dashed line marks the epoch of $V$-band light curve peak.
    Photospheric velocities, inferred from the absorption minima of Balmer and He{\sc\,I}\,$\lambda$5876 lines, appear to be faster than other spectral features.
    }
    \label{fig:velocity_data}
\end{figure}

\subsection{Velocity evolution}  

\begin{figure}
    \centering
    \includegraphics[width=1\columnwidth]{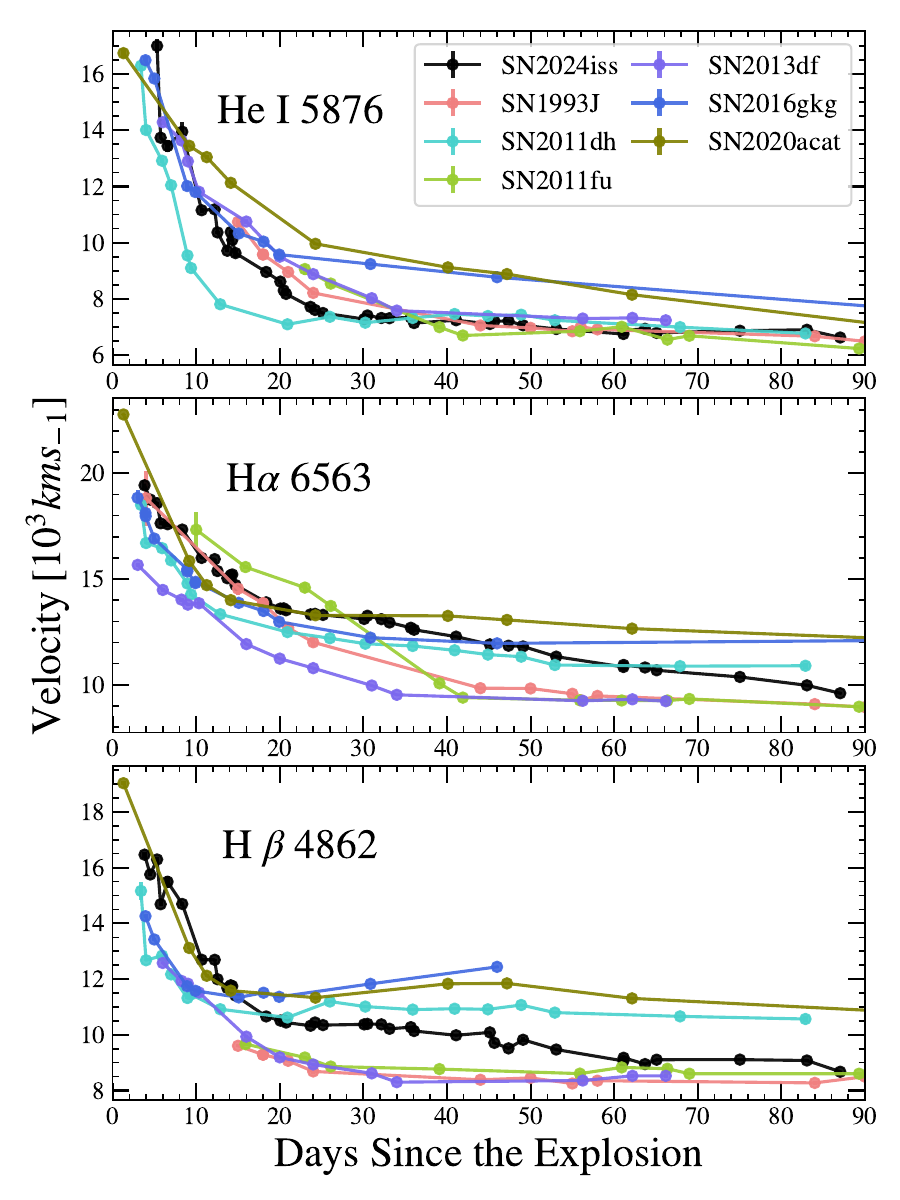}
    \caption{Temporal evolution of 
    He{\sc\,I} $\lambda$5876, $\rm{H}_{\alpha}$ and $\rm{H}_{\beta}$ of 
    SN\,2024iss compared to that measured for
    SNe\,1993J, 2011dh, 2011fu, 2013df, 2016gkg and 2020acat.}
    \label{fig:velocity_comparision}
\end{figure}

The photospheric velocities of SN\,2024iss as measured from $\rm{H}_{\alpha}$, $\rm{H}_{\beta}$, $\rm{H}_{\gamma}$, He{\sc\,I} $\lambda$5876, He{\sc\,I} $\lambda$6678, Fe{\sc\,II} $\lambda$5018, Fe{\sc\,II} $\lambda$5169, Ca{\sc\,II} H\&K $\lambda\lambda$3943, 3969 and Ca{\sc\,II} $\lambda\lambda$8498, 8542, 8662 lines are shown in Figure \ref{fig:velocity_data}. For each spectral feature that shows a P~Cygni profile, we subtracted the pseudo-continuum and fitted the residual with a Gaussian function. We then inferred the expansion velocity from the blueshifted absorption minimum.
After the $V$-band maximum, velocities inferred from the Balmer lines and the He{\sc\,I} $\lambda$5876 feature show a slower decline, which may be due to the transition of energy source from adiabatic to radiative cooling. The Balmer lines shows a similar velocity evolution, with velocities consistently higher than those of other lines. 
Additionally, Ca{\sc\,II} lines also exhibit relatively high expansion velocities, possibly due to element mixing \citep{Fransson1989}. 
At t$>$ 25 days, measuring Ca{\sc\,II} velocity is difficult because of emission-dominated profiles. The Fe{\sc\,II} $\lambda$5169 line velocities are key tracers of the receding photosphere in the
expanding ejecta \citep{Dessart2005}. We measure a velocity of $\sim 7500 \pm 500$\,km\,s$^{-1}$ from the Fe{\sc\,II} $\lambda$5169 line at around the $V$-band peak of SN\,2024iss. 
We also note that the P~Cygni profile of the Fe{\sc\,II} $\lambda5169$ feature suffered from increased contamination as other lines emerge, introducing larger uncertainties in profile fitting and velocity estimation.
In Figure~\ref{fig:velocity_comparision} we compare the velocity evolution of SN\,2024iss with that of other SNe IIb.

\subsection{Comparison with Other Type IIb SNe
}  
\label{subsec:spectra_com}

\begin{figure*}
    \centering
    \includegraphics[width=\textwidth]{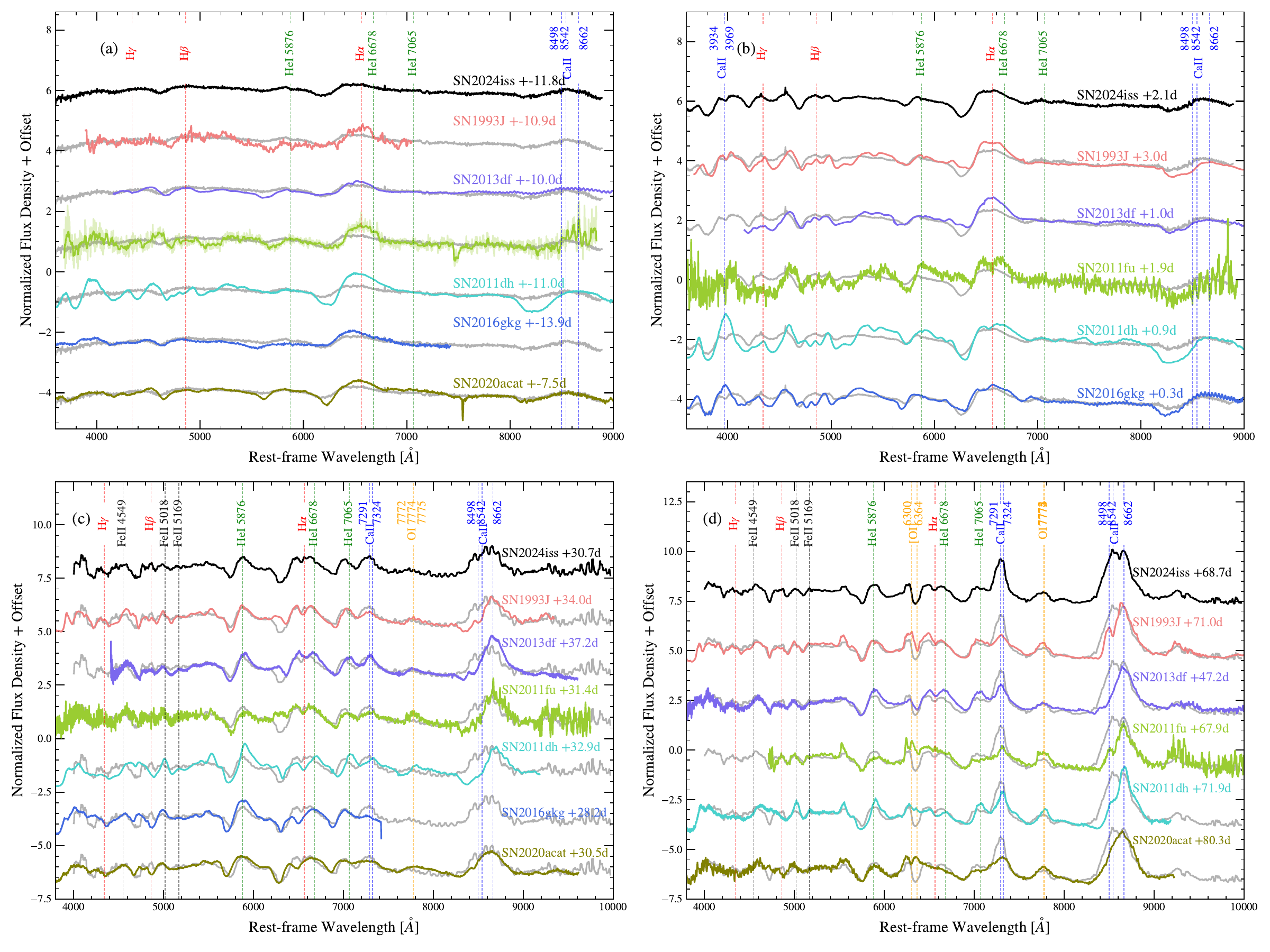}
    \caption{Spectra of SN\,2024iss at -12(a), 2(b), 31(c), and 69(d) days relative to the $V$-band light-curve peak, compared to other well-studied SNe IIb at similar phases. Phases with respect to the $V$-band maximum are labeled. All spectra are shown in the rest-frame and normalized. The vertical dashed lines mark the positions of certain spectral features at zero rest-frame velocity. For comparison, the gray curve underlying the spectrum of each comparison SN shows the shifted spectrum of SN\,2024iss that matches their mean flux within the displayed wavelength range.
    }
    \label{fig:four_phase_spectra_comparision}
\end{figure*}

Figure~\ref{fig:four_phase_spectra_comparision} compares the spectra of SN\,2024iss with those of
SNe\,1993J \citep{Barbon1995,Matheson2000}, 2011dh \citep{Arcavi2011,Ergon_2014}, 2011fu \citep{Kumar2013,Morales_Garoffolo_2015,Shivvers2019}, 2013df \citep{Szalai2016,Shivvers2019}, 2016gkg \citep{Kilpatrick2017} and 2020acat \citep{Medler2022} at similar phases.
All comparison spectra were downloaded from the WISEREP\footnote{https://www.wiserep.org} \citep{Yaron2012} and have been corrected for the redshifts of the host galaxies and the reported reddenings.
Basic information of the comparison SNe is compiled in Table \ref{table:comparison_info}.

At early phases, SNe\,2024iss and 1993J exhibit similar $\rm{H}_{\alpha}$ and $\rm{H}_{\beta}$ profiles, while SN\,1993J displays a more prominent He{\sc\,I}~$\lambda$5876.
The Ca{\sc\,II}\,NIR triplet profile of SN\,2024iss shows a striking resemblance to that of SN\,2020acat.
Additionally, we note that among all the spectra of SNe IIb compared in Fig.~\ref{fig:four_phase_spectra_comparision}~(a), SNe\,2011dh and 2011fu exhibit more prominent line profiles at about ten days before $V$-band maximum, indicating a prompt spectral evolution.

At around the $V$-band maximum, as shown in Fig.~\ref{fig:four_phase_spectra_comparision}~(b), both SNe\,2024iss and 1993J exhibit a flat-topped emission component of H$\alpha$, which may likely be caused by the strengthening He{\sc\,I} $\lambda$6678.
In contrast, SN\,2013df displays a single-peaked H$\alpha$ emission, while SNe\,2011dh and 2011fu exhibit double-peaked profiles, likely indicating a rather prompt mixing of helium into the H-rich envelope. 
SN\,2011fu shows the most prominent He{\sc\,I} $\lambda$5876 feature among the sample. 
Additionally, SNe\,2024iss and 2013df are characterized by shallower Ca{\sc\,II} $\lambda\lambda$8498, 8542, 8662 absorptions relative to other comparison SNe.
We also note that both SNe\,2024iss and 1993J display a double-peaked feature near 4000 \AA, whereas SNe\,2011fu and 2016gkg show a single Ca{\sc\,II} H\&K $\lambda\lambda$3943, 3969 emission peak. 
Finally, SN\,2024iss shows a single-peaked line near $4800$\,\AA\, which can be attributed to H$\beta$, where other comparison SNe display line complexes.

In Fig.~\ref{fig:four_phase_spectra_comparision}~(c) we compare the spectrum of SN\,2024iss at $\sim 31$ days after the $V$-band peak to those of other comparison SNe at a similar phase.
SN\,2024iss exhibits broader Ca{\sc\,II} $\lambda\lambda$8498, 8542, 8662 absorption in comparison with other SNe. Furthermore, unlike the significantly weakened H$\alpha$ absorption component observed in SNe\,1993J and 2013df, such a feature still remains prominent in SNe\,2024iss, 2016gkg, and 2020acat.

At $t \approx 67$ days after the $V$-band maximum light, as shown in Fig.~\ref{fig:four_phase_spectra_comparision}~(d), the iron features became more prominent, indicating a transition to the nebular phase, when the continuum spectral emission becomes dominated by a blend of Fe-group features. The strength of the Balmer lines are decreasing while several He{\sc\,I} lines remain strong, highlighting the type IIb nature of the SN. Additionally, the spectrum of SN\,2024iss exhibits a double-peaked Ca{\sc\,II} $\lambda\lambda$8498, 8542, 8662 profile. Among the comparison sample, only SN\,1993J shows a similar feature at a similar phase.
At earlier phases, these lines are blended due to line broadening. As the ejecta cool down, individual components separate and become distinguishable. The evolution of the double-peaked Ca{\sc\,II} NIR profile of SN\,2024iss during the nebular phase remains an intriguing question.
Additionally, similar to SNe\,1993J and 2013df, SN\,2024iss still shows H$\alpha$ at this phase, which was no longer identified in the spectra of SN\,2011fe, 2011dh, and 2020acat obtained at a similar epoch.
The presence of hydrogen at this phase suggests elemental mixing or a relatively massive hydrogen envelope. 
Finally, we note that SN\,2024iss also shows the strongest [Ca{\sc\,II}] $\lambda\lambda$7291, 7323 emission feature relative to the comparison SNe.

\section{Shock-Cooling Model Fitting} \label{sec:shock_cooling_model}

Among various models developed to describe the shock cooling emission, we adopt the one by \citet{Waxman2017} (hereafter \citetalias{Waxman2017}) to fit the multiband light curves of SN\,2024iss up to $t \approx 5$ days after the explosion. The key parameters in \citetalias{Waxman2017} include the envelope radius ($R$), the envelope mass ($M_{\rm env}$), the shock velocity ($v_{s,8.5}$), the product of stellar mass and a numerical factor related to envelope structure ($f_{\rho}M$) and the first-light time ($t_0$).

\citetalias{Waxman2017} assumes a polytropic density profile, where polytropic index $n = 3/2$ corresponds to the case of RSGs with convective envelope and $n = 3$ corresponds to the case of BGSs with radiative envelope. Building on the framework of \citet{Rabinak2011}, \citetalias{Waxman2017} introduced an exponential suppression factor in luminosity. This modification extends the validity of the model to a few days after the explosion. According to the different polytropic index ($n = 3/2[3]$), the bolometric luminosity is given by (following the formulation recast by \citet{Arcavi_2017}):
\begin{multline}
    L_\mathrm{RW} = 2.0[2.1]\times10^{42} \\
    \times\bigg{[}\frac{v_{s,8.5}t^{2}}{f_{\rho}M\kappa_{0.34}}\bigg{]}^{-0.086[-0.175]}~\frac{v_{s,8.5}^{2}R_{13}}{\kappa_{0.34}}~\mathrm{erg\,s^{-1}} 
\end{multline}

\begin{multline}
    L/L_\mathrm{RW} = 0.94[0.79]\times \\
    \mathrm{exp}\bigg{[}-\left(\frac{1.67[4.57]t}{(19.5\kappa_{0.34}M_{e}v_{s,8.5}^{-1})^{0.5}}\right)^{0.8[0.73]}\bigg{]}
\end{multline}
where $t$ is time since explosion in days, $v_{s,8.5}$ is shock velocity in $10^{8.5}\ \rm{cm~s^{-1}}$, $R_{13}$ is envelope radius in $10^{13}$ cm, $\kappa_{0.34}$ is opacity in $0.34~\rm{cm}^{2}g^{-1}$, $M$ is ejecta mass (core mass $M_{c}$ $+$ envelope mass $M_{e}$) in solar masses, and when $R_c/R \ll 1$, $f_{\rho}$ can be approximated as:
\begin{equation}
f_{\rho} \approx
\begin{cases}
(M_e/M_c)^{0.5}, n=3/2 \\
0.08(M_e/M_c), n=3
\end{cases}
\end{equation}

The temporal evolution of the color temperature is given as:
\begin{multline}
    T = 2.05[1.96]\times 10^{4} \\
    \times \bigg{[}\frac{v_{s,8.5}^{2}t^{2}}{f_{\rho}M\kappa_{0.34}}\bigg{]}^{0.027[0.016]}\left(\frac{R_{13}}{\kappa_{0.34}}\right)^{0.25} t^{-0.5} K
\end{multline}

Assuming that the continuum emission of SN\,2024iss can be well approximated by a blackbody spectrum, we can calculate the photospheric radius from the Stefan-Boltzmann law and fit the model to the multiband photometry.
We fit the $swift$ UVOT $uvw2$, $uvm2$, $uvw1$, $B$, $V$, $g$, $r$ and $i$-band light curves of SN\,2024iss with the open-source package \texttt{Light Curve Fitting} \citep{Hosseinzadeh2020}, implementing the analytic expression given by~\citetalias{Waxman2017}, with an index $n = 3/2$.
We restrict the fit to data obtained within five days after the estimated time of 
explosion, as beyond this point the energy input from the radioactive decay of \isotope[56]{Ni}
exceeds 30\% (see Section \ref{sec:Ni_mass_estimate}), and therefore biases the results. Within 5 days after explosion, the blackbody temperature satisfies the criterion $T > 0.7\,\mathrm{eV}$, ensuring the validity of the \citetalias{Waxman2017} model during the period.
The fit was performed using a Markov Chain Monte Carlo (MCMC) routine with 30 walkers and 2000 steps (1000 steps for burn-in). The additional scatter term ($\sigma$) was included to account for intrinsic variability and deviations from the blackbody assumption. 

\begin{figure*}
    \centering
    \includegraphics[width=\textwidth]{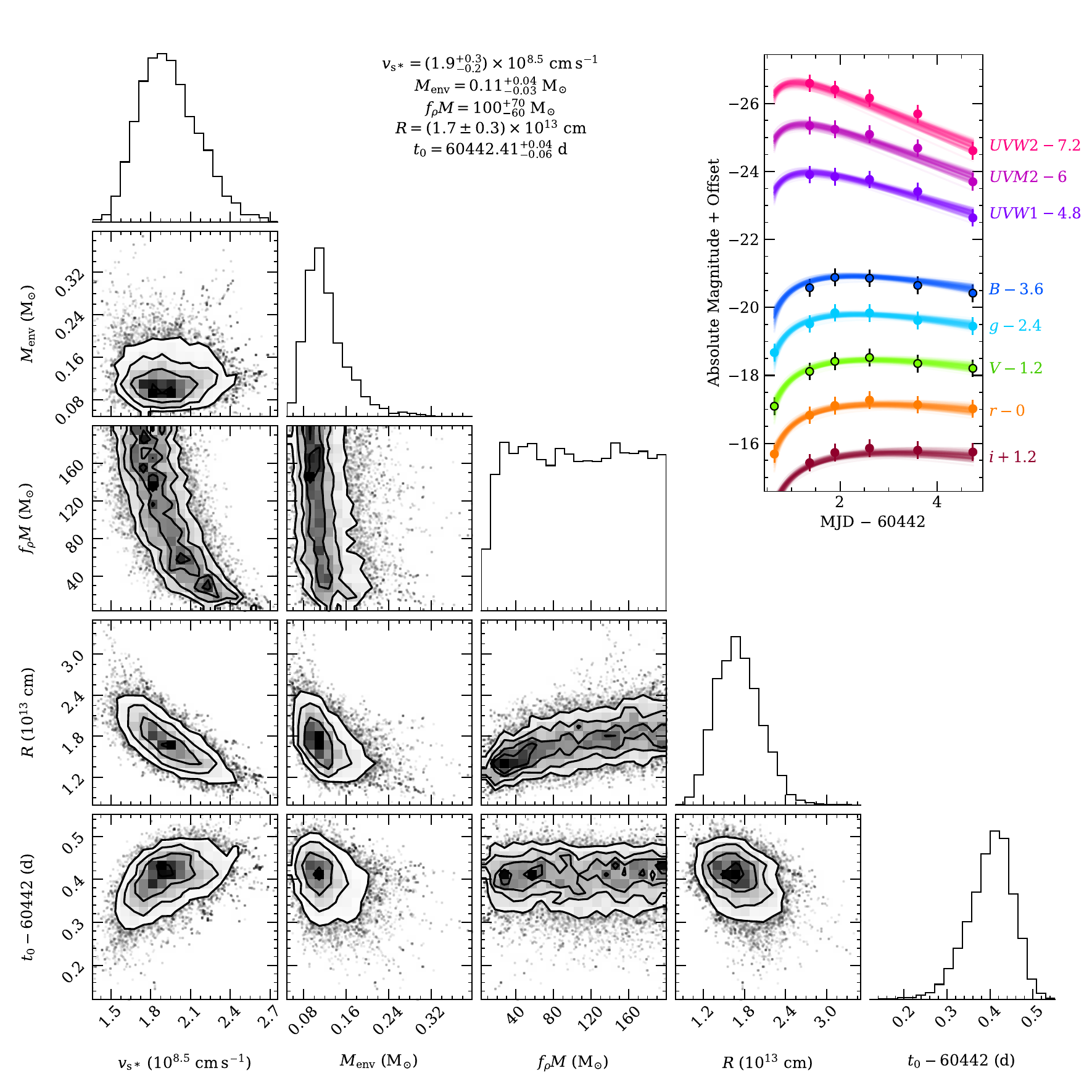}
    \caption{The Corner plot that presents the posterior distributions of the \citetalias{Waxman2017} shock-cooling light-curve model of SN\,2024iss.
    The fit to the early light curves was restricted to five days after the estimated time of the SN explosion. The best-fit parameters are listed at the top of the figure. The upper-right inset presents the model light curves that best fit the multiband photometry, as displayed by color-coded curves and symbols, respectively.
    }
    \label{fig:shock_cooling}
\end{figure*}

The MCMC parameters distributions and the best-fit model multiband light curves are presented in Figure~\ref{fig:shock_cooling}.
The parameters that best describe the shock-cooling phase are the following:
$\mathrm{R} = 244 \pm43~R_{\odot}$, $M_{\rm{env}} = 0.11 \pm 0.04~M_{\odot}$, $v_{s,8.5} = (1.9 \pm 0.3)\times 10^{8.5}\ \rm{cm~s^{-1}}$, $f_{\rho}M = 100 \pm 65~M_{\odot}$ and $t_{\rm{exp}} = 60442.41 \pm 0.05~\rm{d}$. The model light curves agree fairly well with the observations.
The inferred envelope mass and radius of SN\,2024iss are larger than that of SN\,2016gkg (i.e., $\sim$0.03\,$M_{\odot}$, $\sim$141\,$R_{\odot}$; \citealt{Piro_2021}), indicating a more extended envelope than Type cIIb SNe. 
The envelope mass of SN\,2024iss is lower than  that of SN\,1993J ($\sim$0.2\,$M_{\odot}$; \citealt{Woosley1994}), suggesting that its envelope is less extended than some Type eIIb SNe.

\section{X-ray Constraints on the Circumstellar Medium} \label{sec:X_ray}

\begin{figure}
    \centering
    \includegraphics[width=1\columnwidth]{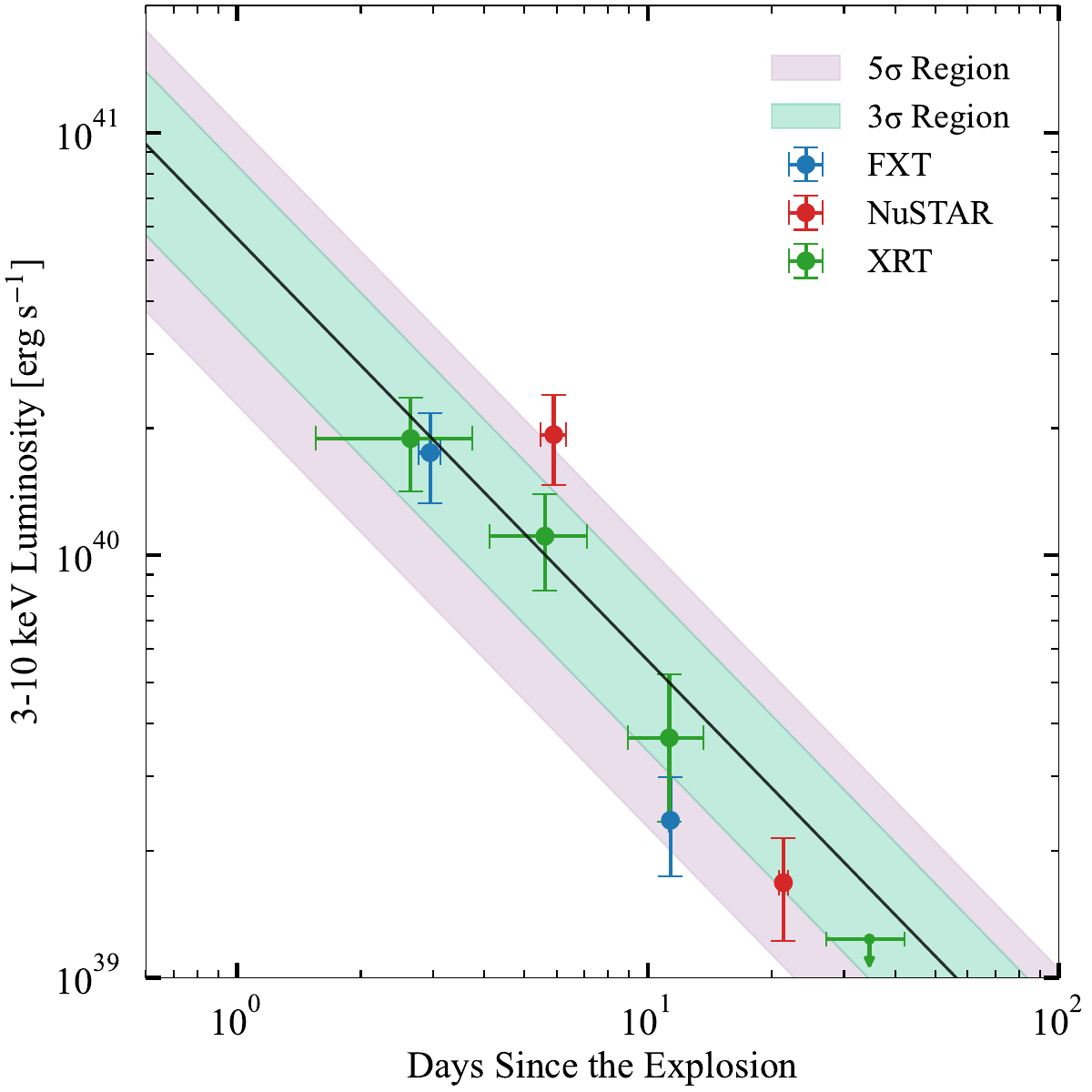}
    \caption{
    The unabsorbed X-ray luminosity of SN\,2024iss observed in the 3--10\,keV band. 
    The red, green, and blue data points represent the observations conducted with the   
    NuSTAR, Swift-XRT, and EP-FXT, respectively. The black line shows the best-fit thermal bremsstrahlung model, with a mass-loss rate of $1.6 \times 10^{-5}\ M_{\odot}\ \rm{yr}^{-1}$ and a shock velocity of $9.5\times 10^8\ \rm{cm~s^{-1}}$. The green and purple shaded areas indicate the 3--$\sigma$ and 5--$\sigma$ confidence regions, respectively.
    }
    \label{fig:luminosity_X_ray}
\end{figure}
 
Figure \ref{fig:luminosity_X_ray} shows the unabsorbed 3$-$10 keV X-ray luminosity of SN\,2024iss. The errors were estimated by adding the uncertainties in the X-ray photometry and the distance modulus in quadrature.
As the temperature of post-shock gas is higher than $2 \times 10^{7}$ K, thermal bremsstrahlung emission in the forward shock dominates over the cooling \citep{Fransson1996, Chevalier2017}. Assuming a homogeneous wind with $v_{\rm wind}=10$\,km\,s$^{-1}$ according to~\citet{Fransson1996}, 
our best-fit result is consistent with a mass-loss rate of $\dot{M} = 1.6 \times 10^{-5}\ M_{\odot}\ \rm{yr}^{-1}$ and a shock velocity of $v_{s}^{\rm X-ray}\approx$ $9.5 \times 10^8\ \rm{cm~s^{-1}}$. 
Based on the 3$\sigma$ and 5$\sigma$ confidence intervals of the 3--10\,keV luminosity as indicated by the green and purple shaded regions in Fig. \ref{fig:luminosity_X_ray}, the mass-loss rate is constrained as 1.2$-$1.9 $\times 10^{-5}$ and 1.0$-$2.2 $\times 10^{-5}\ M_{\odot}\ \rm{yr}^{-1}$, respectively. The mass loss rate inferred for the X-ray observations of SN\,2024iss is comparable to those  of SNe\,1993J  (i.e., $4 \times 10^{-5}\ M_{\odot}\ \rm{yr}^{-1}$, \citealt{Fransson1996}) and 2013df (i.e., $8 \times 10^{-5}\ M_{\odot}\ \rm{yr}^{-1}$, \citealt{Kamble2016}). The shock velocity inferred from the X-ray light curve of SN\,2024iss is consistent with that estimated by the shock-cooling model as discussed in Section \ref{sec:shock_cooling_model}, namely $v_{s}=(6.0\pm0.9)\times10^{8}$\,cm\,s$^{-1}$, but is much smaller than the result inferred from the early time $\rm{H}_{\alpha}$ velocity (i.e.,  $v_{H_{\alpha}} = 1.94 \times 10^9 \rm{cm~s^{-1}}$). 

To further investigate the CSM properties, we also used the method of emission measure (EM) to constrain the density structure of the CSM:  
\begin{equation}
    EM = \int n_e n_H dV \approx \frac{\rho(r)^2 V_{\rm{CSM}}}{\mu_e \mu_H m_p}
\end{equation}

where $\rho(r)$ represents the mass density of the CSM at radius $r$, 
\( V_{\rm{CSM}} \approx 4 \pi r^2 \Delta r \) is the volume of the shocked CSM shell, with \( \Delta r \approx 0.1r \) denoting the thickness of the forward-shocked CSM. We adopt a mean molecular weight of $\mu=0.61$ for the ionized CSM with solar metallicity, where \( \mu_e \approx 1.3 \) and \( \mu_H \approx 1.15 \). The EM parameter can be derived from the "Norm" column in Table~\ref{table:Log of X-ray spectral parameter}. 

\begin{figure}
    \centering
    \includegraphics[width=1\columnwidth]{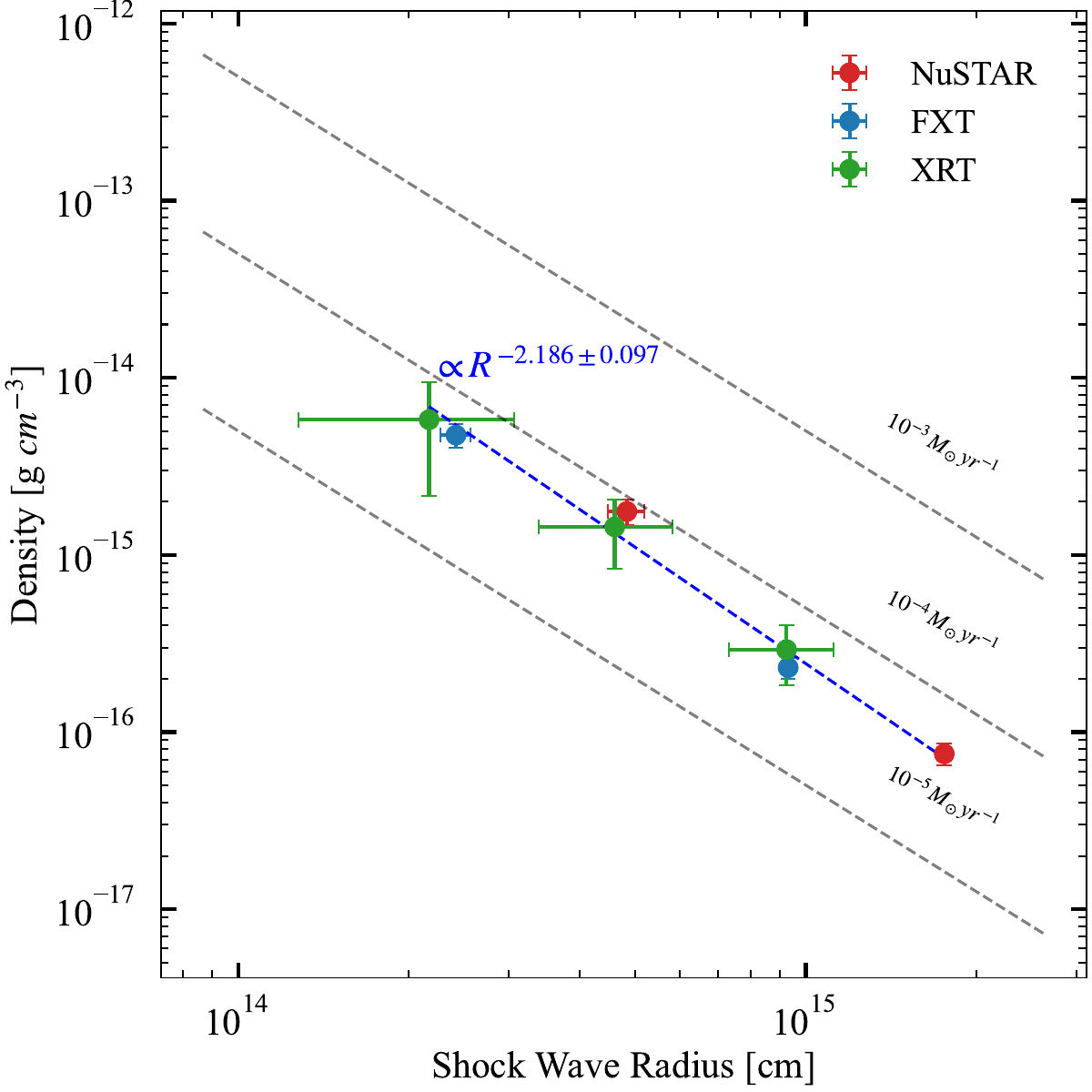}
    \caption{The CSM density profile as inferred from the emission measure.
    Data from NuSTAR, Swift-XRT, and EP-FXT are indicated by the red, green, and blue circles, respectively. 
    The gray dashed lines represent the wind-like density profiles corresponding to mass-loss rates of $\dot{M} = 10^{-5},\ 10^{-4},\ 10^{-3}\ M_{\odot}\ \rm{yr}^{-1}$. The blue dashed line denotes the power-law fit to the X-ray light curve. The inferred radial density profile $\rho_{\rm CSM} \propto R^{-2.19}$ is incompatible with a homogeneous mass-loss wind. 
    }
    \label{fig:density_EM_X_ray}
\end{figure}

Figure \ref{fig:density_EM_X_ray} shows the radial density profile of the CSM around SN\,2024iss ($\rho_{\rm CSM}$) as inferred from the EM method, together with that of the homogeneous wind with different mass-loss rates. The former lies between the \(10^{-5}\) and \(10^{-4} \ M_{\odot} \ \rm{yr}^{-1}\) curves of steady mass loss, with a mean value of $(5.55 \pm 1.57) \times 10^{-5}~M_{\odot}~\rm{yr}^{-1}$. This estimate is higher than the value derived from the X-ray light curve of SN\,2024iss. However, it is comparable to the mass loss rate of $8\times10^{-5}$\,M$_{\odot}$\,yr$^{-1}$ estimated for SN\,2013df, assuming a wind velocity of 10\,km\,s$^{-1}$~\citep{Kamble2016}.
The discrepancy may be due to the unusually low intrinsic neutral hydrogen column density of SN\,2024iss, namely $N_{\rm H}$ $< 8 \times 10^{19}~\rm{cm}^{-2}$ after subtracting the Galactic contribution.
The low column density would introduce a degeneracy in the X-ray spectral fitting and thus overestimate the mass-loss rate compared to the free-free emission model. A power-law fit to this profile yields \( \rho_{\rm{CSM}} \propto R^{-2.19\pm0.10} \), 
which is nearly consistent with the canonical \( R^{-2} \) expected for a steady wind. 

We also note that the inferred $N_{\rm H}$ at t $\sim$1.6 days after the SN explosion exhibited a low value of $\sim 9 \times 10^{19}$\,cm$^{-2}$, which is equal to the combined value of the Galactic and the SN host.
The negligible absorption from the CSM may indicate that the forward shock has already traversed the entire CSM at this phase. 
Assuming a constant shock velocity of \( v_{\rm{sh}} \approx 10^{9} \ \rm{cm \, s^{-1}} \), we infer that the CSM was confined within a radius of $R\lesssim1.3\times10^{14}$\,cm.
With a wind velocity of \( 10 \ \rm{km \, s^{-1}} \), the rather compact CSM suggests that the progenitor had experienced significant mass ejection roughly four years before the explosion. 

\begin{figure}
    \centering
    \includegraphics[width=1\columnwidth]{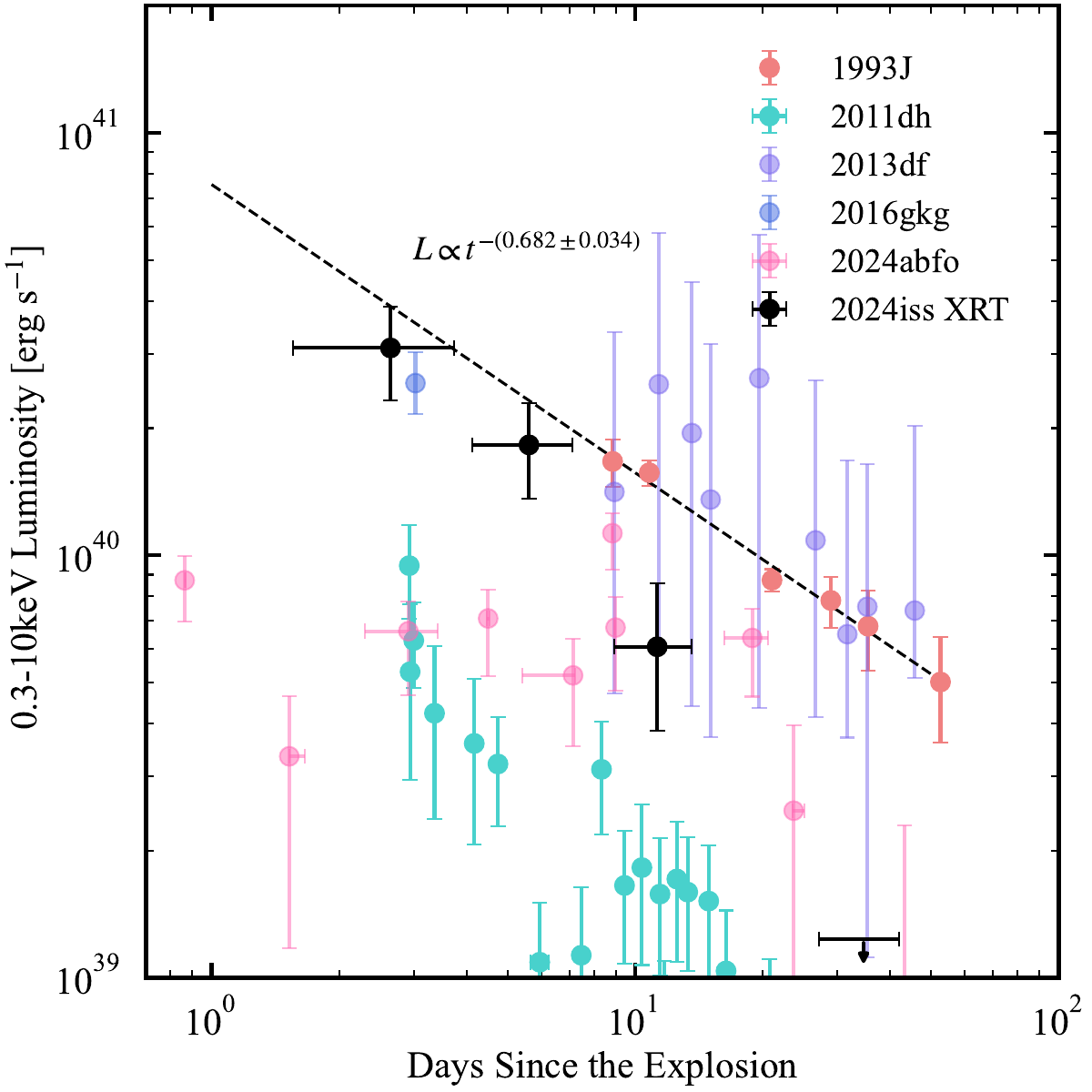}
    \caption{The 0.3--10\,keV X-ray luminosity evolution of SN\,2024iss, compared with those of
    SNe\,1993J \citep{2009Chandra}, 2011dh \citep{Soderberg2012}, 2013df \citep{Kamble2016}, 2016gkg \citep{Vikram2025}, and 2024abfo \citep{reguitti2025}. 
    While most of the comparison data are in the 0.3--10\,keV band, the fluxes of SNe\,1993J and 2013df were measured in the 0.3--8 keV band.
    }
    \label{fig:luminosity_X_ray_compare}
\end{figure}

In Figure~\ref{fig:luminosity_X_ray_compare}, we compare the 0.3--10 Kev X-ray luminosity evolution of SN\,2024iss with those of several other Type IIb SNe, including SNe\,1993J \citep{2009Chandra}, 2011dh \citep{Soderberg2012}, 2013df \citep{Kamble2016}, 2016gkg \citep{Vikram2025} and 2024abfo \citep{reguitti2025}. The X-ray luminosity of SNe\,1993J and 2013df were estimated in 0.3--8 keV band. The uncertainty introduced from different band can be neglected, as the X-ray emission contributes little above 8 keV.
Among the comparison sample, SN\,1993J is found to follow a power-law decline with an index of $\sim-$0.68 (see the black dashed line in Figure~\ref{fig:luminosity_X_ray_compare}). While SN\,2024iss, SN\,1993J, and SN\,2016gkg exhibit an overall higher X-ray luminosity in comparison with SN\,2011dh and SN\,2024abfo. 

The high X-ray luminosity of the former three events can be attributed to an extended hydrogen-rich envelope and dense wind-like CSM \citep{Kamble2016}. In contrast, those Type IIb SNe that are faint in X-ray band may arise from more compact progenitors with a weaker ejecta-CSM interaction. Following the framework of \citet{Chevalier_2010}, we classify SN\,2024iss as Type eIIb, which is compatible with its double-peaked optical light curve and high mass-loss rate, as observed in SNe\,1993J, 2013df and 2011hs.

\section{Discussions} \label{sec:Discussion}

\subsection{The progenitor of SN\,2024iss}
The best fit to the Arnett model based on the second light curve peak of SN\,2024iss suggests an ejecta mass of $M_{\rm ej}=1.27 \pm 0.34\, M_{\odot}$ (see Section~\ref{sec:Ni_mass_estimate}).
Assuming a typical mass for the neutron star remnant of $1.4\, M_{\odot}$~\citep{Thorsett1999}, 
we estimate a helium core mass of 
$M_{remnant}+M_{ej}+M_{env} = 2.8\, M_{\odot}$. 
Using the progenitor grid of \citet{Sukhbold_2016}, this indicates a main-sequence mass of $M_{ZAMS} = 9-11\, M_{\odot}$, which is similar to that inferred for SN\,2016gkg~\citep{Kilpatrick_2022} and compatible with the predicted mass range of Type IIb progenitors in a binary system~\citep{Yoon_2017}.
By fitting the first peak in the optical light curves with the shock-cooling model presented by~\citetalias{Piro_2021}, we infer that the mass and radius of the hydrogen envelope of SN\,2024iss are $M_{\rm env}=0.11\pm0.04$\,M$_{\odot}$ and $R_{\rm env}=224\pm43$\,R$_{\odot}$ (see Section~\ref{sec:shock_cooling_model}). 
The inferred mass and radius favors for a YSG progenitor star. In contrast, models involving single Wolf-Rayet stars are not able to produce an ejecta mass lower than 5~$M_{\odot}$ \citep{Lyman2016}. 
A binary system may naturally account for the partially depleted hydrogen envelope required for the progenitor of SN\,2024iss.

\subsection{SN\,2024iss: Bridging the Gap Between subclasses of Type eIIb and Type cIIb}

Note that in the classification scheme of~\citet{Yoon_2017}, the envelope mass of SN\,2024iss is close to the threshold of 0.15\,M$_{\odot}$ adopted to distinguish between Types eIIb and cIIb. 
For comparison, the compact Type IIb SN\,2016gkg has an envelope mass of $\sim$ 0.03\,M$_{\odot}$ \citep{Piro_2021}, while the extended Type IIb SN\,1993J exhibits a much larger envelope mass of $\sim$ 0.2\,M$_{\odot}$ \citep{Woosley1994}. Similarly, the envelope radius of SN\,2024iss is larger than that of the compact SN\,2008ax ($\sim$~40~R$_{\odot}$; \citealt{Folatelli2015}), but smaller than that of SN\,1993J ($\sim$ 575~R$_{\odot}$; \citealt{Woosley1994}). 
Additionally, as illustrated in Fig.~\ref{fig:luminosity_X_ray_compare}, the X-ray luminosity of SN\,2024iss is overall higher than that of Type IIb SNe with a less extended envelope, e.g., SNe\,2011dh and 2016gkg, but lower than that of SN\,1993J, whose progenitor had a more massive and extended hydrogen envelope.
Taken together, the envelope properties and X-ray emission suggest that SN\,2024iss likely represents a transitional object lying between the compact and extended subtypes of Type IIb SNe.

\begin{figure}
    \centering
    \includegraphics[width=1\columnwidth]{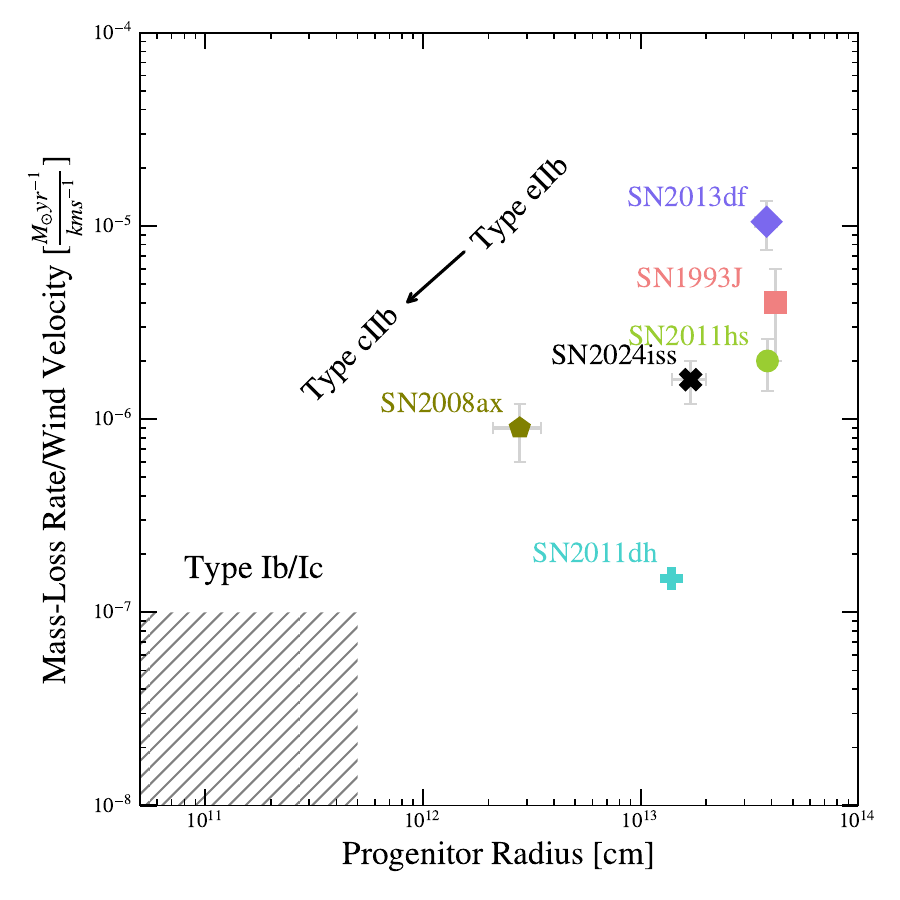}
    \caption{Comparison between the mass-loss rate normalized by wind velocity ($\dot{M}/v_{\rm w}$) and the progenitor radius. The shaded region indicates typical CSM densities of Type Ib/c SNe\citep{Chevalier2006}, though these estimates may suffer from systematic uncertainties of up to an order of magnitude \citep{Maeda2012}. Color-coded symbols represent the estimated values for different SNe as labeled.
    The upper-right to lower-left arrow illustrates a possible evolutionary sequence from Type~eIIb to cIIb and eventually to Type Ib/Ic SNe.}
    \label{fig:CSM_envelope_comparision}
\end{figure}

To further investigate the properties of SN\,2024iss and the CSM-envelope connection, following \citet{Maeda2015}, we plot in Figure~\ref{fig:CSM_envelope_comparision} the mass-loss rate per unit wind velocity ($\dot{M}/v_{\rm w}$) against the envelope radius. The choice of the former parameter minimizes the uncertainties due to the assumption of wind velocities.
The mass-loss rates are estimated from the X-ray analyses of SNe\,1993J \citep{Fransson1996}, 2008ax \citep{Roming_2009}, 2011dh \citep{Maeda_2014}, and 2013df \citep{Kamble2016}. The mass-loss rate of SN 2013df was adopted from \citet {Bufano2014}, based on an analysis of its radio light curves.
The progenitor radius was inferred through a comparison between the SED yielded from archival pre-explosion images obtained by the {\it Hubble Space Telescope} and stellar evolution tracks, i.e., SNe\,1993J \citep{Maund2004}, 2011dh \citep{Maund2011}, 2013df \citep{Dyk2014}, and 2008ax \citep{Folatelli2015}.
The progenitor radius of SN\,2011hs was inferred by modeling its bolometric light curve~\citep{Bufano2014b}.
We adopt the progenitor radii of Type Ib/c SNe from \citet{Maeda2012}. The mass-loss rates of Type Ib/c SNe were inferred from their radio light curves, assuming a typical wind velocity of $\sim$1000\,km\,s$^{-1}$ \citep{Chevalier2006}.

As shown in Fig. \ref{fig:CSM_envelope_comparision}, SN\,2024iss lies in between the eIIb and cIIb groups.
This supports the empirical relation proposed by \citet{Maeda2015}, where more extended progenitors tend to have experienced a stronger mass loss. Although SN\,2024iss seems to have a smaller progenitor radius compared to that of
SN\,2011dh, this discrepancy may stem from systematic differences in envelope radii
as estimated from the light curve fitting and comparison between the progenitor SED and stellar evolution models.
Moreover, there exists a transitional subtype between SNe IIL and SNe IIb, Type II short-plateau SNe. These events show plateaus lasting only several tens of days, and are characterized by relatively massive hydrogen envelopes ($\sim1.7 \,M_\odot$) and high mass-loss rates of $\sim10^{-2}\ M_{\odot}\ \mathrm{yr}^{-1}$, as inferred for the sample presented by \citet{Hiramatsu_2021}. Furthermore, \citet{farah2025d} classified transitional IIb events lying between Type II short-plateau SNe and Type IIb SNe, with envelope masses of around $0.6-0.8\,M_\odot$. We expect Type II short-plateau SNe and these transitional Type IIb SNe to emerge in the upper-right region of Fig. \ref{fig:CSM_envelope_comparision}, offering further insights into the relationship between CSM and envelope in hydrogen-rich SNe. 

\subsection{Implications of the Low Ejecta Mass of SN\,2024iss} 
\label{subsec:low_ejecta_mass}

The Arnett model fit to the bolometric light curve of SN\,2024iss suggests an ejecta mass of 1.27$\pm$0.34\,M$_{\odot}$ and a \isotope[56]{Ni} mass of 0.117$\pm$0.013\,M$_{\odot}$.
While the latter is consistent with the average value derived for SNe IIb,
the ejecta mass of SN\,2024iss is lower than all SNe compiled in \citet{Taddia2018}.
The faint and fast Type IIb SN\,2011hs exhibits a similar ejecta mass and explosion energy compared to that of SN\,2024iss, but with a significantly lower \isotope[56]{Ni} mass \citep{Bufano2014}. 

The secondary peak of SN\,2024iss reached $M_V=-17.43\pm 0.26$, which is consistent with the average of $-17.4$~mag from the sample in \citep{Taddia2018}. However, it appears to be a faster decliner as indicated by its $\Delta m_{15}(V) = 1.13$\,mag (see  Section~\ref{sec:Ni_mass_estimate}), compared to the mean value of 0.93\,mag inferred for SNe IIb~\citep{Taddia2018}.
The late-time decay rate also approaches $2$~mag~per~100~days, one of the steepest in the sample. This is consistent with the trend reported by~\citet{Taddia2018}, where objects with larger $\Delta m_{15}(V)$ also exhibit a steeper late-time decline. \citet{Taddia2018} explained this trend by noting that objects with higher $E_{\mathrm{K}}/M_{\mathrm{ej}}$ ratios are less efficient in trapping gamma-rays, and therefore exhibit a steeper late-time decline and a narrower light-curve peak. 

\citet{Prentice_2018} found that the ejecta masses of SNe IIb exhibited a bimodal distribution peaking at 1.9 and 3.9\,M$_{\odot}$, as represented by SNe\,1993J ($M_{\rm ej}\approx2.2$\,M$_{\odot}$,~\citealp{Lyman2016}) and 2016gkg  ($M_{\rm ej}\approx3.4$\,M$_{\odot}$,~\citealp{Bersten_2018}), respectively.
Such a bimodal mass distribution of the ejecta may indicate a discontinuity of the intrinsic properties of Type IIb SNe with one group having more extended and the other having a rather compact envelope.
The ejecta mass of SN\,2024iss falls within the lower ejecta mass regime, consistent with its Type eIIb classification.

\citet{Ayala2025} also found that although the inferred \isotope[56]{Ni} masses of Type eIIb and cIIb SNe are similar, the former group tends to have a lower average ejecta mass.
\citet{Das_2023} studied a sample of Ca-rich SNe IIb, which exhibit low ejecta masses and are inferred to originate from progenitors with $M_{\rm{ZAMS}} < 12~M_{\odot}$. They suggest that progenitors with smaller initial masses may develop larger envelope radii. According to \citet{Dewi2002,Dewi2003}, a progenitor star with such an extended envelope
may expand sufficiently to fill their Roche lobes again and undergo additional episodes of mass transfer. This process can further strip the envelope and enrich the CSM, thus compatible with low ejecta mass observed in SN\,2024iss.

Taken together, these results indicate that Type eIIb and cIIb SNe may arise from progenitors with similar initial ZAMS masses but diverge in their final envelope and ejecta properties. The interaction with its companion star may provide one natural explanation. In particular, an enhanced CSM enrichment during the common-envelope phase may take place before the SN explosion.
\citet{Maeda_2023} emphasized that instead of the initial mass of the exploding star, the binary separation and the orbital period
may play a more critical role in determining the observational properties of SN explosion.

It is important to note, however, that the Arnett model assumes a constant opacity and fixed photospheric velocity, which may introduce systematic uncertainties in the derived parameters. For instance, adopting a photospheric velocity of $\rm 15,000\,km\,s^{-1}$ following \citet{Yamanaka2025}, the Arnett model yields a larger ejecta mass of $M_{\rm{ej}} = 2.64 \pm 0.53~M_{\odot}$, but the corresponding kinetic energy 
($E_{k} = 3.54 \pm 0.83 \times 10^{51} \rm{erg}$) would be unrealistically high for a Type IIb SN. Alternatively, if we adopt an opacity of \(\kappa = 0.2\,\mathrm{cm^2\,g^{-1}}\) as suggested by \citet{Nagy_2016} and maintain the same photospheric velocity of $\rm 15,000\,km\,s^{-1}$, the fit yields $M_{\rm{ej}} = 1.23 \pm 0.23~M_{\odot}$ and $E_{k} = 1.64 \pm 0.3 \times 10^{51} \rm{erg}$. Taken together, these estimates imply an ejecta mass in the range of 1.23--2.64\,$M_{\odot}$. However, considering the steep decline of the radioactive tail, a lower ejecta mass appears more plausible for SN\,2024iss. With additional data obtained during the nebular phase, it should be possible to better constrain the ejecta mass following \citet{Wheeler_2015} and \citet{Haynie_2023}.

\subsection{The mass-loss history of SN\,2024iss: a confined structure}

As discussed in Section \ref{sec:X_ray}, the column density of neutral hydrogen estimated from the X-ray spectrum of SN\,2024iss suggests that by the time of the first detection, the CSM was already engulfed by the SN ejecta. The extension of the former is therefore confined within
a radius of $\sim 1.3 \times 10^{14}$ cm. Assuming a wind velocity of $\rm 10\,km\,s^{-1}$, this constrains the mass ejection to take place no earlier than four
years before the explosion. For a progenitor star with a ZAMS mass of 9-11 $M_{\odot}$, the episode of such a mass-loss wind coincides with the carbon shell burning phase or the onset of neon burning \citep{Woosley2002}. If the eruptive mass loss originates from instabilities intrinsic to the star, it may be triggered by mechanisms such as pulsation-driven superwinds \citep{Yoon_2010}, turbulent convection fluctuations \citep{DavidArnett_2011}, wave-driven mass loss during late burning stages \citep{Shiode_2014, Wu_2021}, or core neutrino emission \citep{Moriya2014}. Alternatively, if the eruption results from a binary interaction, it supports the late Case B mass transfer scenario for Type eIIb progenitors proposed by \citet{Yoon_2017}, in which mass loss continues up to the time of the SN explosion.

The confined CSM structure inferred from the early X-ray observations of SN\,2024iss may be a ubiquitous configuration for certain subtypes of Type IIb SNe. As suggested by \citet{Gal-Yam2014}, an
enhanced mass loss during the final year prior to the explosion can inflate the apparent photospheric radius of the progenitor, potentially causing a Type cIIb to appear as having an extended envelope. This effect may explain the discrepancy between the progenitor radii inferred from the stellar evolution model fitting to the SED of the progenitor star detected in archival images and the shock-cooling model fitting to the early bolometric light curves of the SN.

Although current observations indicate
that SN\,2024iss has a confined CSM, the presence of CSM further away remains uncertain. The boxy hydrogen structure observed in the nebular spectra of SN\,1993J suggests a hydrogen-rich CSM at large radii \citep{Matheson2000}. Given the similarities between SN\,1993J and SN\,2024iss, the nebular spectra of SN\,2024iss\footnote{The relevant data and analysis will be presented in a forthcoming paper.} may enable a more comprehensive probe of the mass loss history of SN\,2024iss .

The synergy of estimating the mass loss rate and the envelope properties of Type IIb SNe may enable stringent constraints on 
the framework of binary interactions \citep{Ouchi_2017}. The progenitor system of SN 2024iss must satisfy two conditions: the envelope shall retain sufficient mass, and the mass-loss rate should be sufficiently high over the final years to build up a rather confined and dense CSM content.

\section{Conclusion} \label{sec:Conclusion}

We present optical, UV and X-ray observations of SN\,2024iss, a Type IIb SN with a prominent double peak in optical bands and bright thermal X-ray emission. We summarize our key findings as follows:

1. The optical light curves of SN\,2024iss exhibit a double-peak feature.
The first $V$-band peak, which can be attributed to shock cooling, occurred at $\sim$2.4\,days after the explosion and reached a peak magnitude of $M_V = -17.33 \pm 0.26$. The second peak, powered by \isotope[56]{Ni} decay, reached a peak magnitude of $M_V = -17.43 \pm 0.26$ at $\sim 18.7$\,days after the SN explosion.
The photometric follow-up campaign on SN\,2024iss was carried out starting from $\sim$0.87 days after the last non-detection, which corresponds to about 0.44 days after the SN explosion. The prompt observations make SN\,2024iss one of the most well-sampled SNe IIb during the early phases. 

2. We measure a photospheric velocity of SN\,2024iss to be $\sim 19,400$\,km\,s$^{-1}$ from the absorption minimum of the P~Cygni profile of H$\alpha$ in the earliest spectrum obtained at day 3.82. The velocity decreased to $\sim 13,600\,km\,s^{-1}$ after the secondary peak of the $V$-band light curve. The H$\alpha$ and H$\beta$ features remain detectable in our last spectrum obtained at $t \approx 87$ days after the SN explosion, indicating the presence of a hydrogen-rich ejecta, which is similar to that observed in SN\,1993J~\citep{Filippenko1994}. 

3. We fit the second peak of the pseudo-bolometric light curve of SN\,2024iss with an Arnett-like model and adopt a photospheric velocity of 7500~km~s$^{-1}$ based on the Fe{\sc\,II} $\lambda$5169 line measured around the V-band maximum. The best-fitting parameters are $M_{\rm{Ni}} = 0.117 \pm 0.013~M_{\odot}$, $M_{\rm{ej}} = 1.272 \pm 0.343~M_{\odot}$ and $E_{k} = 0.427 \pm 0.115 \times 10^{51} \rm{erg}$. The inferred \isotope[56]{Ni} mass is typical for Type IIb SNe, while the relatively low ejecta mass is compatible
with the rapid decline in luminosity, as indicated by the large $\Delta m_{15}$ value of 1.13~mag, compared to an average of 0.93~mag for Type IIb SNe \citep{Taddia2018}.

4. We fit the multiband shock-cooling peak of SN\,2024iss using the \citetalias{Piro_2021} model, which provides a plausible description of the data. The best-fit parameters are: $R_{\rm{env}} = 244\pm43~R_{\odot}$, $M_{\rm{env}} = 0.11 \pm 0.04~M_{\odot}$, and $v_{\rm{s}} = (1.67 \pm 0.07)\times 10^{4}$~km~s$^{-1}$. 

5. SN\,2024iss exhibits bright thermal bremsstrahlung emission from the forward shock. The strength of the emission is comparable to that observed in SNe\,1993J and 2013df. Assuming a wind velocity of 10\,km\,s$^{-1}$, the best fit to the X-ray light curve of SN\,2024iss indicates
a mass loss rate of $\dot{M} = 1.6 \times 10^{-5}\ M_{\odot}\ \rm{yr}^{-1}$ and a
shock velocity of $9.5 \times 10^8\ \rm{cm~s^{-1}}$. An independent estimate based on the EM gives a slightly higher mass-loss rate of $(5.55 \pm 1.57) \times 10^{-5}~M_{\odot}~\rm{yr}^{-1}$.

6. The low column density of neutral hydrogen of SN\,2024iss placed a stringent constraints on the extension of the CSM within a radius of \( R \lesssim 1.3 \times 10^{14}~\rm{cm} \). Assuming a wind velocity of \( 10~\rm{km~s^{-1}} \), this indicates that a significant mass ejection may have occurred within about four years prior to the SN explosion.

The combination of the optical and X-ray observations provides a powerful approach to probe the physical properties of SESNe.
In particular, the prompt X-ray emissions from SNe obtained within the first hours after the explosion, as facilitated by wide-field sky surveys such as EP~\citep{Yuan2022}, will enable a systematic characterization of their immediate environment and shed light on the mass-loss process and explosion mechanisms of SESNe.
 

\section{Acknowledgements} \label{sec:Acknowledgements}

We thank Yi Yang for helpful suggestions on the writing and for valuable discussions. 
This work is supported by the National Science Foundation of China (NSFC grants 12288102 and 12033003), the Tencent Xplorer Prize, and the Ma Huateng Foundation. The LCO group is supported by U.S. National Science Foundation (NSF) grants AST-1911151 and AST-2308113.
Jujia Zhang is supported by the National Key Research and Development Program of China (NKRDPC) with grant 2021YFA1600404, the NSFC(grants 12173082, 12333008, and 12225304), the Yunnan Fundamental Research Projects (YFRP; grants 202501AV070012, 202401BC070007, 202201AT070069, and 202501AS070005), the Top-notch Young Talents Program of Yunnan Province, the Light of West China Program provided by the Chinese Academy of Sciences (CAS), and the International Centre of Supernovae, Yunnan Key Laboratory (grant 202302AN360001). WXL is supported by NSFC (12120101003 and 12373010), the National Key R\&D Program (2022YFA1602902 and 2023YFA1607804), and Strategic Priority Research Program of CAS (XDB0550100 and XDB0550000).

Time-domain research by the University of Arizona team and D.J.S. is supported by NSF grants AST-2108032, AST-2308181, AST-2407566, and AST-2432036, and the Heising-Simons Foundation under grant 2020-1864.
A.P., A.R.,  and G.V. acknowledge support from the PRIN-INAF 2022 project ``Shedding light on the nature of gap transients: from the observations to the model.''
A.R. also acknowledges financial support from the GRAWITA Large Program Grant (PI P. D'Avanzo). Y.-Z. Cai is supported by NSFC (grant 12303054), the NKRDPC (grant 2024YFA1611603), the YFRP (grants 202401AU070063, 202501AS070078), and the International Centre of Supernovae, Yunnan Key Laboratory (grant 202302AN360001). J.V. is supported by the NKFIH-OTKA Grant K142534.
Maokai Hu is funded by China Postdoctoral Science Foundation (Certificate Number 2025T180872).
This work is sponsored by the Natural Science Foundation of Xinjiang Uygur Autonomous Region under grant 2024D01D32, Tianshan Talent Training Program grants 2023TSYCLJ0053 and 2023TSYCCX0101, the Central Guidance for Local Science and Technology Development Fund under grant ZYYD2025QY27, and the NSFC (grants 12433007, 12373038).
A.V.F.'s research group at UC Berkeley acknowledges financial assistance from the Christopher R. Redlich Fund, as well as donations 
from Gary and Cynthia Bengier, Clark and Sharon Winslow, Alan Eustace and Kathy Kwan, William Draper, Timothy and Melissa Draper, Briggs and Kathleen Wood, Sanford Robertson (W.Z. is a Bengier-Winslow-Eustace Specialist in Astronomy, T.G.B. is a Draper-Wood-Robertson Specialist in Astronomy), and numerous other donors.       
 
This study uses observations obtained with the Hobby-Eberly Telescope, which is a joint project of the University of Texas at Austin, the Pennsylvania State University, Ludwig-Maximilians-Universit{\"a}t M{\"u}nchen, and Georg-August-Universit{\"a}t G{\"o}ttingen. The HET is named in honor of its principal benefactors, William P. Hobby and Robert E. Eberly. The Low Resolution Spectrograph 2 (LRS2) was developed and funded by the University of Texas at Austin McDonald Observatory and Department of Astronomy and by Pennsylvania State University. We thank the Leibniz-Institut f{\"u}r Astrophysik Potsdam (AIP) and the Institut f{\"u}r Astrophysik G{\"o}ttingen (IAG) for their contributions to the construction of the integral field units. 

Based in part on observations obtained as part of (GN-2024A-Q-403; P.I. Jennifer Andrews) at the international Gemini Observatory, a program of the U.S. NSF NOIRLab, which is managed by the Association of Universities for Research in Astronomy (AURA) under a cooperative agreement with the U.S. NSF on behalf of the Gemini Observatory partnership: the NSF (United States), National Research Council (Canada), Agencia Nacional de Investigaci\'{o}n y Desarrollo (Chile), Ministerio de Ciencia, Tecnolog\'{i}a e Innovaci\'{o}n (Argentina), Minist\'{e}rio da Ci\^{e}ncia, Tecnologia, Inova\c{c}\~{o}es e Comunica\c{c}\~{o}es (Brazil), and Korea Astronomy and Space Science Institute (Republic of Korea). These observations were processed using DRAGONS (Data Reduction for Astronomy from Gemini Observatory North and South). This work was enabled by observations made from the Gemini North telescope, located within the Maunakea Science Reserve and adjacent to the summit of Maunakea. We are grateful for the privilege of observing the Universe from a place that is unique in both its astronomical quality and its cultural significance. 
 
 The LBT is an international collaboration among institutions in the United States, Italy, and Germany. LBT Corporation Members are The University of Arizona on behalf of the Arizona Board of Regents; Istituto Nazionale di Astrofisica, Italy; LBT Beteiligungsgesellschaft, Germany, representing the Max-Planck Society, The Leibniz Institute for Astrophysics Potsdam, and Heidelberg University; The Ohio State University, and The Research Corporation, on behalf of The University of Notre Dame, University of Minnesota, and University of Virginia.
 A major upgrade of the Kast spectrograph on the Shane 3 m telescope at        
Lick Observatory, led by Brad Holden, was made possible through generous gifts from the Heising-Simons Foundation, William and Marina Kast, and the University of California Observatories.  Research at Lick Observatory is partially supported by a generous gift from Google. 

This paper made use of the modsCCDRed data-reduction code developed in part with funds provided by NSF grants AST-9987045 and AST-1108693. Some observations reported here were obtained at the MMT Observatory, a joint facility of the University of Arizona and the Smithsonian Institution. We acknowledge the observational data taken at Einstein Probe (EP), a space mission supported by the Strategic Priority Program on Space Science of the Chinese Academy of Sciences, in collaboration with ESA, MPE, and CNES (grant XDA15310000), the Strategic Priority Research Program of the Chinese Academy of Sciences (grant XDB0550200) and the 
NKRDPC (grant 2022YFF0711500). We also acknowledge the observational data taken with Swift and NuSTAR.

%

\bibliographystyle{aa}
\bibliography{reference}

\begin{thebibliography}{155}
\expandafter\ifx\csname natexlab\endcsname\relax\def\natexlab#1{#1}\fi

\bibitem[{{Arcavi} {et~al.}(2011){Arcavi}, {Gal-Yam}, {Yaron}, {Sternberg}, {Rabinak}, {Waxman}, {Kasliwal}, {Quimby}, {Ofek}, {Horesh}, {Kulkarni}, {Filippenko}, {Silverman}, {Cenko}, {Li}, {Bloom}, {Sullivan}, {Nugent}, {Poznanski}, {Gorbikov}, {Fulton}, {Howell}, {Bersier}, {Riou}, {Lamotte-Bailey}, {Griga}, {Cohen}, {Hachinger}, {Polishook}, {Xu}, {Ben-Ami}, {Manulis}, {Walker}, {Maguire}, {Pan}, {Matheson}, {Mazzali}, {Pian}, {Fox}, {Gehrels}, {Law}, {James}, {Marchant}, {Smith}, {Mottram}, {Barnsley}, {Kandrashoff}, \& {Clubb}}]{Arcavi2011}
{Arcavi}, I., {Gal-Yam}, A., {Yaron}, O., {et~al.} 2011, \apjl, 742, L18

\bibitem[{Arcavi {et~al.}(2017)Arcavi, Hosseinzadeh, Brown, Smartt, Valenti, Tartaglia, Piro, Sanchez, Nicholls, Monard, Howell, McCully, Sand, Tonry, Denneau, Stalder, Heinze, Rest, Smith, \& Bishop}]{Arcavi_2017}
Arcavi, I., Hosseinzadeh, G., Brown, P.~J., {et~al.} 2017, The Astrophysical Journal Letters, 837, L2

\bibitem[{{Arnett}(1982)}]{Arnett1982}
{Arnett}, W.~D. 1982, \apj, 253, 785

\bibitem[{{Ayala} {et~al.}(2025){Ayala}, {Anderson}, {Pignata}, {Foerster}, {Smartt}, {Rest}, {Solar}, {Erasmus}, {Dastidar}, {Ramirez}, \& {Pineda-Garcia}}]{Ayala2025}
{Ayala}, B., {Anderson}, J.~P., {Pignata}, G., {et~al.} 2025, arXiv e-prints, arXiv:2503.05909

\bibitem[{{Bai} {et~al.}(2020){Bai}, {Feng}, {Zhang}, {Niu}, {Eskandar}, {Pu}, {Ma}, {Liu}, {Jiang}, {Ma}, {Esamdin}, \& {Wang}}]{Bai2020}
{Bai}, C.-H., {Feng}, G.-J., {Zhang}, X., {et~al.} 2020, Research in Astronomy and Astrophysics, 20, 211

\bibitem[{Barbieri {et~al.}(1994)Barbieri, Bhatia, Bonoli, Bortoletto, Ciani, Conconi, D'Alessandro, Fantinel, Mancini, Maurizio, Ortolani, Pucillo, Rafanelli, Ragazzoni, Zambon, \& Zitelli}]{Barbieri1994}
Barbieri, C., Bhatia, R.~K., Bonoli, C., {et~al.} 1994, in Advanced Technology Optical Telescopes V, ed. L.~M. Stepp, Vol. 2199, International Society for Optics and Photonics (SPIE), 10 -- 21

\bibitem[{{Barbon} {et~al.}(1995){Barbon}, {Benetti}, {Cappellaro}, {Patat}, {Turatto}, \& {Iijima}}]{Barbon1995}
{Barbon}, R., {Benetti}, S., {Cappellaro}, E., {et~al.} 1995, \aaps, 110, 513

\bibitem[{{Bellm} {et~al.}(2019){Bellm}, {Kulkarni}, {Graham}, {Dekany}, {Smith}, {Riddle}, {Masci}, {Helou}, {Prince}, {Adams}, {Barbarino}, {Barlow}, {Bauer}, {Beck}, {Belicki}, {Biswas}, {Blagorodnova}, {Bodewits}, {Bolin}, {Brinnel}, {Brooke}, {Bue}, {Bulla}, {Burruss}, {Cenko}, {Chang}, {Connolly}, {Coughlin}, {Cromer}, {Cunningham}, {De}, {Delacroix}, {Desai}, {Duev}, {Eadie}, {Farnham}, {Feeney}, {Feindt}, {Flynn}, {Franckowiak}, {Frederick}, {Fremling}, {Gal-Yam}, {Gezari}, {Giomi}, {Goldstein}, {Golkhou}, {Goobar}, {Groom}, {Hacopians}, {Hale}, {Henning}, {Ho}, {Hover}, {Howell}, {Hung}, {Huppenkothen}, {Imel}, {Ip}, {Ivezi{\'c}}, {Jackson}, {Jones}, {Juric}, {Kasliwal}, {Kaspi}, {Kaye}, {Kelley}, {Kowalski}, {Kramer}, {Kupfer}, {Landry}, {Laher}, {Lee}, {Lin}, {Lin}, {Lunnan}, {Giomi}, {Mahabal}, {Mao}, {Miller}, {Monkewitz}, {Murphy}, {Ngeow}, {Nordin}, {Nugent}, {Ofek}, {Patterson}, {Penprase}, {Porter}, {Rauch}, {Rebbapragada}, {Reiley}, {Rigault}, {Rodriguez}, {van Roestel}, {Rusholme}, {van
  Santen}, {Schulze}, {Shupe}, {Singer}, {Soumagnac}, {Stein}, {Surace}, {Sollerman}, {Szkody}, {Taddia}, {Terek}, {Van Sistine}, {van Velzen}, {Vestrand}, {Walters}, {Ward}, {Ye}, {Yu}, {Yan}, \& {Zolkower}}]{2019PASP..131a8002B}
{Bellm}, E.~C., {Kulkarni}, S.~R., {Graham}, M.~J., {et~al.} 2019, \pasp, 131, 018002

\bibitem[{{Benetti} {et~al.}(1994){Benetti}, {Patat}, {Turatto}, {Contarini}, {Gratton}, \& {Cappellaro}}]{Benetti1994}
{Benetti}, S., {Patat}, F., {Turatto}, M., {et~al.} 1994, \aap, 285, L13

\bibitem[{Bersten {et~al.}(2018)Bersten, Folatelli, Garc{\'i}a, Van~Dyk, Benvenuto, Orellana, Buso, S{\'a}nchez, Tanaka, Maeda, Filippenko, Zheng, Brink, Cenko, de~Jaeger, Kumar, Moriya, Nomoto, Perley, Shivvers, \& Smith}]{Bersten_2018}
Bersten, M.~C., Folatelli, G., Garc{\'i}a, F., {et~al.} 2018, Nature, 554, 497

\bibitem[{{Breeveld} {et~al.}(2011){Breeveld}, {Landsman}, {Holland}, {Roming}, {Kuin}, \& {Page}}]{Breeveld2011AIPC.1358..373B}
{Breeveld}, A.~A., {Landsman}, W., {Holland}, S.~T., {et~al.} 2011, in American Institute of Physics Conference Series, Vol. 1358, Gamma Ray Bursts 2010, ed. J.~E. {McEnery}, J.~L. {Racusin}, \& N.~{Gehrels}, 373--376

\bibitem[{{Brennan} \& {Fraser}(2022)}]{Brennan2022}
{Brennan}, S.~J. \& {Fraser}, M. 2022, \aap, 667, A62

\bibitem[{{Brown} {et~al.}(2013){Brown}, {Baliber}, {Bianco}, {Bowman}, {Burleson}, {Conway}, {Crellin}, {Depagne}, {De Vera}, {Dilday}, {Dragomir}, {Dubberley}, {Eastman}, {Elphick}, {Falarski}, {Foale}, {Ford}, {Fulton}, {Garza}, {Gomez}, {Graham}, {Greene}, {Haldeman}, {Hawkins}, {Haworth}, {Haynes}, {Hidas}, {Hjelstrom}, {Howell}, {Hygelund}, {Lister}, {Lobdill}, {Martinez}, {Mullins}, {Norbury}, {Parrent}, {Paulson}, {Petry}, {Pickles}, {Posner}, {Rosing}, {Ross}, {Sand}, {Saunders}, {Shobbrook}, {Shporer}, {Street}, {Thomas}, {Tsapras}, {Tufts}, {Valenti}, {Vander Horst}, {Walker}, {White}, \& {Willis}}]{Brown2013}
{Brown}, T.~M., {Baliber}, N., {Bianco}, F.~B., {et~al.} 2013, \pasp, 125, 1031

\bibitem[{{Bufano} {et~al.}(2014{\natexlab{a}}){Bufano}, {Pignata}, {Bersten}, {Mazzali}, {Ryder}, {Margutti}, {Milisavljevic}, {Morelli}, {Benetti}, {Cappellaro}, {Gonzalez-Gaitan}, {Romero-Ca{\~n}izales}, {Stritzinger}, {Walker}, {Anderson}, {Contreras}, {de Jaeger}, {F{\"o}rster}, {Gutierrez}, {Hamuy}, {Hsiao}, {Morrell}, {Olivares E.}, {Paillas}, {Parker}, {Pian}, {Pickering}, {Sanders}, {Stockdale}, {Turatto}, {Valenti}, {Fesen}, {Maza}, {Nomoto}, {Phillips}, \& {Soderberg}}]{Bufano2014}
{Bufano}, F., {Pignata}, G., {Bersten}, M., {et~al.} 2014{\natexlab{a}}, \mnras, 439, 1807

\bibitem[{{Bufano} {et~al.}(2014{\natexlab{b}}){Bufano}, {Pignata}, {Bersten}, {Mazzali}, {Ryder}, {Margutti}, {Milisavljevic}, {Morelli}, {Benetti}, {Cappellaro}, {Gonzalez-Gaitan}, {Romero-Ca{\~n}izales}, {Stritzinger}, {Walker}, {Anderson}, {Contreras}, {de Jaeger}, {F{\"o}rster}, {Gutierrez}, {Hamuy}, {Hsiao}, {Morrell}, {Olivares E.}, {Paillas}, {Parker}, {Pian}, {Pickering}, {Sanders}, {Stockdale}, {Turatto}, {Valenti}, {Fesen}, {Maza}, {Nomoto}, {Phillips}, \& {Soderberg}}]{Bufano2014b}
{Bufano}, F., {Pignata}, G., {Bersten}, M., {et~al.} 2014{\natexlab{b}}, \mnras, 439, 1807

\bibitem[{{Burrows} {et~al.}(2005){Burrows}, {Hill}, {Nousek}, {Kennea}, {Wells}, {Osborne}, {Abbey}, {Beardmore}, {Mukerjee}, {Short}, {Chincarini}, {Campana}, {Citterio}, {Moretti}, {Pagani}, {Tagliaferri}, {Giommi}, {Capalbi}, {Tamburelli}, {Angelini}, {Cusumano}, {Br{\"a}uninger}, {Burkert}, \& {Hartner}}]{Burrows2005}
{Burrows}, D.~N., {Hill}, J.~E., {Nousek}, J.~A., {et~al.} 2005, \ssr, 120, 165

\bibitem[{{Cardelli} {et~al.}(1989){Cardelli}, {Clayton}, \& {Mathis}}]{Cardelli_1989}
{Cardelli}, J.~A., {Clayton}, G.~C., \& {Mathis}, J.~S. 1989, \apj, 345, 245

\bibitem[{{Chandra} {et~al.}(2009){Chandra}, {Dwarkadas}, {Ray}, {Immler}, \& {Pooley}}]{2009Chandra}
{Chandra}, P., {Dwarkadas}, V.~V., {Ray}, A., {Immler}, S., \& {Pooley}, D. 2009, \apj, 699, 388

\bibitem[{{Chen} {et~al.}(2020){Chen}, {Cui}, {Han}, {Wang}, {Yang}, {Wang}, {Li}, {Ma}, {Xu}, {Lu}, {Chen}, {Tang}, {Yuan}, {Friedrich}, {Meidinger}, {Keil}, {Burwitz}, {Eder}, {Hartmann}, {Nandra}, {Keereman}, {Santovincenzo}, {Vernani}, {Bianucci}, {Valsecchi}, {Wang}, {Wang}, {Wang}, {Li}, {Sheng}, {Qiang}, {Shi}, {Chao}, {Song}, {Zhang}, {Huo}, {Wang}, {Cong}, {Yang}, {Hou}, {Zhao}, {Zhao}, {Chen}, {Li}, {Zhang}, {Luo}, {Xu}, {Li}, {Zhang}, {Bi}, {Zhu}, {Yu}, {Chen}, {Lv}, {Lu}, \& {Zhang}}]{2020SPIE11444E..5BC}
{Chen}, Y., {Cui}, W., {Han}, D., {et~al.} 2020, in Society of Photo-Optical Instrumentation Engineers (SPIE) Conference Series, Vol. 11444, Space Telescopes and Instrumentation 2020: Ultraviolet to Gamma Ray, ed. J.-W.~A. {den Herder}, S.~{Nikzad}, \& K.~{Nakazawa}, 114445B

\bibitem[{{Chevalier} \& {Fransson}(2006)}]{Chevalier2006}
{Chevalier}, R.~A. \& {Fransson}, C. 2006, \apj, 651, 381

\bibitem[{Chevalier \& Fransson(2017)}]{Chevalier2017}
Chevalier, R.~A. \& Fransson, C. 2017, Thermal and Non-thermal Emission from Circumstellar Interaction (Springer International Publishing), 875--937

\bibitem[{Chevalier \& Soderberg(2010)}]{Chevalier_2010}
Chevalier, R.~A. \& Soderberg, A.~M. 2010, The Astrophysical Journal, 711, L40

\bibitem[{Chonis {et~al.}(2014)Chonis, Hill, Lee, Tuttle, \& Vattiat}]{chonis_lrs2:_2014}
Chonis, T.~S., Hill, G.~J., Lee, H., Tuttle, S.~E., \& Vattiat, B.~L. 2014, in Ground-Based and {{Airborne Instrumentation}} for {{Astronomy V}}, Vol. 9147 ({International Society for Optics and Photonics}), 91470A

\bibitem[{Das {et~al.}(2023)Das, Kasliwal, Fremling, Yang, Schulze, Sollerman, Sit, De, Tzanidakis, Perley, Anand, Andreoni, Barbarino, Brudge, Drake, Gal-Yam, Laher, Karambelkar, Kulkarni, Masci, Medford, Polin, Reedy, Riddle, Sharma, Smith, Yan, Yang, \& Yao}]{Das_2023}
Das, K.~K., Kasliwal, M.~M., Fremling, C., {et~al.} 2023, The Astrophysical Journal, 959, 12

\bibitem[{David~Arnett \& Meakin(2011)}]{DavidArnett_2011}
David~Arnett, W. \& Meakin, C. 2011, The Astrophysical Journal, 741, 33

\bibitem[{{Dessart} {et~al.}(2024){Dessart}, {Guti{\'e}rrez}, {Ercolino}, {Jin}, \& {Langer}}]{Dessart2024}
{Dessart}, L., {Guti{\'e}rrez}, C.~P., {Ercolino}, A., {Jin}, H., \& {Langer}, N. 2024, \aap, 685, A169

\bibitem[{{Dessart} \& {Hillier}(2005)}]{Dessart2005}
{Dessart}, L. \& {Hillier}, D.~J. 2005, \aap, 439, 671

\bibitem[{{Dewi} \& {Pols}(2003)}]{Dewi2003}
{Dewi}, J.~D.~M. \& {Pols}, O.~R. 2003, \mnras, 344, 629

\bibitem[{{Dewi} {et~al.}(2002){Dewi}, {Pols}, {Savonije}, \& {van den Heuvel}}]{Dewi2002}
{Dewi}, J.~D.~M., {Pols}, O.~R., {Savonije}, G.~J., \& {van den Heuvel}, E.~P.~J. 2002, \mnras, 331, 1027

\bibitem[{Dong {et~al.}(2024)Dong, Valenti, Ashall, Williamson, Sand, Van~Dyk, Filippenko, Jha, Lundquist, Modjaz, Andrews, Jencson, Hosseinzadeh, Pearson, Kwok, Boland, Hsiao, Smith, Elias-Rosa, Srivastav, Smartt, Fulton, Zheng, Brink, Shahbandeh, Bostroem, Hoang, Janzen, Mehta, Meza, Shrestha, Wyatt, Auchettl, Burns, Farah, Galbany, Padilla~Gonzalez, Haislip, Hinkle, Howell, De~Jaeger, Kouprianov, Kumar, Lu, McCully, Moran, Morrell, Newsome, Pellegrino, Polin, Reichart, Shappee, Stritzinger, Terreran, \& Tucker}]{Dong_2024}
Dong, Y., Valenti, S., Ashall, C., {et~al.} 2024, The Astrophysical Journal, 974, 316

\bibitem[{Dwarkadas(2025)}]{Vikram2025}
Dwarkadas, V.~V. 2025, On the X-ray Emission From Supernovae, and Implications for the Mass-Loss Rates of their Progenitor Stars

\bibitem[{Ergon {et~al.}(2014)Ergon, Sollerman, Fraser, Pastorello, Taubenberger, Elias-Rosa, Bersten, Jerkstrand, Benetti, Botticella, Fransson, Harutyunyan, Kotak, Smartt, Valenti, Bufano, Cappellaro, Fiaschi, Howell, Kankare, Magill, Mattila, Maund, Naves, Ochner, Ruiz, Smith, Tomasella, \& Turatto}]{Ergon_2014}
Ergon, M., Sollerman, J., Fraser, M., {et~al.} 2014, Astronomy \& Astrophysics, 562, A17

\bibitem[{Evans {et~al.}(2009)Evans, Beardmore, Page, Osborne, O'Brien, Willingale, Starling, Burrows, Godet, Vetere, Racusin, Goad, Wiersema, Angelini, Capalbi, Chincarini, Gehrels, Kennea, Margutti, Morris, Mountford, Pagani, Perri, Romano, \& Tanvir}]{Evans2009}
Evans, P.~A., Beardmore, A.~P., Page, K.~L., {et~al.} 2009, Monthly Notices of the Royal Astronomical Society, 397, 1177

\bibitem[{{Evans, P. A.} {et~al.}(2007){Evans, P. A.}, {Beardmore, A. P.}, {Page, K. L.}, {Tyler, L. G.}, {Osborne, J. P.}, {Goad, M. R.}, {O'Brien, P. T.}, {Vetere, L.}, {Racusin, J.}, {Morris, D.}, {Burrows, D. N.}, {Capalbi, M.}, {Perri, M.}, {Gehrels, N.}, \& {Romano, P.}}]{Evans2007}
{Evans, P. A.}, {Beardmore, A. P.}, {Page, K. L.}, {et~al.} 2007, A\&A, 469, 379

\bibitem[{{Fabricant} {et~al.}(2019){Fabricant}, {Fata}, {Epps}, {Gauron}, {Mueller}, {Zajac}, {Amato}, {Barberis}, {Bergner}, {Brennan}, {Brown}, {Chilingarian}, {Geary}, {Kradinov}, {McLeod}, {Smith}, \& {Woods}}]{Fabricant_2019}
{Fabricant}, D., {Fata}, R., {Epps}, H., {et~al.} 2019, \pasp, 131, 075004

\bibitem[{Fan {et~al.}(2015)Fan, Bai, Zhang, Wang, Chang, Xin, \& Zhang}]{Fan_2015}
Fan, Y.-F., Bai, J.-M., Zhang, J.-J., {et~al.} 2015, Research in Astronomy and Astrophysics, 15, 918

\bibitem[{Farah {et~al.}(2025{\natexlab{a}})Farah, Howell, Hiramatsu, McCully, Andrews, Newsome, Gonzalez, Pellegrino, Berger, Blanchard, Gomez, Kumar, Bostroem, Ni, Gagliano, \& Ravi}]{farah2025d}
Farah, J.~R., Howell, D.~A., Hiramatsu, D., {et~al.} 2025{\natexlab{a}}, When IIb Ceases To Be: Bridging the Gap Between IIb and Short-plateau Supernovae

\bibitem[{Farah {et~al.}(2025{\natexlab{b}})Farah, Howell, Terreran, Irani, Morag, Pellegrino, McCully, Newsome, Gonzalez, Bostroem, Hosseinzadeh, Andrews, Prust, \& Hiramatsu}]{Farah_2025}
Farah, J.~R., Howell, D.~A., Terreran, G., {et~al.} 2025{\natexlab{b}}, The Astrophysical Journal, 984, 60

\bibitem[{Filippenko(1997)}]{Filippenko1997}
Filippenko, A.~V. 1997, Annual Review of Astronomy and Astrophysics, 35, 309

\bibitem[{{Filippenko} {et~al.}(1994){Filippenko}, {Matheson}, \& {Barth}}]{Filippenko1994}
{Filippenko}, A.~V., {Matheson}, T., \& {Barth}, A.~J. 1994, \aj, 108, 2220

\bibitem[{{Filippenko} {et~al.}(1993){Filippenko}, {Matheson}, \& {Ho}}]{1993ApJ...415L.103F}
{Filippenko}, A.~V., {Matheson}, T., \& {Ho}, L.~C. 1993, \apjl, 415, L103

\bibitem[{{Fixsen} {et~al.}(1996){Fixsen}, {Cheng}, {Gales}, {Mather}, {Shafer}, \& {Wright}}]{Fixsen_1996}
{Fixsen}, D.~J., {Cheng}, E.~S., {Gales}, J.~M., {et~al.} 1996, \apj, 473, 576

\bibitem[{{Foight} {et~al.}(2016){Foight}, {G{\"u}ver}, {{\"O}zel}, \& {Slane}}]{Foight2016}
{Foight}, D.~R., {G{\"u}ver}, T., {{\"O}zel}, F., \& {Slane}, P.~O. 2016, \apj, 826, 66

\bibitem[{{Folatelli} {et~al.}(2015){Folatelli}, {Bersten}, {Kuncarayakti}, {Benvenuto}, {Maeda}, \& {Nomoto}}]{Folatelli2015}
{Folatelli}, G., {Bersten}, M.~C., {Kuncarayakti}, H., {et~al.} 2015, \apj, 811, 147

\bibitem[{Foreman-Mackey {et~al.}(2013)Foreman-Mackey, Hogg, Lang, \& Goodman}]{Foreman_Mackey_2013}
Foreman-Mackey, D., Hogg, D.~W., Lang, D., \& Goodman, J. 2013, Publications of the Astronomical Society of the Pacific, 125, 306

\bibitem[{{Fransson} \& {Chevalier}(1989)}]{Fransson1989}
{Fransson}, C. \& {Chevalier}, R.~A. 1989, \apj, 343, 323

\bibitem[{{Fransson} {et~al.}(1996){Fransson}, {Lundqvist}, \& {Chevalier}}]{Fransson1996}
{Fransson}, C., {Lundqvist}, P., \& {Chevalier}, R.~A. 1996, \apj, 461, 993

\bibitem[{{Fukugita} {et~al.}(1996){Fukugita}, {Ichikawa}, {Gunn}, {Doi}, {Shimasaku}, \& {Schneider}}]{Fukugita1996}
{Fukugita}, M., {Ichikawa}, T., {Gunn}, J.~E., {et~al.} 1996, \aj, 111, 1748

\bibitem[{Gal-Yam(2017)}]{Gal_Yam_2017}
Gal-Yam, A. 2017, Observational and Physical Classification of Supernovae (Springer International Publishing), 195--237

\bibitem[{{Gal-Yam} {et~al.}(2014){Gal-Yam}, {Arcavi}, {Ofek}, {Ben-Ami}, {Cenko}, {Kasliwal}, {Cao}, {Yaron}, {Tal}, {Silverman}, {Horesh}, {De Cia}, {Taddia}, {Sollerman}, {Perley}, {Vreeswijk}, {Kulkarni}, {Nugent}, {Filippenko}, \& {Wheeler}}]{Gal-Yam2014}
{Gal-Yam}, A., {Arcavi}, I., {Ofek}, E.~O., {et~al.} 2014, \nat, 509, 471

\bibitem[{{Gehrels} {et~al.}(2004){Gehrels}, {Chincarini}, {Giommi}, {Mason}, {Nousek}, {Wells}, {White}, {Barthelmy}, {Burrows}, {Cominsky}, {Hurley}, {Marshall}, {M{\'e}sz{\'a}ros}, {Roming}, {Angelini}, {Barbier}, {Belloni}, {Campana}, {Caraveo}, {Chester}, {Citterio}, {Cline}, {Cropper}, {Cummings}, {Dean}, {Feigelson}, {Fenimore}, {Frail}, {Fruchter}, {Garmire}, {Gendreau}, {Ghisellini}, {Greiner}, {Hill}, {Hunsberger}, {Krimm}, {Kulkarni}, {Kumar}, {Lebrun}, {Lloyd-Ronning}, {Markwardt}, {Mattson}, {Mushotzky}, {Norris}, {Osborne}, {Paczynski}, {Palmer}, {Park}, {Parsons}, {Paul}, {Rees}, {Reynolds}, {Rhoads}, {Sasseen}, {Schaefer}, {Short}, {Smale}, {Smith}, {Stella}, {Tagliaferri}, {Takahashi}, {Tashiro}, {Townsley}, {Tueller}, {Turner}, {Vietri}, {Voges}, {Ward}, {Willingale}, {Zerbi}, \& {Zhang}}]{Gehrels2004}
{Gehrels}, N., {Chincarini}, G., {Giommi}, P., {et~al.} 2004, \apj, 611, 1005

\bibitem[{{Georgy}(2012)}]{Georgy2012}
{Georgy}, C. 2012, \aap, 538, L8

\bibitem[{{Graham} {et~al.}(2019){Graham}, {Kulkarni}, {Bellm}, {Adams}, {Barbarino}, {Blagorodnova}, {Bodewits}, {Bolin}, {Brady}, {Cenko}, {Chang}, {Coughlin}, {De}, {Eadie}, {Farnham}, {Feindt}, {Franckowiak}, {Fremling}, {Gezari}, {Ghosh}, {Goldstein}, {Golkhou}, {Goobar}, {Ho}, {Huppenkothen}, {Ivezi{\'c}}, {Jones}, {Juric}, {Kaplan}, {Kasliwal}, {Kelley}, {Kupfer}, {Lee}, {Lin}, {Lunnan}, {Mahabal}, {Miller}, {Ngeow}, {Nugent}, {Ofek}, {Prince}, {Rauch}, {van Roestel}, {Schulze}, {Singer}, {Sollerman}, {Taddia}, {Yan}, {Ye}, {Yu}, {Barlow}, {Bauer}, {Beck}, {Belicki}, {Biswas}, {Brinnel}, {Brooke}, {Bue}, {Bulla}, {Burruss}, {Connolly}, {Cromer}, {Cunningham}, {Dekany}, {Delacroix}, {Desai}, {Duev}, {Feeney}, {Flynn}, {Frederick}, {Gal-Yam}, {Giomi}, {Groom}, {Hacopians}, {Hale}, {Helou}, {Henning}, {Hover}, {Hillenbrand}, {Howell}, {Hung}, {Imel}, {Ip}, {Jackson}, {Kaspi}, {Kaye}, {Kowalski}, {Kramer}, {Kuhn}, {Landry}, {Laher}, {Mao}, {Masci}, {Monkewitz}, {Murphy}, {Nordin}, {Patterson}, {Penprase},
  {Porter}, {Rebbapragada}, {Reiley}, {Riddle}, {Rigault}, {Rodriguez}, {Rusholme}, {van Santen}, {Shupe}, {Smith}, {Soumagnac}, {Stein}, {Surace}, {Szkody}, {Terek}, {Van Sistine}, {van Velzen}, {Vestrand}, {Walters}, {Ward}, {Zhang}, \& {Zolkower}}]{2019PASP..131g8001G}
{Graham}, M.~J., {Kulkarni}, S.~R., {Bellm}, E.~C., {et~al.} 2019, \pasp, 131, 078001

\bibitem[{{Hachinger} {et~al.}(2012){Hachinger}, {Mazzali}, {Taubenberger}, {Hillebrandt}, {Nomoto}, \& {Sauer}}]{Hachinger2012}
{Hachinger}, S., {Mazzali}, P.~A., {Taubenberger}, S., {et~al.} 2012, \mnras, 422, 70

\bibitem[{Hamuy {et~al.}(2009)Hamuy, Deng, Mazzali, Morrell, Phillips, Roth, Gonzalez, Thomas-Osip, Krzeminski, Contreras, Maza, González, Huerta, Folatelli, Chornock, Filippenko, Persson, Freedman, Koviak, Suntzeff, \& Krisciunas}]{Hamuy2009}
Hamuy, M., Deng, J., Mazzali, P.~A., {et~al.} 2009, The Astrophysical Journal, 703, 1612

\bibitem[{{Harrison} {et~al.}(2013){Harrison}, {Craig}, {Christensen}, {Hailey}, {Zhang}, {Boggs}, {Stern}, {Cook}, {Forster}, {Giommi}, {Grefenstette}, {Kim}, {Kitaguchi}, {Koglin}, {Madsen}, {Mao}, {Miyasaka}, {Mori}, {Perri}, {Pivovaroff}, {Puccetti}, {Rana}, {Westergaard}, {Willis}, {Zoglauer}, {An}, {Bachetti}, {Barri{\`e}re}, {Bellm}, {Bhalerao}, {Brejnholt}, {Fuerst}, {Liebe}, {Markwardt}, {Nynka}, {Vogel}, {Walton}, {Wik}, {Alexander}, {Cominsky}, {Hornschemeier}, {Hornstrup}, {Kaspi}, {Madejski}, {Matt}, {Molendi}, {Smith}, {Tomsick}, {Ajello}, {Ballantyne}, {Balokovi{\'c}}, {Barret}, {Bauer}, {Blandford}, {Brandt}, {Brenneman}, {Chiang}, {Chakrabarty}, {Chenevez}, {Comastri}, {Dufour}, {Elvis}, {Fabian}, {Farrah}, {Fryer}, {Gotthelf}, {Grindlay}, {Helfand}, {Krivonos}, {Meier}, {Miller}, {Natalucci}, {Ogle}, {Ofek}, {Ptak}, {Reynolds}, {Rigby}, {Tagliaferri}, {Thorsett}, {Treister}, \& {Urry}}]{2013ApJ...770..103H}
{Harrison}, F.~A., {Craig}, W.~W., {Christensen}, F.~E., {et~al.} 2013, \apj, 770, 103

\bibitem[{Haynie \& Piro(2023)}]{Haynie_2023}
Haynie, A. \& Piro, A.~L. 2023, The Astrophysical Journal, 956, 98

\bibitem[{{Henden} {et~al.}(2016){Henden}, {Templeton}, {Terrell}, {Smith}, {Levine}, \& {Welch}}]{Henden2016}
{Henden}, A.~A., {Templeton}, M., {Terrell}, D., {et~al.} 2016, {VizieR Online Data Catalog: AAVSO Photometric All Sky Survey (APASS) DR9 (Henden+, 2016)}, VizieR On-line Data Catalog: II/336. Originally published in: 2015AAS...22533616H

\bibitem[{Hiramatsu {et~al.}(2021)Hiramatsu, Howell, Moriya, Goldberg, Hosseinzadeh, Arcavi, Anderson, Gutiérrez, Burke, McCully, Valenti, Galbany, Fang, Maeda, Folatelli, Hsiao, Morrell, Phillips, Stritzinger, Suntzeff, Gromadzki, Maguire, Müller-Bravo, \& Young}]{Hiramatsu_2021}
Hiramatsu, D., Howell, D.~A., Moriya, T.~J., {et~al.} 2021, The Astrophysical Journal, 913, 55

\bibitem[{{Hook} {et~al.}(2004){Hook}, {J{\o}rgensen}, {Allington-Smith}, {Davies}, {Metcalfe}, {Murowinski}, \& {Crampton}}]{Hook2004}
{Hook}, I.~M., {J{\o}rgensen}, I., {Allington-Smith}, J.~R., {et~al.} 2004, \pasp, 116, 425

\bibitem[{{Hosseinzadeh} \& {Gomez}(2020)}]{Hosseinzadeh2020}
{Hosseinzadeh}, G. \& {Gomez}, S. 2020, {Light Curve Fitting}

\bibitem[{{Huang} {et~al.}(2012){Huang}, {Li}, {Wang}, {Shang}, {Zhang}, {Hu}, {Qiu}, \& {Jiang}}]{Huang_2012}
{Huang}, F., {Li}, J.-Z., {Wang}, X.-F., {et~al.} 2012, Research in Astronomy and Astrophysics, 12, 1585

\bibitem[{{Johnson} {et~al.}(1966){Johnson}, {Mitchell}, {Iriarte}, \& {Wisniewski}}]{Johnson1966}
{Johnson}, H.~L., {Mitchell}, R.~I., {Iriarte}, B., \& {Wisniewski}, W.~Z. 1966, Communications of the Lunar and Planetary Laboratory, 4, 99

\bibitem[{{Jordi} {et~al.}(2006){Jordi}, {Grebel}, \& {Ammon}}]{Jordi2006}
{Jordi}, K., {Grebel}, E.~K., \& {Ammon}, K. 2006, \aap, 460, 339

\bibitem[{{Kamble} {et~al.}(2016){Kamble}, {Margutti}, {Soderberg}, {Chakraborti}, {Fransson}, {Chevalier}, {Powell}, {Milisavljevic}, {Parrent}, \& {Bietenholz}}]{Kamble2016}
{Kamble}, A., {Margutti}, R., {Soderberg}, A.~M., {et~al.} 2016, \apj, 818, 111

\bibitem[{Kilpatrick {et~al.}(2022)Kilpatrick, Coulter, Foley, Piro, Rest, Rojas-Bravo, \& Siebert}]{Kilpatrick_2022}
Kilpatrick, C.~D., Coulter, D.~A., Foley, R.~J., {et~al.} 2022, The Astrophysical Journal, 936, 111

\bibitem[{{Kilpatrick} {et~al.}(2017){Kilpatrick}, {Foley}, {Abramson}, {Pan}, {Lu}, {Williams}, {Treu}, {Siebert}, {Fassnacht}, \& {Max}}]{Kilpatrick2017}
{Kilpatrick}, C.~D., {Foley}, R.~J., {Abramson}, L.~E., {et~al.} 2017, \mnras, 465, 4650

\bibitem[{{Kochanek} {et~al.}(2017){Kochanek}, {Shappee}, {Stanek}, {Holoien}, {Thompson}, {Prieto}, {Dong}, {Shields}, {Will}, {Britt}, {Perzanowski}, \& {Pojma{\'n}ski}}]{Kochanek2017}
{Kochanek}, C.~S., {Shappee}, B.~J., {Stanek}, K.~Z., {et~al.} 2017, \pasp, 129, 104502

\bibitem[{{Kumar} {et~al.}(2013){Kumar}, {Pandey}, {Sahu}, {Vinko}, {Moskvitin}, {Anupama}, {Bhatt}, {Ordasi}, {Nagy}, {Sokolov}, {Sokolova}, {Komarova}, {Kumar}, {Bose}, {Roy}, \& {Sagar}}]{Kumar2013}
{Kumar}, B., {Pandey}, S.~B., {Sahu}, D.~K., {et~al.} 2013, \mnras, 431, 308

\bibitem[{{Labrie} {et~al.}(2019){Labrie}, {Anderson}, {C{\'a}rdenes}, {Simpson}, \& {Turner}}]{Labrie2019}
{Labrie}, K., {Anderson}, K., {C{\'a}rdenes}, R., {Simpson}, C., \& {Turner}, J. E.~H. 2019, in Astronomical Society of the Pacific Conference Series, Vol. 523, Astronomical Data Analysis Software and Systems XXVII, ed. P.~J. {Teuben}, M.~W. {Pound}, B.~A. {Thomas}, \& E.~M. {Warner}, 321

\bibitem[{{Law} {et~al.}(2009){Law}, {Kulkarni}, {Dekany}, {Ofek}, {Quimby}, {Nugent}, {Surace}, {Grillmair}, {Bloom}, {Kasliwal}, {Bildsten}, {Brown}, {Cenko}, {Ciardi}, {Croner}, {Djorgovski}, {van Eyken}, {Filippenko}, {Fox}, {Gal-Yam}, {Hale}, {Hamam}, {Helou}, {Henning}, {Howell}, {Jacobsen}, {Laher}, {Mattingly}, {McKenna}, {Pickles}, {Poznanski}, {Rahmer}, {Rau}, {Rosing}, {Shara}, {Smith}, {Starr}, {Sullivan}, {Velur}, {Walters}, \& {Zolkower}}]{2009PASP..121.1395L}
{Law}, N.~M., {Kulkarni}, S.~R., {Dekany}, R.~G., {et~al.} 2009, \pasp, 121, 1395

\bibitem[{{Li} {et~al.}(2024){Li}, {Hu}, {Li}, {Yang}, {Wang}, {Yan}, {Hu}, {Zhang}, {Mao}, {Riise}, {Gao}, {Sun}, {Liu}, {Xiong}, {Wang}, {Mo}, {Iskandar}, {Xi}, {Xiang}, {Wang}, {Sun}, {Zhang}, {Chen}, {Lin}, {Guo}, {Liu}, {Cai}, {Zhou}, {Zhao}, {Chen}, {Zheng}, {Li}, {Zhang}, {Xu}, {Lyu}, {Castro-Tirado}, {Chufarin}, {Potapov}, {Ionov}, {Korotkiy}, {Nazarov}, {Sokolovsky}, {Hamann}, \& {Herman}}]{Li2024}
{Li}, G., {Hu}, M., {Li}, W., {et~al.} 2024, \nat, 627, 754

\bibitem[{Liu {et~al.}(2025)Liu, Zhang, Yu, Du, Li, Wu, \& Dai}]{liu2025}
Liu, L.-D., Zhang, Y.-H., Yu, Y.-W., {et~al.} 2025, TransFit: An Efficient Framework for Transient Light-Curve Fitting with Time-Dependent Radiative Diffusion

\bibitem[{{Lyman} {et~al.}(2016){Lyman}, {Bersier}, {James}, {Mazzali}, {Eldridge}, {Fraser}, \& {Pian}}]{Lyman2016}
{Lyman}, J.~D., {Bersier}, D., {James}, P.~A., {et~al.} 2016, \mnras, 457, 328

\bibitem[{Maeda {et~al.}(2022)Maeda, Chandra, Moriya, Reguitti, Ryder, Matsuoka, Michiyama, Pignata, Hiramatsu, Bostroem, Kundu, Kuncarayakti, Bersten, Pooley, Lee, Patnaude, Rodríguez, \& Folatelli}]{Maeda_2023}
Maeda, K., Chandra, P., Moriya, T.~J., {et~al.} 2022, The Astrophysical Journal, 942, 17

\bibitem[{{Maeda} {et~al.}(2015){Maeda}, {Hattori}, {Milisavljevic}, {Folatelli}, {Drout}, {Kuncarayakti}, {Margutti}, {Kamble}, {Soderberg}, {Tanaka}, {Kawabata}, {Kawabata}, {Yamanaka}, {Nomoto}, {Kim}, {Simon}, {Phillips}, {Parrent}, {Nakaoka}, {Moriya}, {Suzuki}, {Takaki}, {Ishigaki}, {Sakon}, {Tajitsu}, \& {Iye}}]{Maeda2015}
{Maeda}, K., {Hattori}, T., {Milisavljevic}, D., {et~al.} 2015, \apj, 807, 35

\bibitem[{Maeda {et~al.}(2014)Maeda, Katsuda, Bamba, Terada, \& Fukazawa}]{Maeda_2014}
Maeda, K., Katsuda, S., Bamba, A., Terada, Y., \& Fukazawa, Y. 2014, The Astrophysical Journal, 785, 95

\bibitem[{{Maeda} {et~al.}(2012){Maeda}, {Moriya}, {Kawabata}, {Tanaka}, {Tominaga}, \& {Nomoto}}]{Maeda2012}
{Maeda}, K., {Moriya}, T., {Kawabata}, K., {et~al.} 2012, \memsai, 83, 264

\bibitem[{{Matheson} {et~al.}(2000){Matheson}, {Filippenko}, {Ho}, {Barth}, \& {Leonard}}]{Matheson2000}
{Matheson}, T., {Filippenko}, A.~V., {Ho}, L.~C., {Barth}, A.~J., \& {Leonard}, D.~C. 2000, \aj, 120, 1499

\bibitem[{{Maund} {et~al.}(2011){Maund}, {Fraser}, {Ergon}, {Pastorello}, {Smartt}, {Sollerman}, {Benetti}, {Botticella}, {Bufano}, {Danziger}, {Kotak}, {Magill}, {Stephens}, \& {Valenti}}]{Maund2011}
{Maund}, J.~R., {Fraser}, M., {Ergon}, M., {et~al.} 2011, \apjl, 739, L37

\bibitem[{{Maund} \& {Smartt}(2009)}]{Maund2009}
{Maund}, J.~R. \& {Smartt}, S.~J. 2009, Science, 324, 486

\bibitem[{{Maund} {et~al.}(2004){Maund}, {Smartt}, {Kudritzki}, {Podsiadlowski}, \& {Gilmore}}]{Maund2004}
{Maund}, J.~R., {Smartt}, S.~J., {Kudritzki}, R.~P., {Podsiadlowski}, P., \& {Gilmore}, G.~F. 2004, \nat, 427, 129

\bibitem[{{Medler} {et~al.}(2022){Medler}, {Mazzali}, {Teffs}, {Ashall}, {Anderson}, {Arcavi}, {Benetti}, {Bostroem}, {Burke}, {Cai}, {Charalampopoulos}, {Elias-Rosa}, {Ergon}, {Galbany}, {Gromadzki}, {Hiramatsu}, {Howell}, {Inserra}, {Lundqvist}, {McCully}, {M{\"u}ller-Bravo}, {Newsome}, {Nicholl}, {Padilla Gonzalez}, {Paraskeva}, {Pastorello}, {Pellegrino}, {Pessi}, {Reguitti}, {Reynolds}, {Roy}, {Terreran}, {Tomasella}, \& {Young}}]{Medler2022}
{Medler}, K., {Mazzali}, P.~A., {Teffs}, J., {et~al.} 2022, \mnras, 513, 5540

\bibitem[{Medler {et~al.}(2021)Medler, Mazzali, Teffs, Prentice, Ashall, Amenouche, Anderson, Burke, Chen, Galbany, Gromadzki, Gutiérrez, Hiramatsu, Howell, Inserra, Kankare, McCully, Müller-Bravo, Nicholl, Pellegrino, \& Sollerman}]{Medler_2021}
Medler, K., Mazzali, P.~A., Teffs, J., {et~al.} 2021, Monthly Notices of the Royal Astronomical Society, 506, 1832

\bibitem[{Miller \& Stone(1994)}]{miller_stone_1994}
Miller, J. \& Stone, R. 1994, The Kast Double Spectrograph

\bibitem[{{Morag} {et~al.}(2023){Morag}, {Sapir}, \& {Waxman}}]{Morag2023}
{Morag}, J., {Sapir}, N., \& {Waxman}, E. 2023, \mnras, 522, 2764

\bibitem[{{Morales-Garoffolo} {et~al.}(2014){Morales-Garoffolo}, {Elias-Rosa}, {Benetti}, {Taubenberger}, {Cappellaro}, {Pastorello}, {Klauser}, {Valenti}, {Howerton}, {Ochner}, {Schramm}, {Siviero}, {Tartaglia}, \& {Tomasella}}]{Morales2014}
{Morales-Garoffolo}, A., {Elias-Rosa}, N., {Benetti}, S., {et~al.} 2014, \mnras, 445, 1647

\bibitem[{Morales-Garoffolo {et~al.}(2015)Morales-Garoffolo, Elias-Rosa, Bersten, Jerkstrand, Taubenberger, Benetti, Cappellaro, Kotak, Pastorello, Bufano, Domínguez, Ergon, Fraser, Gao, García, Howell, Isern, Smartt, Tomasella, \& Valenti}]{Morales_Garoffolo_2015}
Morales-Garoffolo, A., Elias-Rosa, N., Bersten, M., {et~al.} 2015, Monthly Notices of the Royal Astronomical Society, 454, 95

\bibitem[{{Moriya}(2014)}]{Moriya2014}
{Moriya}, T.~J. 2014, \aap, 564, A83

\bibitem[{Nagy \& Vink\'o(2016)}]{Nagy_2016}
Nagy, A.~P. \& Vink\'o, J. 2016, Astronomy \&amp; Astrophysics, 589, A53

\bibitem[{{Nakar} \& {Piro}(2014)}]{Nakar2014}
{Nakar}, E. \& {Piro}, A.~L. 2014, \apj, 788, 193

\bibitem[{Niu {et~al.}(2024)Niu, Sun, \& Liu}]{Niu_2024}
Niu, Z., Sun, N.-C., \& Liu, J. 2024, The Astrophysical Journal Letters, 970, L9

\bibitem[{Niu {et~al.}(2025)Niu, Sun, Maund, Guo, Li, Sun, \& Liu}]{Niu_2025}
Niu, Z., Sun, N.-C., Maund, J.~R., {et~al.} 2025, Discovery of a variable yellow supergiant progenitor for the Type IIb SN 2024abfo

\bibitem[{{Oke} \& {Gunn}(1983)}]{1983ApJ...266..713O}
{Oke}, J.~B. \& {Gunn}, J.~E. 1983, \apj, 266, 713

\bibitem[{{Orellana} \& {Bersten}(2022)}]{Orellana2022}
{Orellana}, M. \& {Bersten}, M.~C. 2022, \aap, 667, A92

\bibitem[{Ouchi \& Maeda(2017)}]{Ouchi_2017}
Ouchi, R. \& Maeda, K. 2017, The Astrophysical Journal, 840, 90

\bibitem[{{Piro}(2015)}]{Piro_2015}
{Piro}, A.~L. 2015, \apjl, 808, L51

\bibitem[{Piro {et~al.}(2021)Piro, Haynie, \& Yao}]{Piro_2021}
Piro, A.~L., Haynie, A., \& Yao, Y. 2021, The Astrophysical Journal, 909, 209

\bibitem[{Podsiadlowski(1992)}]{Podsiadlowski_1992}
Podsiadlowski, P. 1992, Publications of the Astronomical Society of the Pacific, 104, 717

\bibitem[{{Pogge}(2019)}]{Pogge2019}
{Pogge}, R. 2019, {rwpogge/modsCCDRed 2.0}

\bibitem[{{Pogge} {et~al.}(2010){Pogge}, {Atwood}, {Brewer}, {Byard}, {Derwent}, {Gonzalez}, {Martini}, {Mason}, {O'Brien}, {Osmer}, {Pappalardo}, {Steinbrecher}, {Teiga}, \& {Zhelem}}]{Pogge2010}
{Pogge}, R.~W., {Atwood}, B., {Brewer}, D.~F., {et~al.} 2010, in Society of Photo-Optical Instrumentation Engineers (SPIE) Conference Series, Vol. 7735, Ground-based and Airborne Instrumentation for Astronomy III, ed. I.~S. {McLean}, S.~K. {Ramsay}, \& H.~{Takami}, 77350A

\bibitem[{Prentice {et~al.}(2018)Prentice, Ashall, James, Short, Mazzali, Bersier, Crowther, Barbarino, Chen, Copperwheat, Darnley, Denneau, Elias-Rosa, Fraser, Galbany, Gal-Yam, Harmanen, Howell, Hosseinzadeh, Inserra, Kankare, Karamehmetoglu, Lamb, Limongi, Maguire, McCully, Olivares E, Piascik, Pignata, Reichart, Rest, Reynolds, Rodríguez, Saario, Schulze, Smartt, Smith, Sollerman, Stalder, Sullivan, Taddia, Valenti, Vergani, Williams, \& Young}]{Prentice_2018}
Prentice, S.~J., Ashall, C., James, P.~A., {et~al.} 2018, Monthly Notices of the Royal Astronomical Society, 485, 1559

\bibitem[{Prentice {et~al.}(2016)Prentice, Mazzali, Pian, Gal-Yam, Kulkarni, Rubin, Corsi, Fremling, Sollerman, Yaron, Arcavi, Zheng, Kasliwal, Filippenko, Cenko, Cao, \& Nugent}]{Prentice_2016}
Prentice, S.~J., Mazzali, P.~A., Pian, E., {et~al.} 2016, Monthly Notices of the Royal Astronomical Society, 458, 2973

\bibitem[{{Prochaska} {et~al.}(2020){Prochaska}, {Hennawi}, {Westfall}, {Cooke}, {Wang}, {Hsyu}, {Davies}, {Farina}, \& {Pelliccia}}]{pypeit}
{Prochaska}, J., {Hennawi}, J., {Westfall}, K., {et~al.} 2020, The Journal of Open Source Software, 5, 2308

\bibitem[{{Rabinak} \& {Waxman}(2011)}]{Rabinak2011}
{Rabinak}, I. \& {Waxman}, E. 2011, \apj, 728, 63

\bibitem[{{Ramsey} {et~al.}(1998){Ramsey}, {Adams}, {Barnes}, {Booth}, {Cornell}, {Fowler}, {Gaffney}, {Glaspey}, {Good}, {Hill}, {Kelton}, {Krabbendam}, {Long}, {MacQueen}, {Ray}, {Ricklefs}, {Sage}, {Sebring}, {Spiesman}, \& {Steiner}}]{1998SPIE.3352...34R}
{Ramsey}, L.~W., {Adams}, M.~T., {Barnes}, T.~G., {et~al.} 1998, in Society of Photo-Optical Instrumentation Engineers (SPIE) Conference Series, Vol. 3352, Advanced Technology Optical/IR Telescopes VI, ed. L.~M. {Stepp}, 34--42

\bibitem[{{Rau} {et~al.}(2009){Rau}, {Kulkarni}, {Law}, {Bloom}, {Ciardi}, {Djorgovski}, {Fox}, {Gal-Yam}, {Grillmair}, {Kasliwal}, {Nugent}, {Ofek}, {Quimby}, {Reach}, {Shara}, {Bildsten}, {Cenko}, {Drake}, {Filippenko}, {Helfand}, {Helou}, {Howell}, {Poznanski}, \& {Sullivan}}]{2009PASP..121.1334R}
{Rau}, A., {Kulkarni}, S.~R., {Law}, N.~M., {et~al.} 2009, \pasp, 121, 1334

\bibitem[{Reguitti {et~al.}(2025)Reguitti, Pastorello, Smartt, Valerin, Pignata, Campana, Chen, K., Moran, Mazzali, Duarte, Salmaso, Anderson, Ashall, Benetti, Gromadzki, Gutierrez, Humina, Inserra, Kankare, Kravtsov, Muller-Bravo, Pessi, Sollerman, Young, Chambers, de~Boer, Gao, Huber, Lin, Lowe, Magnier, Minguez, Smith, Smith, Srivastav, Wainscoat, \& Benedet}]{reguitti2025}
Reguitti, A., Pastorello, A., Smartt, S.~J., {et~al.} 2025, SN 2024abfo: a partially stripped SN II from a yellow supergiant

\bibitem[{{Richmond} {et~al.}(1996){Richmond}, {Treffers}, {Filippenko}, \& {Paik}}]{Richmond1996}
{Richmond}, M.~W., {Treffers}, R.~R., {Filippenko}, A.~V., \& {Paik}, Y. 1996, \aj, 112, 732

\bibitem[{{Richmond} {et~al.}(1994){Richmond}, {Treffers}, {Filippenko}, {Paik}, {Leibundgut}, {Schulman}, \& {Cox}}]{Richmond1994}
{Richmond}, M.~W., {Treffers}, R.~R., {Filippenko}, A.~V., {et~al.} 1994, \aj, 107, 1022

\bibitem[{{Roming} {et~al.}(2005){Roming}, {Kennedy}, {Mason}, {Nousek}, {Ahr}, {Bingham}, {Broos}, {Carter}, {Hancock}, {Huckle}, {Hunsberger}, {Kawakami}, {Killough}, {Koch}, {McLelland}, {Smith}, {Smith}, {Soto}, {Boyd}, {Breeveld}, {Holland}, {Ivanushkina}, {Pryzby}, {Still}, \& {Stock}}]{Roming2005}
{Roming}, P. W.~A., {Kennedy}, T.~E., {Mason}, K.~O., {et~al.} 2005, \ssr, 120, 95

\bibitem[{Roming {et~al.}(2009)Roming, Pritchard, Brown, Holland, Immler, Stockdale, Weiler, Panagia, Van~Dyk, Hoversten, Milne, Oates, Russell, \& Vandrevala}]{Roming_2009}
Roming, P. W.~A., Pritchard, T.~A., Brown, P.~J., {et~al.} 2009, The Astrophysical Journal, 704, L118

\bibitem[{Ryder {et~al.}(2018)Ryder, Dyk, Fox, Zapartas, Mink, Smith, Brunsden, Bostroem, Filippenko, Shivvers, \& Zheng}]{Ryder_2018}
Ryder, S.~D., Dyk, S. D.~V., Fox, O.~D., {et~al.} 2018, The Astrophysical Journal, 856, 83

\bibitem[{{Sana} {et~al.}(2012){Sana}, {de Mink}, {de Koter}, {Langer}, {Evans}, {Gieles}, {Gosset}, {Izzard}, {Le Bouquin}, \& {Schneider}}]{Sana2012}
{Sana}, H., {de Mink}, S.~E., {de Koter}, A., {et~al.} 2012, Science, 337, 444

\bibitem[{{Sapir} \& {Waxman}(2017)}]{Waxman2017}
{Sapir}, N. \& {Waxman}, E. 2017, \apj, 838, 130

\bibitem[{{Schlafly} \& {Finkbeiner}(2011)}]{Schlafly_2011}
{Schlafly}, E.~F. \& {Finkbeiner}, D.~P. 2011, \apj, 737, 103

\bibitem[{Shiode \& Quataert(2013)}]{Shiode_2014}
Shiode, J.~H. \& Quataert, E. 2013, The Astrophysical Journal, 780, 96

\bibitem[{{Shivvers} {et~al.}(2019){Shivvers}, {Filippenko}, {Silverman}, {Zheng}, {Foley}, {Chornock}, {Barth}, {Cenko}, {Clubb}, {Fox}, {Ganeshalingam}, {Graham}, {Kelly}, {Kleiser}, {Leonard}, {Li}, {Matheson}, {Mauerhan}, {Modjaz}, {Serduke}, {Shields}, {Steele}, {Swift}, {Wong}, \& {Yuk}}]{Shivvers2019}
{Shivvers}, I., {Filippenko}, A.~V., {Silverman}, J.~M., {et~al.} 2019, \mnras, 482, 1545

\bibitem[{{Silverman} {et~al.}(2009){Silverman}, {Mazzali}, {Chornock}, {Filippenko}, {Clocchiatti}, {Phillips}, {Ganeshalingam}, \& {Foley}}]{Silverman2009}
{Silverman}, J.~M., {Mazzali}, P., {Chornock}, R., {et~al.} 2009, \pasp, 121, 689

\bibitem[{Smith(2014)}]{Smith2014}
Smith, N. 2014, Annual Review of Astronomy and Astrophysics, 52, 487

\bibitem[{Smith(2017)}]{Smith_2017}
Smith, N. 2017, Interacting Supernovae: Types IIn and Ibn (Springer International Publishing), 403--429

\bibitem[{Soderberg {et~al.}(2012)Soderberg, Margutti, Zauderer, Krauss, Katz, Chomiuk, Dittmann, Nakar, Sakamoto, Kawai, Hurley, Barthelmy, Toizumi, Morii, Chevalier, Gurwell, Petitpas, Rupen, Alexander, Levesque, Fransson, Brunthaler, Bietenholz, Chugai, Grindlay, Copete, Connaughton, Briggs, Meegan, von Kienlin, Zhang, Rau, Golenetskii, Mazets, \& Cline}]{Soderberg_2012}
Soderberg, A.~M., Margutti, R., Zauderer, B.~A., {et~al.} 2012, The Astrophysical Journal, 752, 78

\bibitem[{{Soderberg} {et~al.}(2012){Soderberg}, {Margutti}, {Zauderer}, {Krauss}, {Katz}, {Chomiuk}, {Dittmann}, {Nakar}, {Sakamoto}, {Kawai}, {Hurley}, {Barthelmy}, {Toizumi}, {Morii}, {Chevalier}, {Gurwell}, {Petitpas}, {Rupen}, {Alexander}, {Levesque}, {Fransson}, {Brunthaler}, {Bietenholz}, {Chugai}, {Grindlay}, {Copete}, {Connaughton}, {Briggs}, {Meegan}, {von Kienlin}, {Zhang}, {Rau}, {Golenetskii}, {Mazets}, \& {Cline}}]{Soderberg2012}
{Soderberg}, A.~M., {Margutti}, R., {Zauderer}, B.~A., {et~al.} 2012, \apj, 752, 78

\bibitem[{Spergel {et~al.}(2007)Spergel, Bean, Dor{\'e}, Nolta, Bennett, Dunkley, Hinshaw, Jarosik, Komatsu, Page, Peiris, Verde, Halpern, Hill, Kogut, Limon, Meyer, Odegard, Tucker, Weiland, Wollack, \& Wright}]{Spergel_2007}
Spergel, D.~N., Bean, R., Dor{\'e}, O., {et~al.} 2007, The Astrophysical Journal Supplement Series, 170, 377

\bibitem[{{Sravan} {et~al.}(2018){Sravan}, {Marchant}, {Kalogera}, \& {Margutti}}]{Sravan2018}
{Sravan}, N., {Marchant}, P., {Kalogera}, V., \& {Margutti}, R. 2018, \apjl, 852, L17

\bibitem[{Sravan {et~al.}(2020)Sravan, Marchant, Kalogera, Milisavljevic, \& Margutti}]{Sravan_2020}
Sravan, N., Marchant, P., Kalogera, V., Milisavljevic, D., \& Margutti, R. 2020, The Astrophysical Journal, 903, 70

\bibitem[{Steeghs {et~al.}(2022)Steeghs, Galloway, Ackley, Dyer, Lyman, Ulaczyk, Cutter, Mong, Dhillon, O’Brien, Ramsay, Poshyachinda, Kotak, Nuttall, Pallé, Breton, Pollacco, Thrane, Aukkaravittayapun, Awiphan, Burhanudin, Chote, Chrimes, Daw, Duffy, Eyles-Ferris, Gompertz, Heikkilä, Irawati, Kennedy, Killestein, Kuncarayakti, Levan, Littlefair, Makrygianni, Marsh, Mata-Sanchez, Mattila, Maund, McCormac, Mkrtichian, Mullaney, Noysena, Patel, Rol, Sawangwit, Stanway, Starling, Strøm, Tooke, West, White, \& Wiersema}]{Steeghs_2022}
Steeghs, D., Galloway, D.~K., Ackley, K., {et~al.} 2022, Monthly Notices of the Royal Astronomical Society, 511, 2405

\bibitem[{Subrayan {et~al.}(2025)Subrayan, Sand, Bostroem, Jha, Ravi, Schwab, Andrews, Hosseinzadeh, Valenti, Dong, Pearson, Shrestha, Kwok, Hoang, Rho, Park, Yoon, Geball, Haislip, Janzen, Kouprianov, Mehta, Retamal, Reichart, Andrews, Farah, Newsome, Howell, \& McCully}]{Subrayan2025}
Subrayan, B.~M., Sand, D.~J., Bostroem, K.~A., {et~al.} 2025, Early Shock-Cooling Observations and Progenitor Constraints of Type IIb SN 2024uwq

\bibitem[{Sukhbold {et~al.}(2016)Sukhbold, Ertl, Woosley, Brown, \& Janka}]{Sukhbold_2016}
Sukhbold, T., Ertl, T., Woosley, S.~E., Brown, J.~M., \& Janka, H.-T. 2016, The Astrophysical Journal, 821, 38

\bibitem[{{Szalai} {et~al.}(2016){Szalai}, {Vink{\'o}}, {Nagy}, {Silverman}, {Wheeler}, {Dhungana}, {Marion}, {Kehoe}, {Fox}, {S{\'a}rneczky}, {Marschalk{\'o}}, {B{\'\i}r{\'o}}, {Borkovits}, {Heged{\"u}s}, {Szak{\'a}ts}, {Ferrante}, {B{\'a}nyai}, {Hodos{\'a}n}, {Kelemen}, \& {P{\'a}l}}]{Szalai2016}
{Szalai}, T., {Vink{\'o}}, J., {Nagy}, A.~P., {et~al.} 2016, \mnras, 460, 1500

\bibitem[{{Taddia} {et~al.}(2018){Taddia}, {Stritzinger}, {Bersten}, {Baron}, {Burns}, {Contreras}, {Holmbo}, {Hsiao}, {Morrell}, {Phillips}, {Sollerman}, \& {Suntzeff}}]{Taddia2018}
{Taddia}, F., {Stritzinger}, M.~D., {Bersten}, M., {et~al.} 2018, \aap, 609, A136

\bibitem[{{Tartaglia} {et~al.}(2017){Tartaglia}, {Fraser}, {Sand}, {Valenti}, {Smartt}, {McCully}, {Anderson}, {Arcavi}, {Elias-Rosa}, {Galbany}, {Gal-Yam}, {Haislip}, {Hosseinzadeh}, {Howell}, {Inserra}, {Jha}, {Kankare}, {Lundqvist}, {Maguire}, {Mattila}, {Reichart}, {Smith}, {Smith}, {Stritzinger}, {Sullivan}, {Taddia}, \& {Tomasella}}]{Tartaglia2017}
{Tartaglia}, L., {Fraser}, M., {Sand}, D.~J., {et~al.} 2017, \apjl, 836, L12

\bibitem[{{Thorsett} \& {Chakrabarty}(1999)}]{Thorsett1999}
{Thorsett}, S.~E. \& {Chakrabarty}, D. 1999, \apj, 512, 288

\bibitem[{{Tody}(1986)}]{Tody1986}
{Tody}, D. 1986, in Society of Photo-Optical Instrumentation Engineers (SPIE) Conference Series, Vol. 627, Instrumentation in astronomy VI, ed. D.~L. {Crawford}, 733

\bibitem[{{Tody}(1993)}]{Tody1993}
{Tody}, D. 1993, in Astronomical Society of the Pacific Conference Series, Vol.~52, Astronomical Data Analysis Software and Systems II, ed. R.~J. {Hanisch}, R.~J.~V. {Brissenden}, \& J.~{Barnes}, 173

\bibitem[{{Tonry} {et~al.}(2018){Tonry}, {Denneau}, {Heinze}, {Stalder}, {Smith}, {Smartt}, {Stubbs}, {Weiland}, \& {Rest}}]{Tonry2018}
{Tonry}, J.~L., {Denneau}, L., {Heinze}, A.~N., {et~al.} 2018, \pasp, 130, 064505

\bibitem[{Tsvetkov {et~al.}(2012)Tsvetkov, Volkov, Sorokina, Blinnikov, Pavlyuk, \& Borisov}]{tsvetkov2012}
Tsvetkov, D.~Y., Volkov, I.~M., Sorokina, E.~I., {et~al.} 2012, Photometric observations and preliminary modeling of type IIb supernova 2011dh

\bibitem[{{Valenti} {et~al.}(2008){Valenti}, {Elias-Rosa}, {Taubenberger}, {Stanishev}, {Agnoletto}, {Sauer}, {Cappellaro}, {Pastorello}, {Benetti}, {Riffeser}, {Hopp}, {Navasardyan}, {Tsvetkov}, {Lorenzi}, {Patat}, {Turatto}, {Barbon}, {Ciroi}, {Di Mille}, {Frandsen}, {Fynbo}, {Laursen}, \& {Mazzali}}]{Valenti2008}
{Valenti}, S., {Elias-Rosa}, N., {Taubenberger}, S., {et~al.} 2008, \apjl, 673, L155

\bibitem[{{Valenti} {et~al.}(2014){Valenti}, {Sand}, {Pastorello}, {Graham}, {Howell}, {Parrent}, {Tomasella}, {Ochner}, {Fraser}, {Benetti}, {Yuan}, {Smartt}, {Maund}, {Arcavi}, {Gal-Yam}, {Inserra}, \& {Young}}]{Valenti2014}
{Valenti}, S., {Sand}, D., {Pastorello}, A., {et~al.} 2014, \mnras, 438, L101

\bibitem[{Van~Dyk {et~al.}(2014)Van~Dyk, Zheng, Fox, Cenko, Clubb, Filippenko, Foley, Miller, Smith, Kelly, Lee, Ben-Ami, \& Gal-Yam}]{Van_Dyk_2014}
Van~Dyk, S.~D., Zheng, W., Fox, O.~D., {et~al.} 2014, The Astronomical Journal, 147, 37

\bibitem[{{Van Dyk} {et~al.}(2014){Van Dyk}, {Zheng}, {Fox}, {Cenko}, {Clubb}, {Filippenko}, {Foley}, {Miller}, {Smith}, {Kelly}, {Lee}, {Ben-Ami}, \& {Gal-Yam}}]{Dyk2014}
{Van Dyk}, S.~D., {Zheng}, W., {Fox}, O.~D., {et~al.} 2014, \aj, 147, 37

\bibitem[{{Wang} {et~al.}(2019){Wang}, {Bai}, {Fan}, {Mao}, {Chang}, {Xin}, {Zhang}, {Lun}, {Wang}, {Zhang}, {Ying}, {Lu}, {Wang}, {Ji}, {Xiong}, {Yu}, {Ding}, {Ye}, {Xing}, {Yi}, {Xu}, {Zheng}, {Feng}, {He}, {Wang}, {Liu}, {Chen}, {Xu}, {Qin}, {Zhang}, {Tan}, {Li}, {Lou}, {Li}, \& {Liu}}]{Wang2019}
{Wang}, C.-J., {Bai}, J.-M., {Fan}, Y.-F., {et~al.} 2019, Research in Astronomy and Astrophysics, 19, 149

\bibitem[{{Waxman} \& {Katz}(2017)}]{2017hsn..book..967W}
{Waxman}, E. \& {Katz}, B. 2017, in Handbook of Supernovae, ed. A.~W. {Alsabti} \& P.~{Murdin}, 967

\bibitem[{Wheeler {et~al.}(2015)Wheeler, Johnson, \& Clocchiatti}]{Wheeler_2015}
Wheeler, J.~C., Johnson, V., \& Clocchiatti, A. 2015, Monthly Notices of the Royal Astronomical Society, 450, 1295

\bibitem[{{Woosley} {et~al.}(1994){Woosley}, {Eastman}, {Weaver}, \& {Pinto}}]{Woosley1994}
{Woosley}, S.~E., {Eastman}, R.~G., {Weaver}, T.~A., \& {Pinto}, P.~A. 1994, \apj, 429, 300

\bibitem[{{Woosley} {et~al.}(2002){Woosley}, {Heger}, \& {Weaver}}]{Woosley2002}
{Woosley}, S.~E., {Heger}, A., \& {Weaver}, T.~A. 2002, Reviews of Modern Physics, 74, 1015

\bibitem[{{Woosley} {et~al.}(1993){Woosley}, {Langer}, \& {Weaver}}]{Woosley1993}
{Woosley}, S.~E., {Langer}, N., \& {Weaver}, T.~A. 1993, \apj, 411, 823

\bibitem[{{Woosley} {et~al.}(1989){Woosley}, {Pinto}, \& {Hartmann}}]{Woosley1989}
{Woosley}, S.~E., {Pinto}, P.~A., \& {Hartmann}, D. 1989, \apj, 346, 395

\bibitem[{Wu \& Fuller(2020)}]{Wu_2021}
Wu, S. \& Fuller, J. 2020, The Astrophysical Journal, 906, 3

\bibitem[{{Yamanaka} {et~al.}(2025){Yamanaka}, {Nagayama}, \& {Horikiri}}]{Yamanaka2025}
{Yamanaka}, M., {Nagayama}, T., \& {Horikiri}, T. 2025, \pasj [\eprint[arXiv]{2503.05054}]

\bibitem[{{Yaron} \& {Gal-Yam}(2012)}]{Yaron2012}
{Yaron}, O. \& {Gal-Yam}, A. 2012, \pasp, 124, 668

\bibitem[{{Yoon}(2017)}]{Yoon2017a}
{Yoon}, S.-C. 2017, \mnras, 470, 3970

\bibitem[{Yoon \& Cantiello(2010)}]{Yoon_2010}
Yoon, S.-C. \& Cantiello, M. 2010, The Astrophysical Journal Letters, 717, L62

\bibitem[{Yoon {et~al.}(2017)Yoon, Dessart, \& Clocchiatti}]{Yoon_2017}
Yoon, S.-C., Dessart, L., \& Clocchiatti, A. 2017, The Astrophysical Journal, 840, 10

\bibitem[{Yuan {et~al.}(2022)Yuan, Zhang, Chen, \& Ling}]{Yuan2022}
Yuan, W., Zhang, C., Chen, Y., \& Ling, Z. 2022, The Einstein Probe Mission (Springer Nature Singapore), 1--30

\end{thebibliography}

\begin{appendix}




\onecolumn
\section{X-ray and Optical Spectroscopic Observations}\label{A}

\renewcommand{\thetable}{A.\arabic{table}}
\setcounter{table}{0}

\begin{table*}[h]
\centering
\caption{The X-ray Observations of SN\,2024iss}\label{table:Log of X-ray observation}
\begin{tabular}{cccccc}
\hline
Instrument & Start time & End time & Exposure & $\rm{Phase}$\tablefootmark{a} \\ 
& (UTC) & (UTC) & (second) & (day)\\
\hline
Swift-XRT & 2024-05-13 18:18:31.00 & 2024-05-15 22:34:53.00 & 4939.62 & 1.95 $\pm$ 1.09  \\
EP-FXT & 2024-05-14 23:21:13.65 & 2024-05-15 08:08:30.80 & 17375.48 & 2.26 $\pm$ 0.18  \\
Swift-XRT & 2024-05-16 07:54:00.00 & 2024-05-19 07:29:53.00 & 5721.27 & 4.92 $\pm$ 1.49  \\
NuSTAR & 2024-05-17 16:11:09.00 & 2024-05-18 12:51:09.00 & 37069.35 & 5.20 $\pm$ 0.43  \\
Swift-XRT & 2024-05-21 03:10:22.00 & 2024-05-25 19:42:53.00 & 7394.45 & 10.58 $\pm$ 2.34  \\
EP-FXT & 2024-05-23 11:08:18.67 & 2024-05-23 15:10:13.84 & 8924.59 & 10.65 $\pm$ 0.08  \\
NuSTAR & 2024-06-02 00:21:09.00 & 2024-06-03 03:01:09.00 & 50139.94 & 20.67 $\pm$ 0.56  \\
Swift-XRT & 2024-06-08 08:58:29.00 & 2024-06-23 05:45:53.00 & 20605.16 & 33.91 $\pm$ 7.43  \\
Swift-XRT & 2024-11-13 17:42:04.00 & 2024-12-26 15:30:53.00 & 11982.10 & 206.29 $\pm$ 21.45  \\
Swift-XRT & 2025-02-06 19:42:21.00 & 2025-02-20 12:26:54.00 & 6458.06 & 276.77 $\pm$ 6.85  \\
\hline
\end{tabular}
\tablefoot{
\tablefoottext{a}{Time respect to MJD=60,442.21. }
}
\end{table*}

\begin{table*}[h]
\centering
\caption{Spectral property from X-ray Observation of SN\,2024iss}\label{table:Log of X-ray spectral parameter}
\begin{tabular}{ccccccc}
\hline
Instrument & $\rm{Phase}$\tablefootmark{a} & $N_{int}$ & kT & Norm & Flux\tablefootmark{b} &
Unabsorbed Flux\tablefootmark{b}  \\ 
 & (day) & ($10^{22} cm^{-2}$) & (keV) & $10^{-4} cm^{-5}$ & ($erg/cm^{2}/s$) & ($erg/cm^{2}/s$)\\
\hline
Swift-XRT & 1.95 $\pm$ 1.09 & 0.009 & 32.57 & $7.75^{\pm 0.58}_{\pm 0.55}$ & $1.44^{\pm 0.11}_{\pm 0.10} \times 10^{-12}$ & $1.43^{\pm 0.11}_{\pm 0.10} \times 10^{-12}$\\
EP-FXT & 2.26 $\pm$ 0.18 & 0.009 & 32.57 & $7.19^{\pm 0.18}_{\pm 0.18}$ & $1.29^{\pm 0.03}_{\pm 0.03} \times 10^{-12}$ & $1.28^{\pm 0.03}_{\pm 0.03} \times 10^{-12}$\\
Swift-XRT & 4.92 $\pm$ 1.49 & 0.009 & 32.57 & $4.57^{\pm 0.43}_{\pm 0.40}$ & $8.50^{\pm 0.79}_{\pm 0.75} \times 10^{-13}$ & $8.43^{\pm 0.78}_{\pm 0.74} \times 10^{-13}$\\
NuSTAR & 5.20 $\pm$ 0.43 & 0.009 & $32.57^{\pm 6.93}_{\pm 4.65}$ & $7.91^{\pm 0.23}_{\pm 0.23}$ & $2.85^{\pm 0.27}_{\pm 0.21} \times 10^{-12}$ & $2.85^{\pm 0.27}_{\pm 0.21} \times 10^{-12}$\\
Swift-XRT & 10.58 $\pm$ 2.34 & 0.009 & 32.57 & $1.52^{\pm 0.52}_{\pm 0.43}$ & $2.83^{\pm 0.96}_{\pm 0.79} \times 10^{-13}$ & $2.80^{\pm 0.95}_{\pm 0.78} \times 10^{-13}$\\
EP-FXT & 10.65 $\pm$ 0.08 & 0.009 & 32.57 & $0.97^{\pm 0.11}_{\pm 0.10}$ & $1.73^{\pm 0.20}_{\pm 0.19} \times 10^{-13}$ & $1.73^{\pm 0.19}_{\pm 0.18} \times 10^{-13}$\\
NuSTAR & 20.67 $\pm$ 0.56 & 0.009 & 32.57 & $0.69^{\pm 0.09}_{\pm 0.09}$ & $2.49^{\pm 0.33}_{\pm 0.32} \times 10^{-13}$ & $2.49^{\pm 0.33}_{\pm 0.32} \times 10^{-13}$\\
Swift-XRT & 33.91 $\pm$ 7.43 & 0.009 & 32.57 & $ < 0.24 $ & $ < 4.83 \times 10^{-14} $ & $ < 4.79 \times 10^{-14} $\\
Swift-XRT & 206.29 $\pm$ 21.45 & 0.009 & 32.57 & $ < 10.95 $ & $ < 2.04 \times 10^{-12} $ & $ < 2.02 \times 10^{-12} $\\
Swift-XRT & 276.77 $\pm$ 6.85 & 0.009 & 32.57 & $ < 0.29 $ & $ < 5.35 \times 10^{-14} $ & $ < 5.31 \times 10^{-14} $\\
\hline
\end{tabular}
\tablefoot{
\tablefoottext{a}{Time respect to MJD=60,442.21.}\\
\tablefoottext{b}{The absorbed flux and unabsorbed flux are in 0.5-10 keV for FXT, 0.3-10 keV for XRT, and 3-79 keV for NuSTAR. We adopt a fixed plasma temperature to the temperature estimated from the 4th observation in the column
 for all other observations.
Because of the limited photon counts and the restricted energy coverage
of FXT (0.5-10 keV), the plasma temperature (kT) of
SN\,2024iss at each epoch cannot be reliably constrained
independently (see Section~\ref{sec:X-ray_Observations}).
}
}
\end{table*}

\onecolumn

\begin{longtable}{lllllll}
\caption{Log of Optical Spectroscopy of SN\,2024iss}\\
\label{table:Log of spectra}

UT & MJD & Phase\footnotemark{} & Telescope & Instrument & Exposure Time (s) & Range (\AA)\\ 
\hline
\endfirsthead
\caption{Continued}\\
\hline
UT & MJD & Phase & Telescope & Inst. & Exp.Time (s) & Range (\AA) \\ 
\hline
\endhead
\hline
\endfoot
2024-05-13 12:37 & 60443.53 & 1.32 & XLT & BFOSC & 2100 & 3769-8911\\
2024-05-14 03:17 & 60444.14 & 1.93 & MMT & Binospec & 1800 & 3823-9214\\
2024-05-14 11:53 & 60444.50 & 2.29 & FTS & FLOYDS & 900 & 3300-10180\\
2024-05-14 14:00 & 60444.58 & 2.37 & XLT & BFOSC & 2100 & 3770-8909\\
2024-05-14 16:52 & 60444.70 & 2.49 & LJT & YFOSC & 600 & 3616-8920\\
2024-05-15 08:17 & 60445.35 & 3.14 & Lick & Kast & 300 & 3636-10752\\
2024-05-15 13:09 & 60445.55 & 3.34 & XLT & BFOSC & 1800 & 3776-8917\\
2024-05-16 00:44 & 60446.03 & 3.82 & TNG & LRS & 300 & 3450-10510\\
2024-05-16 17:05 & 60446.71 & 4.50 & LJT & YFOSC & 600 & 3615-8920\\
2024-05-17 12:44 & 60447.53 & 5.32 & XLT & BFOSC & 1800 & 3770-8909\\
2024-05-17 21:08 & 60447.88 & 5.67 & GT & B\&C & 1800 & 5775-6975\\
2024-05-17 22:54 & 60447.95 & 5.74 & GT & B\&C & 1200 & 3315-7889\\
2024-05-18 18:48 & 60448.78 & 6.57 & LJT & YFOSC & 900 & 3612-8917\\
2024-05-20 12:57 & 60450.54 & 8.33 & XLT & BFOSC & 1800 & 3770-8909\\
2024-05-22 20:36 & 60452.86 & 10.65 & GT & B\&C & 1800 & 3314-7889\\
2024-05-24 10:05 & 60454.42 & 12.21 & FTN & FLOYDS & 600 & 3300-10180\\
2024-05-24 18:07 & 60454.76 & 12.55 & LJT & YFOSC & 900 & 3613-8918\\
2024-05-25 22:26 & 60455.93 & 13.72 & GT & B\&C & 1800 & 3314-7887\\
2024-05-26 08:06 & 60456.34 & 14.13 & FTN & FLOYDS & 600 & 3300-10180\\
2024-05-26 12:34 & 60456.52 & 14.31 & XLT & BFOSC & 2100 & 3776-8919\\
2024-05-26 21:49 & 60456.91 & 14.70 & GT & B\&C & 1200 & 3316-7889\\
2024-05-29 06:30 & 60459.27 & 17.06 & HET & LRS2 & 1400 & 3640-10500\\
2024-05-29 07:26 & 60459.31 & 17.10 & MMT & Binospec & 1350 & 3826-9215\\
2024-05-30 07:32 & 60460.31 & 18.11 & LBT & MODS1B & 2400 & 3200-8918\\
2024-05-30 13:58 & 60460.58 & 18.37 & XLT & BFOSC & 1800 & 3774-8911\\
2024-06-01 06:38 & 60462.28 & 20.07 & FTN & FLOYDS & 600 & 3300-10180\\
2024-06-01 16:15 & 60462.68 & 20.47 & LJT & YFOSC & 900 & 3615-8920\\
2024-06-01 23:00 & 60462.96 & 20.75 & GT & B\&C & 900 & 3316-7890\\
2024-06-04 07:00 & 60465.29 & 23.08 & MMT & Binospec & 1500 & 3823-9215\\
2024-06-04 21:14 & 60465.89 & 23.68 & GT & B\&C & 1200 & 3317-7891\\
2024-06-05 10:12 & 60466.43 & 24.22 & FTN & FLOYDS & 600 & 3300-10180\\
2024-06-06 09:00 & 60467.38 & 25.17 & Lick & Kast & 300 & 3636-10750\\
2024-06-11 05:15 & 60472.22 & 30.01 & Bok & B\&C Spec & 2400 & 4000-8000\\
2024-06-11 07:13 & 60472.30 & 30.09 & FTN & FLOYDS & 600 & 3300-10180\\
2024-06-11 16:41 & 60472.70 & 30.49 & LJT & YFOSC & 1000 & 3612-8918\\
2024-06-13 08:41 & 60474.36 & 32.15 & Lick & Kast & 400 & 3626-10756\\
2024-06-14 & 60475.00 & 32.79 & GNT & GMOS & 240 & 3300-10180\\
2024-06-16 21:10 & 60477.88 & 35.67 & GT & B\&C & 1800 & 3317-7889\\
2024-06-17 06:31 & 60478.27 & 36.06 & FTN & FLOYDS & 900 & 3300-10180\\
2024-06-22 & 60483.00 & 40.79 & GNT & GMOS & 240 & 3300-10180\\
2024-06-26 08:08 & 60487.34 & 45.13 & FTN & FLOYDS & 900 & 3300-10180\\
2024-06-26 21:02 & 60487.88 & 45.67 & GT & B\&C & 1200 & 3316-7890\\
2024-06-28 13:42 & 60489.57 & 47.36 & XLT & BFOSC & 2100 & 3777-8907\\
2024-06-30 07:23 & 60491.31 & 49.10 & FTN & FLOYDS & 900 & 3300-10180\\
2024-07-04 06:42 & 60495.28 & 53.07 & FTN & FLOYDS & 900 & 3300-10180\\
2024-07-08 07:46 & 60499.32 & 57.11 & FTN & FLOYDS & 900 & 3300-10180\\
2024-07-12 & 60503.00 & 60.79 & GNT & GMOS & 240 & 3300-10180\\
2024-07-12 07:37 & 60503.32 & 61.11 & FTN & FLOYDS & 900 & 3300-10180\\
2024-07-14 21:37 & 60505.90 & 63.69 & GT & B\&C & 1800 & 3318-7889\\
2024-07-16 06:54 & 60507.29 & 65.08 & FTN & FLOYDS & 900 & 3300-10180\\
2024-07-26 06:01 & 60517.25 & 75.04 & FTN & FLOYDS & 1800 & 3300-10180\\
2024-08-02 04:25 & 60524.18 & 81.97 & Lick & Kast & 1200 & 3624-10720\\
2024-08-03 06:21 & 60525.26 & 83.05 & FTN & FLOYDS & 1800 & 3300-10180\\
2024-08-07 06:02 & 60529.25 & 87.04 & FTN & FLOYDS & 1800 & 3300-10180\\
\end{longtable}
\tablefoot{
\tablefoottext{a}{Time respect to MJD=60,442.21.}\\
}

\end{appendix}
\end{document}